\documentclass[aps,prb,twocolumn,amsmath,footinbib,amssymb,superscriptaddress,longbibliography]{revtex4-2}
\usepackage{graphicx}
\usepackage{braket}
\usepackage{mathrsfs}
\usepackage{hyperref}
\usepackage{balance}
\usepackage{color}
\usepackage{float}
\usepackage[caption=false]{subfig}

\begin{document}

\title{Quantum Inductance as a Phase-Sensitive Probe of Fermion Parity Switching in Majorana Nanowires}

\author{Binayyak B. Roy}
\affiliation{Department of Physics and Astronomy, Clemson University, Clemson, SC 29634, USA}

\author{Jay D. Sau}
\affiliation{Department of Physics, University of Maryland, College Park, MD 20742, USA}

\author{Sumanta Tewari}
\affiliation{Department of Physics and Astronomy, Clemson University, Clemson, SC 29634, USA}

\begin{abstract}
We study the flux-dependent quantum inductance of a one-dimensional (1D) semiconductor–superconductor (SM–SC) Majorana nanowire coupled to a quantum dot in an interferometric setup. Although quantum capacitance in this setup enables fast fermion parity readout, as has been demonstrated experimentally, it cannot by itself reliably confirm a protected fermion parity switch, a key signature of non-trivial topology and the existence of Majorana zero modes (MZMs). In realistic devices, disorder can produce avoided crossings or narrow double crossings between the two parity sectors that can mimic the behavior of a protected single parity switching, leading to false positives for non-trivial topological behavior. We show that quantum inductance provides a complementary probe that is directly sensitive to the phase structure of the low energy spectrum, allowing us to distinguish genuine fermion-parity crossings from avoided crossings or narrow double crossings. Using a general Lehmann framework applied to both effective models and full microscopic simulations with disorder, we demonstrate that only a true fermion-parity switch produces the characteristic inductive response of a protected crossing. In contrast, topologically trivial avoided crossings or narrow double crossings yield quantum inductance signatures that are markedly different from those of topologically nontrivial fermion parity crossings. Therefore, our results show that combined measurements of quantum capacitance and quantum inductance provide a robust and experimentally accessible means to identify true fermion-parity switches, corresponding to a nontrivial Pfaffian invariant.

\end{abstract}

\maketitle


\begin{figure}
    \centering
    \hspace*{-0.5cm}
    \includegraphics[width=1.1\linewidth]{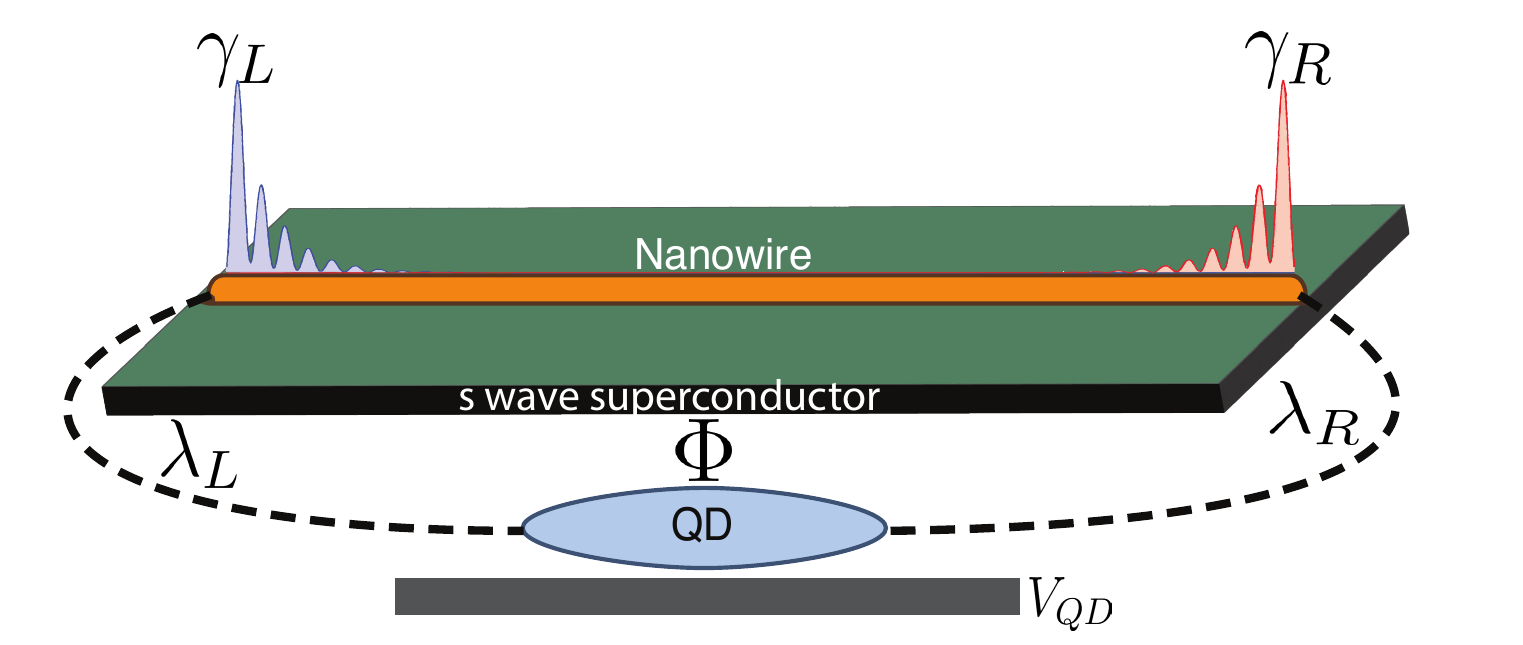}
    \caption{Schematic representation of the semiconductor nanowire in proximity to an $s$-wave superconductor, with both ends of the wire coupled to a quantum dot (QD). The effective coupling to the left and right ends of the nanowire is controlled by the constants $\lambda_L$ and $\lambda_R$, respectively. A back-gate electrode tunes the dot potential $V_{QD}$, while a magnetic flux $\Phi$ threads the loop formed between the nanowire and the QD. In the topological regime, the nanowire hosts a pair of Majorana modes, $\gamma_L$ and $\gamma_R$, that are localized near its opposite ends.}
    \label{fig:SchematicSetup}
\end{figure}

\section{Introduction}
Majorana zero modes (MZMs) have long been proposed as building blocks for topological quantum information processing, owing to their nonlocal fermion parity degree of freedom and the associated protection against local perturbations\cite{Kitaev2003, Nayak2008, wilczek1982quantum, Moore1991, Read2000, Nayak1996}. In the realization involving spin-orbit coupled semiconductor-superconductor (SM-SC) heterostructure, an effectively spinless superconducting phase can be engineered in a Rashba nanowire proximitized by an $s$-wave superconductor under an applied Zeeman field, producing end localized MZMs beyond a critical field set by the induced pairing and chemical potential \cite{sau2010generic, sau2010non, oreg2010helical, lutchyn2010majorana}.
The corresponding nonlocal fermion parity, encoded by the occupancy of a complex fermion formed from two spatially separated Majorana operators, is the fundamental qubit variable; the ability to initialize, manipulate, and read out this parity in an experimentally viable architecture is therefore central to Majorana based approaches to fault-tolerant quantum computing\cite{Kitaev2003}. Because of its conceptual simplicity and its potential for transformative technological applications, the proposal has spurred tremendous experimental activity in the past few years \cite{mourik2012signatures, Deng2012, Das2012, rokhinson2012fractional, churchill2013superconductor, finck2013anomalous, deng2016majorana, zhang2017ballistic, chen2017experimental, nichele2017scaling, albrecht2017transport, o2018hybridization, shen2018parity, 
sherman2017normal, vaitiekenas2018selective, albrecht2016exponential, Yu_2021, zhang2021, kells2012near, PhysRevB.107.245423}.

\begin{figure*}
    \centering
    \includegraphics[width=\textwidth]{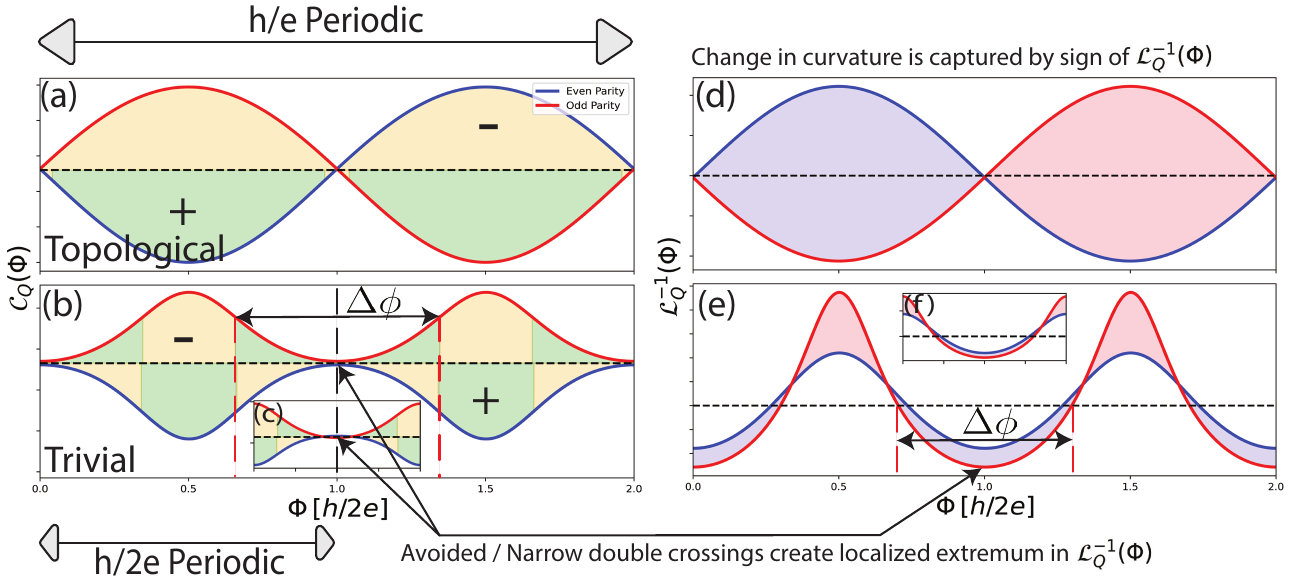}
    \caption{Flux dependence of quantum capacitance and quantum inductance in low energy effective models.(a) $h/e$-periodic quantum capacitance $\mathcal{C}(\Phi)$ for a pair of topological MZMs (model~1 in Sec.~\ref{Model1}), where the ground state switches fermion parity at $\Phi=h/2e$ (even and odd parity branches shown in blue and red, respectively). (b) $h/2e$-periodic $\mathcal{C}(\Phi)$ in the trivial regime for a pair of Andreev bound states localized near the nanowire ends (model~2 in Sec.~\ref{Model2}), illustrating an avoided crossing near $\Phi=h/2e$. (c) Same as (b), but for a trivial narrow double crossing scenario near $\Phi=h/2e$. (d)-(f) Corresponding quantum inductance curves $\mathcal{L}^{-1}(\Phi)$ for the cases shown in (a)-(c), respectively, highlighting the response of the phase derivative operators to the flux driven evolution of the low energy spectrum. In (a)-(c), shaded regions indicate the sign of the local curvature of $\mathcal{C}(\Phi)$ (green: positive; yellow: negative). The phase window $\Delta\phi$ marks the avoided and narrow double crossing region in (b) and (c), which produces a extrema in $\mathcal{L}^{-1}(\Phi)$ in (e) and (f), respectively.}
    \label{fig:schematic_transport_curve}
\end{figure*}

A central challenge is that trivial low energy Andreev bound states (ABSs) can reproduce many qualitative signatures\cite{Mi2014, bagrets2012class, pikulin2012zero, prada2012transport, moore2018two, vuik2018reproducing, stanescu2019robust, added_Loss_2018prb_abs, san2016majorana, ramon2019nonhermitian, Jorge2019supercurrent, ramon2020from, Jorge2021distinguishing} expected of topological MZMs, particularly in realistic devices with disorder, smooth confinement, spatially inhomogeneous potentials, or quantum dot-like regions at the wire ends \cite{PRBKells2012,PRBLiu2017,Moore2018, PRBMoore2018,SciPostVuik2019,PRRPan2020,PRBZeng2022}.
Such ``quasi-Majorana'' or partially separated MZMs may remain near zero energy over extended parameter ranges and can exhibit wavefunction weight near both ends of the device, thereby masquerading as topological Majoranas in probes that are primarily sensitive to low energy spectral weight near the wire ends \cite{SciPostVuik2019,PRRPan2020,PRBZeng2022}. Consequently, establishing the presence of a pair of topologically robust MZMs, and the associated non-local fermion parity switch as a function of an external flux remain non-trivial even when low energy modes with properties resembling those of MZMS are detected in local tunneling experiments at the ends of the wire \cite{ARCMPBeenakker2013,RPPAlicea2012}.


Recent interferometric nanowire devices have introduced flux dependent radio frequency measurements of the quantum capacitance as a promising route for fast fermion parity readout \cite{arXivMicrosoft2024,MSRMicrosoft2024,PubMedMicrosoft2025}.
In this geometry, the two ends of the proximitized nanowire are coupled through a quantum dot (QD) to form a loop threaded by magnetic flux [Fig.~\ref{fig:SchematicSetup}], so that the low energy spectrum becomes a joint function of the flux induced phase $\phi$ and the QD gate potential $V_{QD}$ \cite{arXivMicrosoft2024,PubMedMicrosoft2025}.

In the topological regime, the low energy sector can be described in terms of Majorana end modes ($\gamma_1,\gamma_2$) whose coupling through the QD produces a pair of flux tunable Andreev bound states (ABSs); the occupation of these ABSs distinguishes the even ($c^\dagger\ket{0}=\frac{1}{2}(\gamma_1+i\gamma_2)\ket{0}$) and odd fermion ($c\ket{0}=\frac{1}{2}(\gamma_1-i\gamma_2)\ket{0}$) parity sectors, and tuning flux can drive parity dependent spectral rearrangements \cite{arXivMicrosoft2024,PubMedMicrosoft2025,aasen2025topologicallyfaulttolerantquantumcomputer}.
The measured QD quantum capacitance acquires a parity dependent contribution because it probes the gate curvature of the many body energy, which is a joint function of the flux induced phase $\phi$ and the gate potential $V_{QD}$. Specifically, fixing the gate potential to a resonant value ($V_{QD}=V_{QD}^{(\mathrm{reso})}$) in $\mathcal{C}\propto -\partial^2 E(\phi,V_{QD})/\partial V_{QD}^2$, with $E=\sum_{E_n<0}E_n$ makes it primarily a function of $\phi$ and therefore encodes the flux dependence of the low energy spectrum \cite{arXivMicrosoft2024,PubMedMicrosoft2025}.
In the regime where a Majorana mediated hybridization yields an effectively $h/e$ response in the low energy sector, the even and odd parity traces are shifted relative to one another by approximately half a period, enabling single-shot parity discrimination at microsecond timescales \cite{arXivMicrosoft2024,PubMedMicrosoft2025}.
These results represent an important step toward nondestructive parity readout in high-quality InAs-Al hybrid platforms \cite{arXivMicrosoft2024,PubMedMicrosoft2025,ScienceMourik2012,NatPhysDas2012,NatureAlbrecht2016,PRLNichele2017,PRBSetiawan2017,PRLNichele2020}.

However, flux dependent capacitance oscillations do not, by themselves, uniquely identify topological MZMs. In particular, microscopic modeling shows that disordered and inhomogeneous nanowires can produce parity dependent quantum capacitance phenomenology that resembles the interferometric ``Majorana'' signal even when the low energy modes are overlapping \cite{PRBSau2025,arXivStanescuTewari2025}. Even in the presence of pronounced $h/e$ periodic components, the quantum capacitance based parity signal is not, by itself, a definitive indicator of a topological phase. Moreover, in the interferometric measurements, fermion parity is not accessed as an independent observable; instead, it is inferred from a bimodal quantum capacitance readout, with two discrete capacitance levels associated with the two parity sectors at a given flux \cite{arXivMicrosoft2024,PubMedMicrosoft2025,MSRMicrosoft2024}. As a result, when the readout is not explicitly conditioned on the parity assignment, the flux dependence reverts to an effectively $h/2e$-periodic profile, and interpreting crossing-like features in the capacitance traces as true parity protected crossings can lead to false positives. For instance, as shown in Fig.~\ref{fig:schematic_transport_curve}(a), in the topological regime with the Majorana zero modes, the flux-dependent capacitance, which qualitatively follows the behavior of $E$ with $\Phi$, has two possible values in the two parity sectors and the ground state fermion parity switches at $\Phi\sim h/2e$. However, if the parity associated with each curve in Fig.~\ref{fig:schematic_transport_curve}(a) is not known, as is the case in recent experiments \cite{MSRMicrosoft2024,arXivMicrosoft2024,PubMedMicrosoft2025}, then the capacitance curves showing an avoided crossing at $\Phi\sim h/2e$, or a narrow double crossing at the same flux (refer to Fig.~\ref{fig:schematic_transport_curve}(b)), will experimentally appear indistinguishable from those with a protected fermion parity switch. Therefore, since avoided crossings and narrow double crossings can occur in the topologically trivial regime, quantum capacitance measurements without parity identification can in principle produce false positives, by incorrectly identifying topologically trivial phase with no Majorana zero modes with non-trivial topology. This motivates the need for additional phase sensitive diagnostics capable of distinguishing true parity protected level crossings from avoided or narrow double crossings. \cite{PRBKells2012,PRBLiu2017,PRBMoore2018,SciPostVuik2019,PRRPan2020,PRBZeng2022}.

We introduce flux dependent quantum inverse inductance\cite{RMPNayak2008}, or, equivalently the reactive part of quantum impedance\cite{PRTewari2012} as a complementary observable that substantially strengthens interferometric diagnostics. Whereas quantum capacitance primarily probes the flux dependence of the low energy spectrum, the quantum inductance is governed by the derivatives of the low energy spectrum with flux $\partial_{\phi}H$ and $\partial_{\phi}^2 H$ and is therefore intrinsically sensitive to how the curvature of the low energy spectrum evolves with flux (see Sec.~\ref{Transport}). For clarity and consistency, we refer to this quantity simply as the quantum inductance throughout the remainder of this work.

Fig.~\ref{fig:schematic_transport_curve} provides a schematic illustration of this complementarity: it summarizes in addition to representative quantum capacitance curves in topological and trivial regimes, including an effectively $h/e$-periodic response with a true crossing and an $h/2e$-periodic response exhibiting either an avoided crossing or a narrow double crossing, the corresponding quantum inductance curves in Figs.~\ref{fig:schematic_transport_curve}(d,e,f). In the case of a true crossing [Fig.~\ref{fig:schematic_transport_curve}(a)], the parity-resolved capacitance traces undergo a single change in curvature across the crossing, which corresponds to a simple sign change (or analogously a crossing between different parity sectors) in the quantum inductance curves at the same flux value [Fig.~\ref{fig:schematic_transport_curve}(d)]. By contrast, for an avoided crossing [Fig.~\ref{fig:schematic_transport_curve}(b)] or a narrow double crossing [Fig.~\ref{fig:schematic_transport_curve}(c)], the curvature changes twice at a pair of inflection points within a narrow flux window, leading to an extremum in the quantum inductance [Fig.~\ref{fig:schematic_transport_curve}(e) and Fig.~\ref{fig:schematic_transport_curve}(f), respectively]. This distinction makes quantum inductance particularly effective at distinguishing protected fermion parity switches as a function of flux, indicating the presence of Majorana zero modes, from avoided crossings and narrow double crossings of the quantum capacitance, that may appear in a topologically trivial phase. We show that even though quantum capacitance curves may look similar in these two cases, the quantum inductance curves exhibit distinctive extrema at avoided crossings and narrow double crossings, enabling unambiguous identification of the topological superconducting phase. 

It is worth emphasizing that phase sensitive electrical response measurements, including quantum admittance and inductance spectroscopy, have a long history in superconducting weak links and superconductor-insulator-superconductor (SIS) junctions \cite{McCumber_1968, PRFu2009, PRWoerkom2017}, where they have been used to probe Andreev bound states, Josephson dynamics, and parity dependent excitations. Experimentally, the relevant observable is the reactive (imaginary) part of the AC admittance extracted from radio-frequency reflectometry measurements\cite{VigneauAPR2023, KurilovichPRB2021}. In the low frequency limit, this reactive response contains both capacitive and inductive contributions, with the latter corresponding to an effective quantum inductance of the device. The quantity computed in this work, $\mathcal{L}^{-1}(\Phi)$, is directly proportional to the inductive component of the linear response admittance and therefore captures the same flux-dependent features accessible in experiment.  

Quantum inductance thus represents a natural extension of the current experimental approaches, as it probes the complementary phase response of the same low energy spectrum using closely related measurement techniques. Despite this, the role of quantum inductance as a diagnostic of fermion parity switching in the specific nanowire-QD interferometric geometry has not been systematically explored. In this paper, using analytically tractable effective models and full microscopic Bogoliubov-de Gennes simulations incorporating spin-orbit coupling, Zeeman field, induced superconductivity, disorder, and a flux tunable QD link, we demonstrate that avoided and narrow double crossings generically generate pronounced extremum features in the quantum inductance curves near the putative crossing flux values, while true crossings yield inductance curves that cross between parity sectors and remain consistent with the $h/e$ parity resolved structure required for protected parity switching.

The remainder of this paper is organized as follows. In Sec.~\ref{sec:eff_model}, we introduce analytically tractable effective models that illustrate how the flux dependence of the low energy spectrum manifests in the quantum capacitance and quantum inductance responses, and establish the characteristic signatures distinguishing true fermion parity–protected crossings from avoided or narrow double crossings. In Sec.~\ref{Transport}, we present the microscopic Bogoliubov-de Gennes model for the SM-SC nanowire coupled to a QD and summarize the linear-response expressions used to compute the quantum capacitance and quantum inductance in the numerical calculations. The results of the microscopic simulations are discussed in Sec.~\ref{sec:results}, where we analyze the flux-dependent responses for clean, weakly disordered, and strongly disordered nanowires, as well as systems with smooth Gaussian confinement. These calculations demonstrate that the combined behavior of quantum capacitance and quantum inductance provides a robust diagnostic for distinguishing genuine fermion parity switches from trivial hybridization effects across a broad range of physical regimes. Additional parameter-space analyses and supporting results are presented in the Appendix.


\section{Effective Models}
\label{sec:eff_model}

Before turning to the microscopic calculations, it is useful to illustrate the essential physical mechanisms using minimal effective models.

\subsection{Model 1: Ideal Majorana zero modes coupled to quantum dot}
\label{Model1}

To model the topological regime, we consider the minimal low energy description of a proximitized nanowire hosting a pair of spatially separated MZMs at its ends. In an interferometric nanowire-QD geometry (see Fig.~\ref{fig:SchematicSetup}), the QD provides a local probe that couples to both Majorana operators $\gamma_L$ and $\gamma_R$ through flux dependent tunneling amplitudes. The low energy physics is therefore governed by three ingredients: (i) a small Majorana splitting $\delta$ arising from finite overlap across the wire, (ii) a tunable QD level $\delta_{QD}$ controlled by the gate potential, and (iii) coherent tunneling processes through the left and right ends of the wire, whose relative phase ($\phi$) is set by the applied magnetic flux. The Hamiltonian \cite{PRBSau2025} below represents the most general quadratic coupling between a single QD fermionic level and two Majorana operators consistent with Hermiticity and flux dependent interference. 

\begin{align}
H_{\mathrm{MZM}} &= 
\delta\,c^\dagger c 
+ \delta_{QD} d^\dagger d
\nonumber \\
&\quad
+ \sqrt{2}\,t_R e^{i\phi}(\gamma_R + \zeta_R \gamma_L)d
\nonumber \\
&\quad
+ \frac{1}{\sqrt{2}} t_L d(\gamma_L + \zeta_L \gamma_R)
+ \text{h.c.}
\end{align}
\noindent The parameters $t_{L,R}$ describe the coupling strengths through each end, while $\zeta_{L,R}$ account for finite leakage of the Majorana wavefunctions across the wire. This minimal model captures the essential physics of flux tunable Majorana mediated hybridization responsible for parity switching and the associated $h/e$-periodic response in the low energy spectrum. Because Majorana operators encode a nonlocal fermionic mode this hamiltonian can be expressed purely in terms of fermion operators and resolved into separate even and odd parity forms

\begin{figure}
    \centering
    \hspace*{-0.7cm}
    \includegraphics[width=0.45\textwidth]{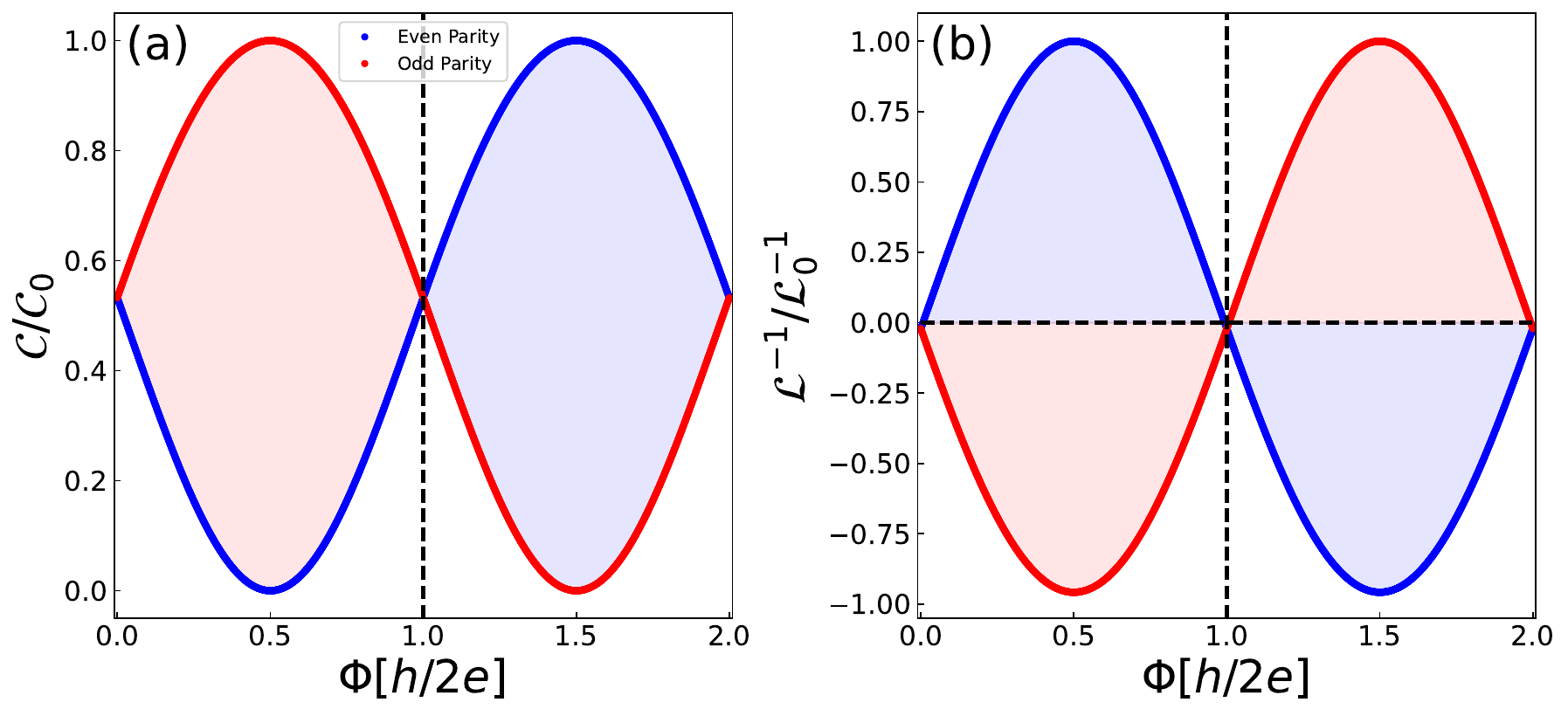}
    \caption{(a) Normalized quantum capacitance $C/C_{0}$ versus magnetic flux $\Phi$ threading the topological superconducting ring, expressed in units of the flux quantum $\Phi_{0}=h/2e$. (b) Normalized quantum inductance $\mathcal{L}^{-1}/\mathcal{L}_{0}^{-1}$ as a function of $\Phi/\Phi_{0}$. Even and odd parity responses are indicated by the blue and red curves, respectively. Both quantum capacitance and quantum inductance exhibit an $h/e$ periodicity, and the even and odd branches cross at $\Phi=\Phi_{0}$, signaling a fermion parity switch.}
    \label{fig:Model1_norm_cap}
\end{figure}

\begin{equation}
    H_{\mathrm{MZM}} = \frac{1}{2}(\delta + \delta_{QD}) + \frac{1}{2}(\delta \pm \delta_{QD})\nu_z + A\nu_+ + A^*\nu_-
\end{equation}

\noindent where the second term is $(\delta+\delta_{QD})$ for even parity and $(\delta-\delta_{QD})$ for odd parity. $A$ is the respective grouping of the tunneling terms based on parity. Here, $\nu_z$ denotes the Pauli $Z$ matrix, and $\nu_{\pm} = \nu_x \pm i\nu_y$, where $\nu_x$ and $\nu_y$ are the Pauli $X$ and $Y$ matrices, respectively.

\begin{equation}
    A = \begin{cases}
        t_Re^{i\pi\phi}(1 - i\zeta_R) + t_L(-i+\zeta_L), & \text{for even parity} \\
        t_Re^{i\pi\phi}(1 + i\zeta_R) + t_L(i+\zeta_L), & \text{for odd parity}
    \end{cases}
\end{equation}

\noindent This lets us compute the energy eigenvalues as follows,

\begin{equation}
    E = \dfrac{\delta+\delta_{QD}}{2}\pm\begin{cases}
        \sqrt{\dfrac{(\delta+\delta_{QD})^2}{4} + \lvert A\rvert^2}, & \text{for even parity} \\
        \sqrt{\dfrac{(\delta-\delta_{QD})^2}{4} + \lvert A\rvert^2}, & \text{for odd parity}
    \end{cases}
\end{equation}

\noindent where $\lvert A\rvert^2$ is given by,

\begin{equation}
    \lvert A\rvert^2 = \begin{cases}
        \lvert t_Re^{i\phi}(1 - i\zeta_R) + t_L(-i+\zeta_L)\rvert^2, & \text{for even parity} \\
    \lvert t_Re^{i\phi}(1 + i\zeta_R) + t_L(i+\zeta_L)\rvert^2, & \text{for odd parity}
    \end{cases}
\end{equation}

\noindent Here the $\phi$ dependence of $\lvert A\rvert^2$ can also be written in terms of a function $f(\phi) = P + Q\cos{\phi} + R\sin{\phi}$. 

Having obtained the analytical expression for the energy spectrum, we can directly evaluate the corresponding response functions. In particular, the quantum capacitance, $C \propto \partial^2 E / \partial \delta_{QD}^2$, and the quantum inductance, $L^{-1} \propto \partial^2 E / \partial \phi^2$, follow from taking the appropriate second derivatives of the energy eigenvalues.

\begin{equation}%
    C = \begin{cases}
        \dfrac{2\lvert A\rvert^2}{\left[(\delta+\delta_{QD})^2 + 4\lvert A\rvert^2\right]^{3/2}}, & \text{for even parity}\\
        \dfrac{2\lvert A\rvert^2}{\left[(\delta-\delta_{QD})^2 + 4\lvert A\rvert^2\right]^{3/2}}, & \text{for odd parity}\\
    \end{cases}%
\end{equation}%
\balance

\begin{widetext}
\begin{equation}
    L^{-1} = \begin{cases}
    \dfrac{1}{2\sqrt{\dfrac{(\delta+\delta_{QD})^2}{4} + f(\phi)}}\dfrac{\partial^2 f}{\partial \phi^2} - \dfrac{1}{4\left(\dfrac{(\delta+\delta_{QD})^2}{4} + f(\phi) \right)^{3/2}}\left(\dfrac{\partial f}{\partial \phi}\right)^2 & \text{for even parity} \\
    \dfrac{1}{2\sqrt{\dfrac{(\delta-\delta_{QD})^2}{4} + f(\phi)}}\dfrac{\partial^2 f}{\partial \phi^2} - \dfrac{1}{4\left(\dfrac{(\delta-\delta_{QD})^2}{4} + f(\phi) \right)^{3/2}}\left(\dfrac{\partial f}{\partial \phi}\right)^2 & \text{for odd parity}
    \end{cases}
\end{equation}
\end{widetext}

Fig.~\ref{fig:Model1_norm_cap} shows the flux dependent oscillations of the normalized quantum capacitance [panel (a)] and quantum inductance [panel (b)] for parity resolved responses at $\delta=0$, $\delta_{QD}=4.0$, $t_L=t_R=0.3$, and $\zeta_L=\zeta_R=0$, corresponding to a genuinely topological regime. The red (blue) curves denote the odd (even) parity sector. As expected for a topological configuration, both the quantum capacitance and quantum inductance exhibit $h/e$ periodic oscillations for each parity sector. Notably, the crossing of the quantum capacitance curves at $\Phi = h/2e$ is accompanied by a corresponding crossing of the quantum inductance curves at the same flux value, with the quantum inductance passing smoothly through zero. The simultaneous occurrence of crossings in both response functions provides a clear signature of a true fermion parity switch, consistent with the nanowire–QD system residing in a topological regime.

\subsection{Model 2: Pair of quasi-MZMs on each end coupled to a QD}
\label{Model2}

To describe a topologically trivial regime capable of producing near zero energy features, we consider an effective model consisting of a pair of quasi-Majorana modes (or partially separated Andreev bound states)\cite{Moore2018, SciPostVuik2019, PhysRevB.96.075161} localized near each end of the nanowire\cite{MSRMicrosoft2024}, represented by fermionic operators $c^\dagger_{L}$ and $c^\dagger_{R}$, with corresponding energy splittings $\epsilon_{L}$ and $\epsilon_{R}$. Unlike the true Majorana case, these modes arise from conventional subgap physics such as smooth confinement or finite size effects and therefore correspond to ordinary fermionic degrees of freedom with finite energy splittings $\epsilon_L$ and $\epsilon_R$. When coupled to a single QD level, described by the operator $d^\dagger$ with onsite energy $\epsilon_d$, these end localized modes hybridize with the dot through flux dependent tunneling processes that carry phase factors $e^{\pm i\phi/2}$ associated with the left and right paths of the interferometer. 

\begin{figure}
    \centering
    \hspace*{-0.7cm}
    \includegraphics[width=0.45\textwidth]{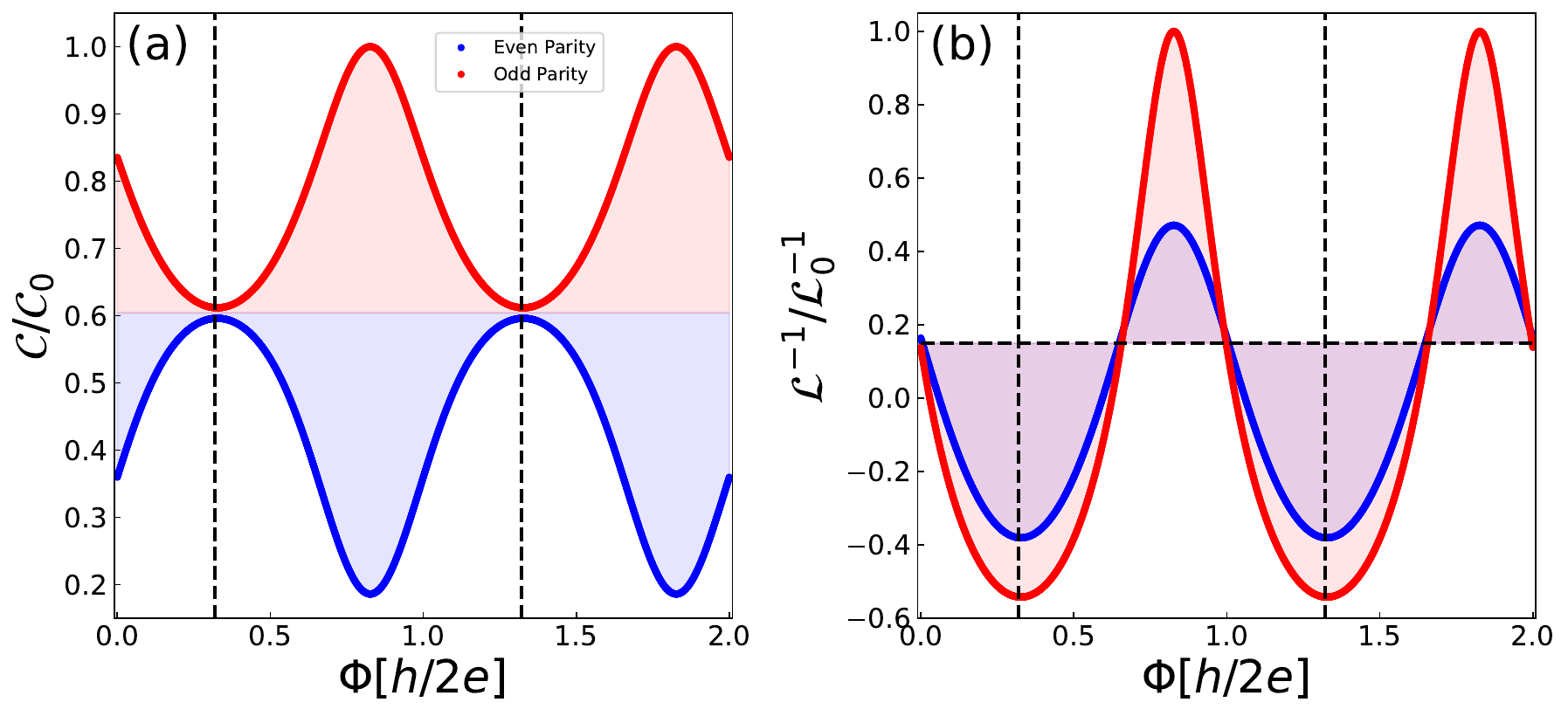}
    \caption{(a) Normalized quantum capacitance $C/C_{0}$ versus magnetic flux $\Phi$, expressed in units of the flux quantum $\Phi_{0}=h/2e$. (b) Normalized quantum inductance $\mathcal{L}^{-1}/\mathcal{L}_{0}^{-1}$ as a function of $\Phi/\Phi_{0}$. Even and odd parity responses are indicated by the blue and red curves, respectively. Both capacitance and quantum inductance are $h/2e$ periodic, and the absence of a true crossing (presence of an avoided crossing) in the capacitance is reflected in the quantum inductance curves by the pair of extrema at the corresponding flux values $\Phi \approx 0.325\,\Phi_{0}$ and $1.325\,\Phi_{0}$.}
    \label{fig:qMZM_C_L_curve}
\end{figure}

\noindent Because the underlying degrees of freedom are conventional fermions rather than spatially separated Majoranas forming a single nonlocal mode, the resulting Hamiltonian is intrinsically $2\pi$ periodic in $\phi$, corresponding to an $h/2e$ flux periodicity. In the regime where the QD is strongly detuned from the nanowire, the QD degree of freedom can be eliminated perturbatively, yielding an effective low energy coupling between the left and right end modes. This reduced Hamiltonian captures the formation of avoided crossings or narrow double crossings in the flux dependent spectrum, providing a controlled trivial regime against which genuine Majorana induced parity switches can be contrasted. The resulting low energy effective Hamiltonian for this system is given by

\begin{align}
    H_{\mathrm{q-MZM}} &= \epsilon_dd^\dagger d + \sum_{s=L,R} \{(A_sc^\dagger_s + B_sc_s)d^\dagger e^{is\phi/2} + h.c.\}\nonumber \\
    & + \epsilon_sc^\dagger_sc_s
\end{align}

\noindent where $A_s$ and $B_s$ denote the coupling strengths between each ABS pair and the quantum dot. The index $s=L,R$ also labels the superconducting phase factors associated with the left ($s=1$) and right ($s=0$) ends of the wire, respectively. Upon performing the gauge transformation $c^\dagger_L \rightarrow c^\dagger_L e^{-i\phi/2}$, the Hamiltonian takes the form

\begin{align}
    H_{\mathrm{q-MZM}} &= \epsilon_dd^\dagger d + \sum_{s=L,R} \{(A_sc^\dagger_s + e^{is\phi}B_sc_s)d^\dagger + h.c.\} \nonumber \\
    & + \epsilon_sc^\dagger_sc_s
\end{align}

\noindent The transformed Hamiltonian is now manifestly $2\pi$ periodic, corresponding to an $h/2e$ flux periodicity. For analytical simplicity, we assume the hierarchy $\epsilon_d \gg A_s, B_s$, which allows the quantum dot degree of freedom to be integrated out perturbatively. This can be accomplished using a Schrieffer-Wolff transformation\cite{BRAVYI20112793}, yielding

\begin{align}
    H_{\mathrm{q-MZM}} &\approx \sum_{s=\{L,R\}} \epsilon_s c_s^\dagger c_s \nonumber \\
    &- \epsilon_d^{-1}\left[(A_L c_L^\dagger + e^{i\phi}B_L c_L)(B_R c_R^\dagger + A_R c_R) + h.c \right]
\end{align}

We now separate the effective Hamiltonian into distinct fermion parity sectors and express it in terms of Pauli matrices. In the even parity sector, the fermion number constraint $c_L^\dagger c_L + c_R^\dagger c_R = 0,2$ implies that $c_L c_R^\dagger = c_L^\dagger c_R = 0$. Under this constraint, the number operators can be written as $c_L^\dagger c_L = c_R^\dagger c_R = (1 - \nu_z)/2$, while the pairing operators satisfy $c_s^\dagger c_s^\dagger = \nu_-$. Here, $\nu_z$ denotes the Pauli $Z$ matrix, and $\nu_{\pm} = \nu_x \pm i\nu_y$, with $\nu_x$ and $\nu_y$ representing the Pauli $X$ and $Y$ matrices, respectively. The effective Hamiltonian in the even parity sector therefore takes the form

\begin{align}
    H_{\mathrm{q-MZM}} & \approx \sum_{s=\{L,R\}} \dfrac{\epsilon_s}{2} (1-\sigma_z) - \epsilon_d^{-1}\left[(A_L B_R +  \nonumber \right. \\
    & \left. e^{-i\phi}A_R^*B_L^*)\nu_{-} + (A_L^*B_R^* + e^{i\phi}A_RB_L)\nu_{+} \right]
\end{align}

\noindent For the odd fermion parity sector $c_L^\dagger c_L + c_R^\dagger c_R = 1$ then $c_L^\dagger c_R^\dagger = c_Lc_R = 0$ enabling us to write $1 - c_R^\dagger c_R = c_L^\dagger c_L = (1 - \nu_z)/2$ and $c_R^\dagger c_L = c_Lc_R^\dagger = \nu_+$, giving us the following Hamiltonian,

\begin{align}
    H_{\mathrm{q-MZM}} & \approx \sum_{s=\{L,R\}} \dfrac{\epsilon_s}{2} (1\pm\sigma_z) - \epsilon_d^{-1}\left[(A_L A_R \right. \nonumber \\ 
    & \left. + e^{-i\phi}B_R^*B_L^*)\nu_{-} + (A_L^*A_R^* + e^{i\phi}B_RB_L)\nu_{+} \right]
\end{align}

\noindent where the first term is $(1+\sigma_z)$ for $s=R$ and $(1-\sigma_z)$ for $s=L$. The Hamiltonian in each case can be condensed into the form $H = \dfrac{1}{2}\left(\epsilon_L+\epsilon_R\right) - \dfrac{1}{2}(\epsilon_L + (-1)^P \epsilon_R)\nu_z - \dfrac{1}{\epsilon_d}\left[A\nu_+ + A^*\nu_-\right]$ ($P=0$ for even parity and $P=1$ for odd). Allowing us to compute the energy eigenvalues as follows,

\begin{equation}
    E = \dfrac{\epsilon_L+\epsilon_R}{2} \pm \begin{cases}
        \sqrt{\dfrac{(\epsilon_L+\epsilon_R)^2}{4} + \dfrac{\lvert A_p\rvert^2}{\epsilon_d^2}} & \text{for even parity} \\
        \sqrt{\dfrac{(\epsilon_L-\epsilon_R)^2}{4} + \dfrac{\lvert A_p\rvert^2}{\epsilon_d^2}} & \text{for odd parity}
    \end{cases}
\end{equation}

\noindent where, 

\begin{equation}
    \lvert A_p\rvert^2 = \begin{cases}
    \lvert A_RB_Le^{i\phi} + A_L^*B_R^*\rvert^2, & \text{for even parity} \\
    \lvert B_RB_Le^{i\phi} + A_L^*A_R^*\rvert^2, & \text{for odd parity}
    \end{cases}
\end{equation}

\noindent Similar to sec. \ref{Model1} $\lvert A_p\rvert^2$ can be expressed as a function $f(\phi) = P + Q\cos{\phi} + R\sin{\phi}$. Allowing us to calculate the capacitance and inductance formulas as follows,

\begin{equation}
    C = \begin{cases}
        \lvert A_p\rvert^2\left(\dfrac{2 + 3\epsilon_d^2(\epsilon_L+\epsilon_R)^2}{\epsilon_d^3\left[((\epsilon_L+\epsilon_R)\epsilon_d)^2 + \lvert A_p\rvert^2\right]^{3/2}}\right), & \text{for even parity} \\
    \lvert A_p\rvert^2\left(\dfrac{2 + 3\epsilon_d^2(\epsilon_L-\epsilon_R)^2}{\epsilon_d^3\left[((\epsilon_L-\epsilon_R)\epsilon_d)^2 + \lvert A_p\rvert^2\right]^{3/2}}\right), & \text{for odd parity}
    \end{cases}
\end{equation}

\balance
\begin{widetext}
    \begin{equation}
        L^{-1} = \begin{cases}
            \dfrac{1}{2\epsilon_d^2\sqrt{\dfrac{(\epsilon_L+\epsilon_R)^2}{4} + \dfrac{f(\phi)}{\epsilon_d^2}}}\dfrac{\partial^2 f}{\partial \phi^2} - \dfrac{1}{4\epsilon_d^4\left(\dfrac{(\epsilon_L+\epsilon_R)^2}{4} + \dfrac{f(\phi)}{\epsilon_d^2} \right)^{3/2}}\left(\dfrac{\partial f}{\partial \phi}\right)^2 & \text{for even parity} \\
            \dfrac{1}{2\epsilon_d^2\sqrt{\dfrac{(\epsilon_L-\epsilon_R)^2}{4} + \dfrac{f(\phi)}{\epsilon_d^2}}}\dfrac{\partial^2 f}{\partial \phi^2} - \dfrac{1}{4\epsilon_d^4\left(\dfrac{(\epsilon_L-\epsilon_R)^2}{4} + \dfrac{f(\phi)}{\epsilon_d^2} \right)^{3/2}}\left(\dfrac{\partial f}{\partial \phi}\right)^2 & \text{for odd parity}
        \end{cases}
    \end{equation}
\end{widetext}

Having established the effective model, we now compare the resulting phase dependence of the quantum capacitance and quantum inductance. The purpose of this comparison is to illustrate how a genuinely topological MZM–QD configuration, discussed in Sec.~\ref{Model1}, differs from a purely trivial ABS based configuration in terms of their flux periodic responses and fermion parity structure. Fig.~\ref{fig:qMZM_C_L_curve} shows the flux dependent oscillations of the normalized quantum capacitance [Fig.~\ref{fig:qMZM_C_L_curve}(a)] and quantum inductance [Fig.~\ref{fig:qMZM_C_L_curve}(b)] for parity resolved responses obtained using the model parameters $\epsilon_L=0.49$, $\epsilon_R=0.33$, $\epsilon_D=3.14$, $A_L=0.72+0.46i$, $A_R=-0.21+1.00i$, $B_L=0.10+0.30i$, and $B_R=0.72+0.53i$. The red (blue) curves correspond to the odd (even) parity sector.

As expected for a topologically trivial configuration, both the quantum capacitance and quantum inductance exhibit $h/2e$ periodic oscillations for each parity sector. Panel (a) in Fig.~\ref{fig:qMZM_C_L_curve} shows that the quantum capacitance responses develop avoided crossings in the intervals $\Phi < h/4e$ and $h/2e < \Phi < 3h/4e$. Crucially in the absence of any knowledge of fermion parity for each capacitance curve, as is the case in experiments \cite{MSRMicrosoft2024,arXivMicrosoft2024,PubMedMicrosoft2025}, the true parity crossing in Fig.~\ref{fig:Model1_norm_cap} and the avoided crossing in Fig.~\ref{fig:qMZM_C_L_curve} would be indistinguishable in quantum capacitance measurements. Remarkably however, the corresponding quantum inductance responses shown in Fig.~\ref{fig:qMZM_C_L_curve}(b) display pronounced extrema at the same values of $\Phi$ where the avoided crossings occur in the capacitance curves, while it shows corresponding crossing between the parity sectors at the same flux value for a true fermion parity switch in the capacitance curves. These extrema provide a clear diagnostic that distinguishes the avoided crossings present here from the true crossings observed in the topological case shown in Fig.~\ref{fig:Model1_norm_cap}, thereby demonstrating the ability of quantum inductance to discriminate between trivial and topological parity switching scenarios. These effective models provide a transparent framework for understanding how the flux dependence of the low energy spectrum manifests in the quantum capacitance and quantum inductance responses. In particular, they show that true fermion parity–protected crossings produce simultaneous sign changes in both observables, while avoided or narrow double crossings in the capacitance response generate pronounced extrema in the quantum inductance. These qualitative features establish quantum inductance as a complementary probe to quantum capacitance for distinguishing genuine parity switches from trivial hybridization effects.


\section{MICROSCOPIC MODEL AND QUANTUM RESPONSE FORMALISM}
\label{Transport}

We now turn to full microscopic Bogoliubov-de Gennes (BdG) calculations based on a realistic SM-SC model that incorporates spin–orbit coupling, an applied Zeeman field, disorder, and coupling to a quantum dot whose energy is tunable by an external gate. While the effective models discussed above capture the essential mechanisms governing the quantum capacitance and quantum inductance responses, a microscopic treatment is required to verify that these signatures persist in realistic nanowire–QD devices with spatially extended states and disorder. The semiconductor nanowire is described by the tight-binding Hamiltonian
\begin{align}
\mathcal{H}_{SM} &=
\sum_{i,\sigma}
\Big[
    -t(c^\dagger_{i\sigma}c_{i+1\sigma} + \text{h.c.})
    + (V_{dis}(i) - \mu)c^\dagger_{i\sigma}c_{i\sigma}
\Big]
\nonumber\\
&\quad
+ \frac{\alpha}{2}\sum_i
\Big[
    c^\dagger_{i\uparrow}c_{i+1\downarrow}
    - c^\dagger_{i\downarrow}c_{i+1\uparrow}
    + \text{h.c.}
\Big] \nonumber \\
&\quad + \Gamma\sum_i(c^\dagger_{i\uparrow}c_{i\downarrow} + \text{h.c.}),
\end{align}
where $i = 0,2,\ldots,N-1$ labels the sites of the one-dimensional lattice, with $N=300$ and lattice constant $a=10$~nm. Here $c^\dagger_{i\sigma}$ ($c_{i\sigma}$) creates (annihilates) an electron with spin $\sigma$ on site $i$. The effective parameters in the model include the nearest-neighbor hopping $t=16.56$~meV, the chemical potential $\mu$, the Rashba spin-orbit coupling strength $\alpha=1.4$~meV, and the Zeeman field $\Gamma$. Disorder in the finite semiconductor wire is modeled through a random onsite potential $V_{dis}(i)$ satisfying $\langle V_{dis} \rangle = 0$ and $\langle V^2_{dis} \rangle = V_0^2$. In our calculations we change the disorder by changing the magnitude of $V_0$.

The effect of proximity induced superconductivity is incorporated via a local self-energy term,
\begin{equation}
    [\Sigma_{SC}(\omega)]_{ii} = -\gamma\frac{\omega+\Delta_0\sigma_y\tau_y}{\sqrt{\Delta^2_0-\omega^2}},
\end{equation}
where $\sigma_y$ and $\tau_y$ denote Pauli matrices acting in spin and particle-hole space, respectively. Here $\Delta_0$ is the gap of the parent $s$-wave superconductor and $\gamma=0.2$~meV characterizes the strength of the SM-SC coupling. We assume that the parent superconducting gap decreases with increasing Zeeman field, taking a maximum value $\Delta(0)=0.3$~meV and collapsing at $\Gamma=1.25$~meV. We work in the static approximation, $\omega^2\ll\Delta_0^2$. Writing the Green's function in the form $G^{-1}(\omega) = \omega \mathbb{I} - H_{SM} - \Sigma_{SC}$, and defining the corresponding effective Hamiltonian\cite{SauPRB2010Lutchyn} as $\mathcal{H}_{eff} = \mathbb{Z} \mathcal{H}_{SM} + \mathbb{Z}\gamma \sigma_y \tau_y$, we obtain
\begin{equation}
\mathcal{H}_{eff} = \mathbb{Z}\cdot\mathcal{H}_{SM} + \mathbb{I}\cdot\Delta\,\sigma_y \tau_y,
\end{equation}
where the quasiparticle residue is given by $\mathbb{Z}=\Delta_0/(\Delta_0+\gamma)$ and the induced superconducting gap by $\Delta=\Delta_0\gamma/(\Delta_0+\gamma)$.

The hybrid nanowire is coupled to a quantum dot (QD) located at site $N$, and is tuned by a gate potential $V_{QD}$, see Fig.~\ref{fig:SchematicSetup}. For simplicity, we retain a single dot level $n_0$ with energy $\epsilon_{n_0}=0$. The couplings of the dot to the left ($0$th) and right ($(N-1)$th) ends of the wire are parameterized by $\lambda_L$ and $\lambda_R$, respectively:
\begin{align}
\mathcal{H}_{QD} &=
\sum_\sigma (V_{QD}-\mu)a^\dagger_\sigma a_\sigma \nonumber
\\
&\quad
+ \sum_\sigma
\Big[
\lambda_L e^{i\phi}
\left(
    -t c^\dagger_{0\sigma}a_\sigma
    - \frac{\alpha}{2}c^\dagger_{0\uparrow}a_\downarrow
    + \text{h.c.}
\right)
\nonumber\\
&\qquad
+
\lambda_R
\left(
    -t c^\dagger_{N-1\sigma}a_\sigma
    + \frac{\alpha}{2}c^\dagger_{N-1\uparrow}a_\downarrow
    + \text{h.c.}
\right)
\Big],
\end{align}
where $a_\sigma^\dagger \equiv c_{N\sigma}^\dagger$ labels the QD state and $\phi = \pi\Phi/\Phi_0$ is the flux tunable phase, with $\Phi$ denoting the magnetic flux threaded through the nanowire-QD loop and $\Phi_0 = h/2e$ the superconducting flux quantum. This Hamiltonian forms the basis for the calculation of quantum capacitance and quantum inductance described below. 

To characterize the phase and gate dependent response of the system we compute quantum capacitance\cite{PubMedMicrosoft2025,PRBSau2025,ArXivStanescu2025} and quantum inductance \cite{PRTewari2012} using linear response theory (see Appendix.~\ref{Derived_L_inverse} for a detailed derivation of quantum inductance formula). We consider the Bogoliubov-de Gennes Hamiltonian 
\begin{equation}
    H(\phi,V_{QD}) = \mathcal{H}_{eff} + \mathcal{H}_{QD}
    \label{eq:Hamil_phi}
\end{equation}
that depends on the magnetic flux $\Phi$ through the interferometric loop via the phase $\phi = \pi\Phi/\Phi_0$ inserted on the quantum dot–nanowire link, as well as on the quantum dot gate potential $V_{QD}$. The quantum capacitance is defined through the linear response susceptibility associated with the charge operator $\widehat{Q} \equiv \partial_{V_{QD}} H$, while quantum inductance is governed by the current operator $\widehat{J} \equiv \partial_\phi H$ and diamagnetic curvature term $\widehat{D} \equiv \partial_\phi^2 H$\cite{MinutilloPRB2024,martinis2004superconductingqubitsphysicsjosephson,ShnyrkovLTP2014}.

Following the standard Lehmann representation\cite{Kubo1957,Mahan2000,Bruus2004}, the dynamic quantum capacitance and quantum inductance at frequency $\omega$ and broadening $\eta$ are given by
\begin{align}
\mathcal{C}(\phi,\omega) &=
    \kappa_C
    \sum_{n,m}
    \frac{
        \left|\bra{n} \partial_{V_{QD}} H \ket{m}\right|^2 (E_m - E_n)
    }{
        (E_m - E_n)^2 - (\omega + i\eta)^2
    },
\label{eq:cap_lehmann}
\\[3pt]
\mathcal{L}^{-1}(\phi,\omega) &=
    \kappa_L \bigg[
        \sum_{n}\bra{n}\partial_{\phi}^2 H\ket{n} \nonumber \\
        & \qquad
        - 2\sum_{n,m}
        \frac{
            \left|\bra{n}\partial_\phi H\ket{m}\right|^2 (E_m - E_n)
        }{
            (E_m - E_n)^2 - (\omega + i\eta)^2
        }
    \bigg].
\label{eq:ind_lehmann}
\end{align}

\noindent Here $\ket{m}$ and $\ket{n}$ denote eigenstates of the BdG Hamiltonian with energies $E_m$ and $E_n$, respectively, ordered such that $m\ge2N+1$ labels empty states in the BdG vacuum and $n\le2N-2$ labels filled states. To compute parity resolved responses, the Lehmann sums in Eqs.~(\ref{eq:cap_lehmann}) and (\ref{eq:ind_lehmann}) are restricted to eigenstates consistent with a fixed fermion parity sector. For the odd- (even-) parity calculation, the index $m$ runs over all empty states with $m \ge 2N+1$ and $m=e(o)$, while the index $n$ runs over all filled states with $n \le 2N-2$ and $n=o(e)$. The even and odd parity ground states ($e$ and $o$) correspond to the two lowest energy BdG modes labeled by $2N-1$ and $2N$, which satisfy $E_{\mathrm{e}} = -E_{\mathrm{o}}$ \cite{ArXivStanescu2025,PRBSau2025} due to particle-hole symmetry. In practice, the exact evaluation of the quantum inductance, which involves the second derivative of the energy with respect to the flux induced phase $\phi = \pi\Phi/\Phi_0$, formally requires summation over the complete BdG spectrum, which is computationally prohibitive. We therefore evaluate the summation in Eq.~\ref{eq:cap_lehmann} Eq.~\ref{eq:ind_lehmann} using a truncated spectrum consisting of a fixed number of eigenstates symmetrically chosen about zero energy. We have verified that retaining 200 eigenstates is sufficient to capture all qualitative features of the capacitance and inductance responses relevant to this work, as increasing the number of included states primarily results in an overall vertical shift of the curves without modifying their flux dependence or parity structure.

The constants $\kappa_C = 0.032\,\mathrm{fF}\cdot\mathrm{meV}$ and $\kappa_L = 1.48\,\mathrm{nH}^{-1}\!\cdot\mathrm{meV}$ in Eq.~\ref{eq:cap_lehmann} and Eq.~\ref{eq:ind_lehmann} convert the microscopic matrix elements into experimentally relevant units for the quantum capacitance and quantum inductance, respectively. Throughout this work, we focus primarily on the low frequency limit $\omega=0$, while including a small finite broadening $\eta = 0.002$~meV to ensure numerical stability and to reflect experimentally relevant linewidths. The quantum inductance is intrinsically sensitive to phase dependent hybridization and level anticrossings, and therefore contains information not captured by $\mathcal{C}(\phi)$ alone. This completes the specification of the microscopic model used in the remainder of this work. The full Bogoliubov–de Gennes Hamiltonian describing the nanowire–QD system is now defined, together with the linear response expressions for the quantum capacitance $\mathcal{C}$ and quantum inductance $\mathcal{L}^{-1}$. These quantities form the basis for all numerical calculations and diagnostics presented in the following sections.

\section{RESULTS}
\label{sec:results}

We now present the flux-dependent quantum capacitance and quantum inductance of the interferometric nanowire-QD device introduced above. In recent experiments, fermion parity is not measured directly; rather, parity information is inferred from a bimodal quantum capacitance readout that arises when the measurement time exceeds the quasiparticle poisoning time $\tau_{\mathrm{qpp}}$, leading to stochastic switching between even and odd sectors \cite{arXivMicrosoft2024,PubMedMicrosoft2025,MSRMicrosoft2024}. Consequently, avoided crossings or narrowly split double crossings in the quantum capacitance profiles between different parity sectors may appear experimentally similar to parity-protected crossings when the readout is not explicitly conditioned on the underlying parity sector, as shown in Fig.~\ref{fig:schematic_transport_curve}.

The central objective of this section is therefore to identify spectral features that distinguish true fermion parity-protected crossings from avoided or narrow double crossings. While quantum capacitance reflects the flux variation of the many-body energy, the quantum inductance probes the flux curvature change and is intrinsically sensitive to sharp curvature changes caused by avoided or narrow double crossing scenarios. As demonstrated in the effective models discussed in Sec.~\ref{sec:eff_model}, avoided and narrow double crossings generate pronounced extremum features in $\mathcal{L}^{-1}(\Phi)$, whereas true parity-protected crossings produce a corresponding crossing between the two parity sectors in the inductance response without such extrema. The combined analysis of $\mathcal{C}(\Phi)$ and $\mathcal{L}^{-1}(\Phi)$ therefore provides a more discriminating diagnostic of fermion parity switching than capacitance alone.

We begin with the clean limit to establish the characteristic behavior of parity-resolved quantum capacitance and inductance in a topological regime. The clean system results serve to calibrate the microscopic simulations against the characteristic flux-dependent features obtained in the effective models, establishing the correspondence between true parity-protected crossings in both $\mathcal{C}(\Phi)$ and $\mathcal{L}^{-1}(\Phi)$ at the same flux values. We then introduce weak disorder ($V_0=0.3$~meV) to examine whether these features remain robust under moderate perturbations. In this regime, we provide explicit examples of both a true parity-protected crossing and a narrow double crossing, demonstrating that the qualitative distinction identified in the effective models (in Sec.~\ref{sec:eff_model}) persists in realistic microscopic simulations. We next consider the strongly disordered regime ($V_0=1.2$~meV), where the topological landscape becomes fragmented. Here the analysis focuses on identifying parameter regions in which the magnitudes of the quantum capacitance and quantum inductance remain experimentally resolvable and assessing whether their flux dependence continues to faithfully distinguish true crossings from avoided or narrow double crossings. Finally, we turn to smooth Gaussian confinement, which provides a controlled setting for engineering quasi-Majorana states analogous to those described in Sec.~\ref{Model2}. In this geometry, we illustrate representative avoided and narrow double crossing scenarios and demonstrate their persistence over extended parameter regimes. The trimmed deviation inductance metric $\mathcal{D}_{\mathcal{L}}$ introduced in this context is subsequently applied to the strongly disordered system, where it helps identify parameter regimes exhibiting avoided or narrow double crossings and enables direct comparison with regimes supporting true crossings. For each case, we analyze whether the joint behavior of $\mathcal{C}(\Phi)$ and $\mathcal{L}^{-1}(\Phi)$ is consistent with a true fermion parity switch. As shown below, agreement between the two observables provides a robust operational criterion for identifying parameter regimes consistent with a negative Pfaffian invariant, or equivalently a negative topological stability (defined later in Sec.~\ref{sec:weak_dis_main}) \cite{PUKitaev2001, PRBSau2025}.

\begin{figure}
    \centering
    \includegraphics[width=1.\linewidth]{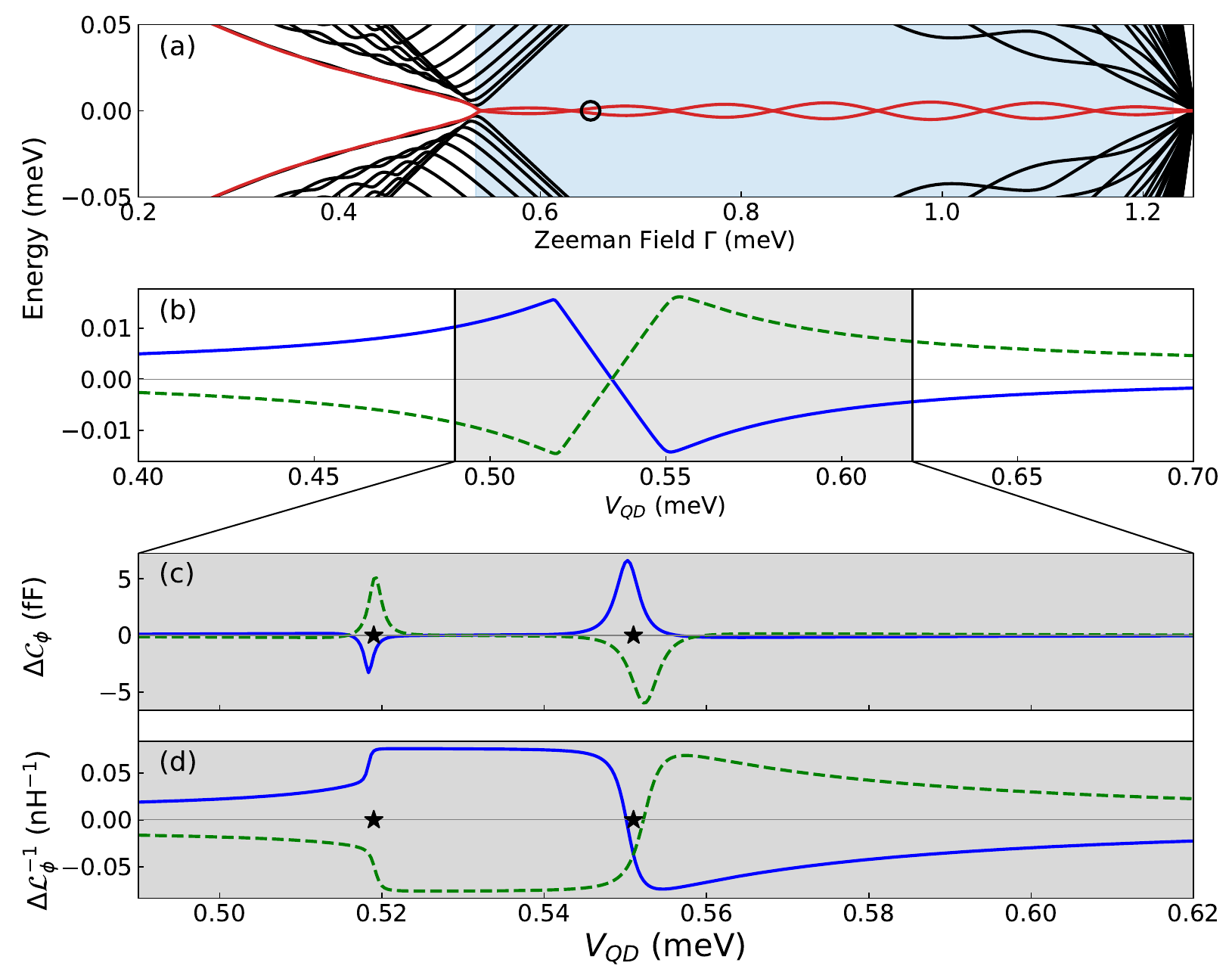}
    \caption{(a) BdG low energy spectrum of the clean nanowire ($V_{0}=0$ meV) in the open boundary configuration, i.e., with the ends of the wire not coupled to the quantum dot (QD). For Zeeman fields exceeding the topological quantum phase transition (TQPT) point $\Gamma_{c}=\sqrt{\mu^{2}+\Delta^{2}}=0.54$ meV, a pair of Majorana zero modes (highlighted in red) emerges and persists until the superconducting gap collapses near $\Gamma\approx1.25$ meV. The light blue shaded region denotes the parameter range where the Pfaffian invariant (refer to Eq.~\ref{eq:pfaff}) is negative, corresponding to the operationally topological regime. The black circle marks the point that is further examined once the wire is coupled to the QD. (b) Low energy modes (fixed fermion parity) as a function of the QD potential $V_{QD}$ for $\Phi=0$ (blue solid line) and $\Phi=h/2e$ (green dashed line), evaluated at the Zeeman field indicated in panel (a). The tunnel couplings are $\lambda_{L}=\lambda_{R}=0.025$. (c) Difference in quantum capacitance between the even and odd parity sectors, $\Delta \mathcal{C}_\Phi = \mathcal{C}_{\mathrm{even}} - \mathcal{C}_{\mathrm{odd}}$, plotted as a function of $V_{QD}$ for the same flux values, $\Phi=0$ (blue solid line) and $\Phi=h/2e$ (green dashed line). (d) Difference in quantum inductance, $\Delta \mathcal{L}_\Phi^{-1}=\mathcal{L}^{-1}_{\mathrm{even}} - \mathcal{L}^{-1}_{\mathrm{odd}}$, for the same parameters. Panels (b)-(d) show the detailed behavior of the system in the vicinity of the point highlighted in panel (a).}
    \label{fig:Clean_sys_topo_regime}
\end{figure}

\subsection{Clean System}
\label{sec:clean_sys}

\begin{figure}
    \centering
    \includegraphics[width=1.\linewidth]{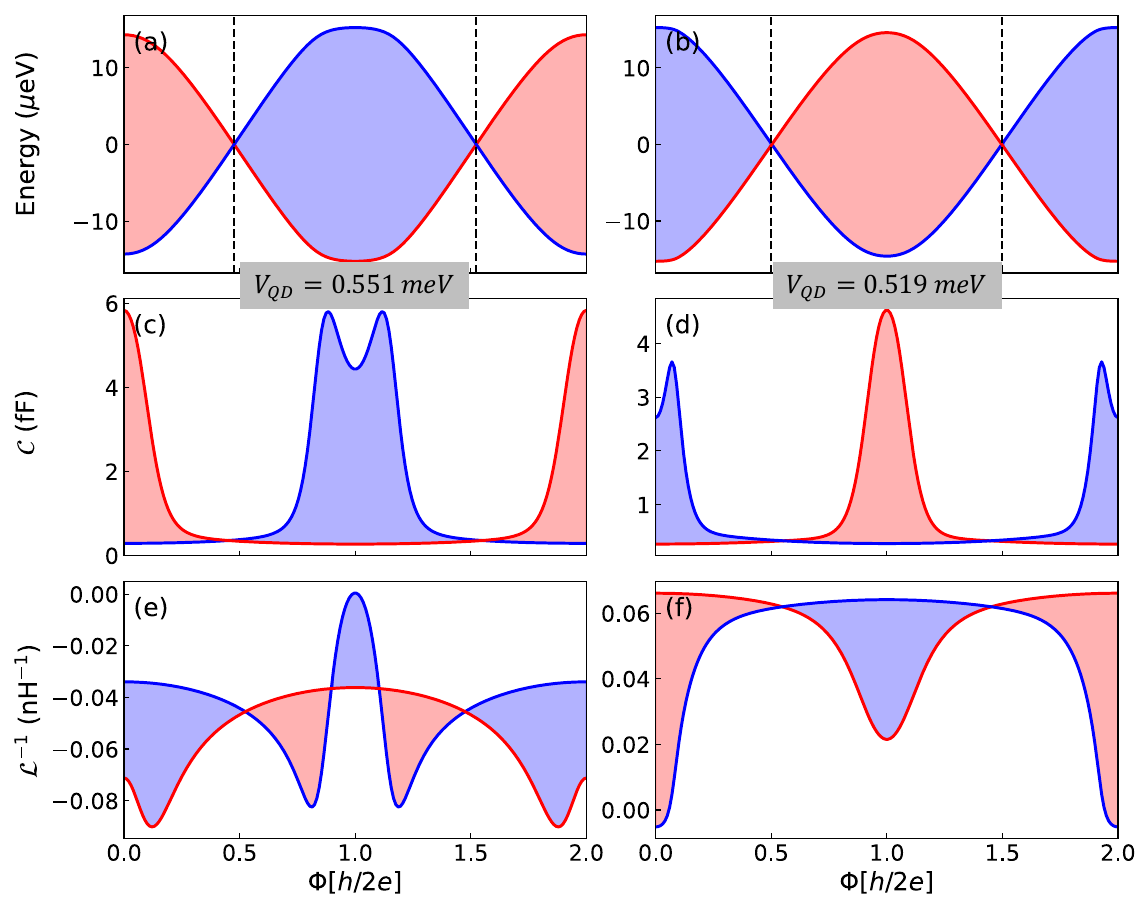}
    \caption{Flux dependent energy spectrum, quantum capacitance, and quantum inductance for a clean nanowire in a topologically nontrivial regime at chemical potential $\mu = 0.5$~meV and coupling $\Gamma = 0.65$~meV, shown by a black circle in Fig.~\ref{fig:Clean_sys_topo_regime}(a). Two representative values of the quantum dot gate potential $V_{\mathrm{QD}}$ are chosen (shown by black stars in Fig.~\ref{fig:Clean_sys_topo_regime}(c) and (d)) to illustrate the dependence of the response on gate tuning. Panels (a) and (b) show the two lowest energy eigenvalues (closest to zero energy) as a function of magnetic flux $\Phi$, exhibiting crossings near $\Phi \approx h/4e$ and $3h/4e$. Panels (c) and (d) show the corresponding capacitance response, while panels (e) and (f) show the quantum inductance response. In both cases, the responses are $h/e$ periodic for each parity sector and display crossings at similar values of $\Phi$. Blue (red) curves denote the even- (odd-) parity branch. The relative ordering of the parity branches for fixed $\Phi$ is interchanged between panels (a),(c),(e) and their counterparts (b),(d),(f), reflecting that the two $V_{\mathrm{QD}}$ values lie on opposite sides of the sign changes in the energy, $\Delta\mathcal{C}$, and $\Delta\mathcal{L}^{-1}$, as shown in Fig.\ref{fig:Clean_sys_topo_regime}(c) and (d).}
    \label{fig:Diff_vqd}
\end{figure}

We begin with the clean limit ($V_{0}=0$~meV) in order to establish the characteristic flux-dependent behavior of the quantum capacitance and quantum inductance in a topological nanowire–QD system. In the absence of disorder, the relation between the low energy spectrum in the topological regime and the corresponding capacitance and inductance responses can be examined in its simplest form. This provides a reference against which the effects of disorder and smooth confinement considered in the following subsections can be assessed. 

Fig.~\ref{fig:Clean_sys_topo_regime}(a) shows the low energy BdG spectrum of a clean $3~\mu$m-long nanowire in the open-boundary configuration, i.e., with the two ends of the wire not coupled to the quantum dot (QD). The red curves highlight the pair of states closest to zero energy, which evolve into Majorana zero modes once the system enters the topological phase. The light blue shaded region denotes the Zeeman field range in which the Pfaffian invariant is negative, corresponding to the operationally topological regime.

The Pfaffian invariant is defined as
\begin{equation}
    \mathcal{Q} = \mathrm{sgn}\left[\mathrm{Pf}[H(\Phi=0)]\,\mathrm{Pf}[H(\Phi=h/2e)]\right],
    \label{eq:pfaff}
\end{equation}
where the Hamiltonians entering Eq.~(\ref{eq:pfaff}) are constructed using periodic $\Phi=0$ and anti-periodic $\Phi=h/2e$ boundary conditions between the first and last sites of the wire. In this definition, the quantum dot is not included and the end-to-end hopping is taken to be unity ($\lambda=1$), corresponding to a direct coupling between the wire ends. 

\begin{figure*}[!htbp]
    \centering
    \includegraphics[height=0.25\textheight,width=\linewidth]{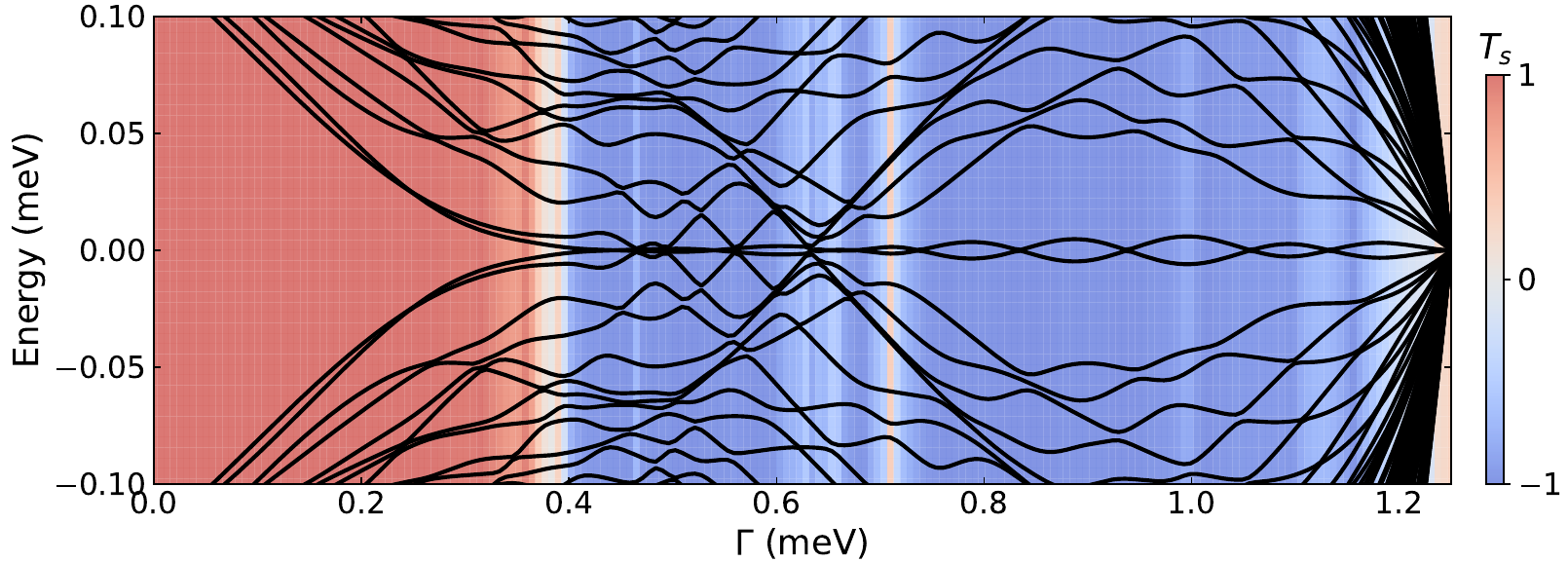}
    \caption{Energy spectrum for a weakly disordered nanowire at chemical potential $\mu = 0.5$~meV with disorder amplitude $V_0 = 0.3$~meV. Black curves show the low energy eigenvalues as a function of the coupling strength $\Gamma$. The background color map represents the topological stability $T_s$ (refer to Eq.~\ref{eq:Topo_stability}), used as a topological indicator in this work, with values close to $T_s = -1$ (blue) indicating an operationally topological regime and values close to $T_s = +1$ (red) indicating a trivial regime. The quantity $T_s$ is computed including the quantum dot with gate potential $V_{\mathrm{QD}} = 0.551$~meV and symmetric tunnel couplings $\lambda_L = \lambda_R = 0.025$ between the wire ends and the quantum dot.}
    \label{fig:Weak_dis_E_spec}
\end{figure*}

We examine the gate potential dependence of the lowest energy modes by selecting parameters corresponding to the topological phase of the nanowire in the open boundary condition, indicated by the black circle in Fig.~\ref{fig:Clean_sys_topo_regime}(a), with chemical potential $\mu=0.5$~meV and Zeeman field $\Gamma=0.65$~meV. Fig.~\ref{fig:Clean_sys_topo_regime}(b) shows the evolution of the lowest energy modes as a function of the quantum dot gate potential $V_{QD}$ at the Zeeman field indicated in panel (a). A pronounced resonance-lobe structure appears near $V_{QD}\sim0.53$~meV, slightly above the chemical potential. In the analysis below, we focus on the gate potential window $V_{QD}\in[0.49,0.62]$~meV that captures this resonance region.

Figs.~\ref{fig:Clean_sys_topo_regime}(c) and (d) display the corresponding parity-resolved differences in quantum capacitance and quantum inductance, defined as $\Delta\mathcal{C}_\Phi=\mathcal{C}_{\mathrm{even}}-\mathcal{C}_{\mathrm{odd}}$ and $\Delta\mathcal{L}_\Phi^{-1}=\mathcal{L}^{-1}_{\mathrm{even}}-\mathcal{L}^{-1}_{\mathrm{odd}}$, evaluated at $\Phi=0$ and $\Phi=h/2e$. As seen in Fig.~\ref{fig:Clean_sys_topo_regime}(c), $\Delta\mathcal{C}_\Phi$ exhibits pronounced features near the resonance regions around $V_{QD} \approx 0.52$~meV and $V_{QD} \approx 0.55$~meV. Notably the opposite signs of $\Delta\mathcal{C}_\Phi$ for $\Phi=0$ and $\Phi = h/2e$ at corresponding values of $V_{QD}$ demonstrate the expected flux induced oscillation of the parity sectors. The qualitative behavior of the quantum inductance closely parallels that of the capacitance, with it reversing sign as $V_{QD}$ is tuned from approximately $0.54$~meV to $0.56$~meV, coinciding with the quantum dot gate potential values at which the lowest energy BdG states switch signs, refer to Fig.~\ref{fig:Clean_sys_topo_regime}(b).

To examine the flux dependence explicitly, we select two representative gate voltages within the dominant resonance regions (marked by black stars in Fig.~\ref{fig:Clean_sys_topo_regime} panels (c) and (d)). The corresponding flux-dependent energy spectra, quantum capacitance, and quantum inductance are shown in Fig.~\ref{fig:Diff_vqd}. Blue (red) curves denote even (odd) parity sectors.

The left set of panels [(a), (c), and (e) in Fig.~\ref{fig:Diff_vqd}] corresponds to $V_{\mathrm{QD}} = 0.551$~meV. Fig.~\ref{fig:Diff_vqd}(a) shows the lowest-energy spectrum as a function of magnetic flux $\Phi$, exhibiting level crossings at $\Phi \sim h/4e$ and $\sim 3h/4e$. Consistent with these crossings, Fig.~\ref{fig:Diff_vqd}(c) shows $h/e$-periodic quantum capacitance responses with crossings between the even and odd parity sectors at the same flux values, and with the two parity curves shifted by $h/2e$. Fig.~\ref{fig:Diff_vqd}(e) shows the corresponding quantum inductance responses, which are likewise $h/e$ periodic, shifted by $h/2e$ between parity sectors, and exhibit crossings at the same flux values. These features are in direct agreement with the expectations from the topological effective model discussed in Sec.~\ref{Model1}.

The right set of panels [(b), (d), and (f) in Fig.~\ref{fig:Diff_vqd}] shows the corresponding responses for $V_{\mathrm{QD}} = 0.519$~meV, which lies in the second resonance region of $\Delta\mathcal{C}_\Phi$. As anticipated from the $V_{\mathrm{QD}}$ dependence of the low energy spectrum in Fig.~\ref{fig:Clean_sys_topo_regime}(b), the parity sectors are interchanged relative to the left panels. The characteristic features of the topological regime are preserved, including $h/e$-periodic responses, flux-shifted oscillations, and crossings at $\Phi \sim h/4e$ and $\sim 3h/4e$.

We note that the even-parity quantum inductance response in Fig.~\ref{fig:Diff_vqd}(e) exhibits a peak at $\Phi = h/2e$. This occurs because, although $V_{\mathrm{QD}}$ was chosen based on the resonance condition for $\Delta\mathcal{C}_\Phi$, it lies close to a gate potential at which $\Delta\mathcal{L}_\Phi^{-1}$ is close to switching sign between $\Phi = 0$ and $\Phi = h/2e$. Importantly, this does not alter the qualitative behavior of the inductance response, which continues to display all features expected in the topological regime. This observation further underscores that the quantum dot gate potential $V_{\mathrm{QD}}$ should not be selected solely based on prominent features in $\Delta\mathcal{C}$, but should also satisfy consistency with the corresponding behavior of $\Delta\mathcal{L}_\Phi^{-1}$, thereby enabling a more reliable identification of fermion parity switching.

Taken together, these results establish the characteristic flux-dependent behavior of the quantum capacitance and quantum inductance in the clean topological regime. The parity-resolved responses exhibit the same qualitative features anticipated from the effective models discussed in Sec.~\ref{sec:eff_model}, including $h/e$-periodic oscillations of each parity sector, an $h/2e$ flux shift between the parity branches, and crossings near $\Phi \sim h/4e$ and $\Phi \sim 3h/4e$. The clean system therefore provides a clear microscopic realization of the parity switching signatures predicted by the effective model analysis.

\subsection{Weakly Disordered System}
\label{sec:weak_dis_main}

To assess robustness under moderate perturbations, we introduce a weak disorder potential with amplitude $V_0 = 0.3$~meV. We plot the energy spectrum for the weakly disordered Hamiltonian given in Eq.~\ref{eq:Hamil_phi} in Fig.\ref{fig:Weak_dis_E_spec}. The energy spectrum is superimposed on a density plot of the topological stability $T_s$, which serves as an indicator of operational topology in the system by accounting for the QD gate potential $V_{QD}$ and the coupling strength to QD, $\lambda_{L,R}$.

\begin{figure*}
    \centering
    \includegraphics[width=\linewidth]{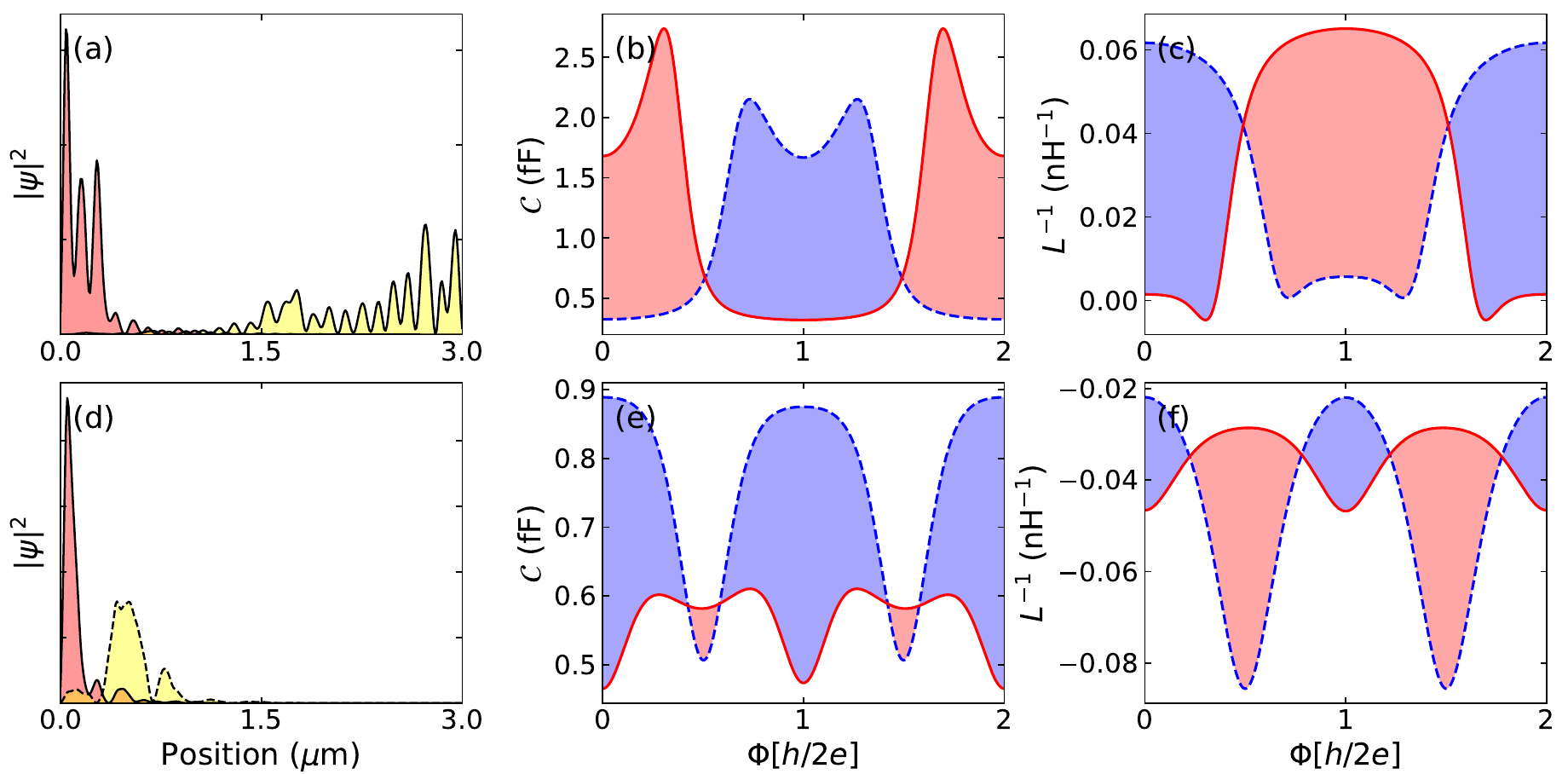}
    \caption{Comparison of topological and trivial regimes for a weakly disordered nanowire–QD system. Panels show (a) the Majorana wavefunction corresponding to a point in the topological (blue) region in Fig.~\ref{fig:Weak_dis_E_spec}, (b) the quantum capacitance, and (c) the quantum inductance as a function of magnetic flux $\Phi$ for a wire of length $L=3~\mu$m with disorder strength $V_0=0.3$~meV at chemical potential $\mu=0.5$~meV, Zeeman field $\Gamma=0.6$~meV, and gate potential $V_{\mathrm{QD}}=0.525$~meV. In this regime, the wavefunctions are localized near the wire ends and both the capacitance and inductance responses exhibit crossings at $\Phi = h/4e$ and $3h/4e$, consistent with a fermion parity switch and a negative topological stability $T_s=-1$ (refer to Eq.~\ref{eq:Topo_stability}). Panels (d)–(f) show the corresponding results for a point in the trivial (red) region in Fig.~\ref{fig:Weak_dis_E_spec}, for $\mu=0.5$~meV, $\Gamma=0.3$~meV, and $V_{\mathrm{QD}}=0.545$~meV, where the topological stability takes the value $T_s=+0.82$. In this regime, the wavefunctions are localized near the left end of the wire (highly overlapping) and both the capacitance and inductance responses are $h/2e$ periodic with a pair of narrow double crossing, and corresponding pair of extrema, respectively. In experimental measurements, the quantum capacitance curves corresponding to panels (b) and (e) would appear similar due to the absence of direct parity information. In contrast, the quantum inductance curves exhibit characteristically distinct features that enable the two cases to be distinguished. Solid red (dashed blue) curves denote the odd- (even-) parity branch in the capacitance and inductance panels.}
    \label{fig:Tr_Tp_in_weak_dis}
\end{figure*}

The topological stability is defined as the product of the Pfaffian invariant (including all the parameters relating to coupling to the QD) and the ratio of the minimum to maximum quasiparticle gap evaluated at $\Phi=0$ and $\Phi=h/2e$. Explicitly,
\begin{equation}
    T_s = \mathcal{Q}\,
    \frac{\min\!\left(E_g(\Phi=0),E_g(\Phi=h/2e)\right)}
    {\max\!\left(E_g(\Phi=0),E_g(\Phi=h/2e)\right)},
    \label{eq:Topo_stability}
\end{equation}
A positive value of this invariant indicates that no fermion parity switch occurs as $\Phi$ is tuned from $0$ to $h/2e$, while a negative value signals the presence of a fermion parity switch indicating the presence of non-trivial topology and Majorana zero modes. The magnitude of $T_s$ further encodes information about the degree of hybridization of the low energy states, with values close to $\pm 1$ corresponding to well developed gaps and values near zero indicating strong hybridization. The red region in Fig.\ref{fig:Weak_dis_E_spec} is indicative of topologically trivial regime (positive $T_s$) while the blue region corresponds to operationally topological regime (negative $T_s$). In addition, the explicit dependence of the topological stability $T_s$ on the QD coupling strength is examined in Appendix~\ref{sec:T_s_dependence}.

In the following analysis, we select two representative parameter sets $(\mu,\Gamma)$ from the phase diagram in Fig.~\ref{fig:Weak_dis_E_spec}: one in the trivial regime prior to the topological quantum phase transition, at $\mu=0.5$~meV and $\Gamma=0.30$~meV, and one in the topological regime beyond the transition, at $\mu=0.5$~meV and $\Gamma=0.66$~meV.

We now examine the quantum capacitance and quantum inductance responses of the weakly disordered system in both topological and trivial regimes to identify features that distinguish between them. Fig.~\ref{fig:Tr_Tp_in_weak_dis} shows the corresponding wavefunctions, quantum capacitance, and quantum inductance for the topological regime (top row, panels (a)–(c)) and the trivial regime (bottom row, panels (d)–(f)). In the topological regime, Fig.~\ref{fig:Tr_Tp_in_weak_dis}(a) shows a pair of well separated, robust Majorana zero mode wavefunctions for disorder strength $V_0=0.3$~meV at chemical potential $\mu=0.5$~meV and Zeeman field $\Gamma=0.6$~meV. The nanowire is coupled to the quantum dot with $\lambda_{L,R}=\lambda=0.025$, and the gate potential is fixed at $V_{QD}=0.525$~meV. The corresponding quantum capacitance, shown in Fig.~\ref{fig:Tr_Tp_in_weak_dis}(b), exhibits $h/e$ periodic oscillations with crossings near $\Phi\sim h/4e$ and $\Phi\sim 3h/4e$, such that the even and odd parity branches are flux shifted by $h/2e$. Fig.~\ref{fig:Tr_Tp_in_weak_dis}(c) shows the corresponding quantum inductance response, which is likewise $h/e$ periodic with crossings at the same flux values, and parity curves out of phase by $h/2e$. The consistency between the capacitance and inductance responses indicates a fermion parity switch, classifying this configuration as operationally topological.

In contrast, Fig.~\ref{fig:Tr_Tp_in_weak_dis}(d) shows partially separated Andreev bound states for the trivial regime at $\mu=0.5$~meV and $\Gamma=0.30$~meV, with the gate potential at $V_{QD}=0.545$~meV and all other parameters unchanged. The corresponding quantum capacitance, shown in Fig.~\ref{fig:Tr_Tp_in_weak_dis}(e), is $h/2e$ periodic and exhibits a pair of narrow double crossings near $\Phi\sim h/4e$ and $\Phi\sim 3h/4e$, resulting in no net interchange of the parity branches over a flux period. The quantum inductance response in Fig.~\ref{fig:Tr_Tp_in_weak_dis}(f) displays the same $h/2e$ periodicity and an even number of crossings (along with the expected pronounced extrema, refer to Sec.\ref{Model2}), consistent with the absence of a fermion parity switch. These features identify this parameter regime as topologically trivial. The structure of the quantum capacitance and quantum inductance responses depends sensitively on both the quantum dot gate potential $V_{QD}$ and the coupling strength between the dot and the nanowire. To systematically assess the influence of these parameters and how the two responses complement each other, we analyze the parity-resolved differences $\Delta\mathcal{C}=\mathcal{C}_{\mathrm{even}}-\mathcal{C}_{\mathrm{odd}}$ and $\Delta\mathcal{L}^{-1}=\mathcal{L}^{-1}_{\mathrm{even}}-\mathcal{L}^{-1}_{\mathrm{odd}}$ as functions of magnetic flux $\Phi$, Zeeman field $\Gamma$, and gate potential $V_{QD}$ in Appendix~\ref{sec:ps_C_L_inv_phi}.

Taken together, these results show that the characteristic quantum capacitance and inductance signatures identified in the effective model analysis remain robust in the presence of weak disorder. While the quantum capacitance profiles alone can display features that resemble parity crossings, the corresponding quantum inductance responses provide a clear distinction between true fermion parity–protected crossings and narrow double crossings. The weakly disordered system therefore provides a realistic setting in which the complementary diagnostic role of quantum inductance becomes evident.

\subsection{Strongly Disordered System}
\label{sec:strong_dis_main}

\begin{figure}
    \centering
    \includegraphics[width=\linewidth]{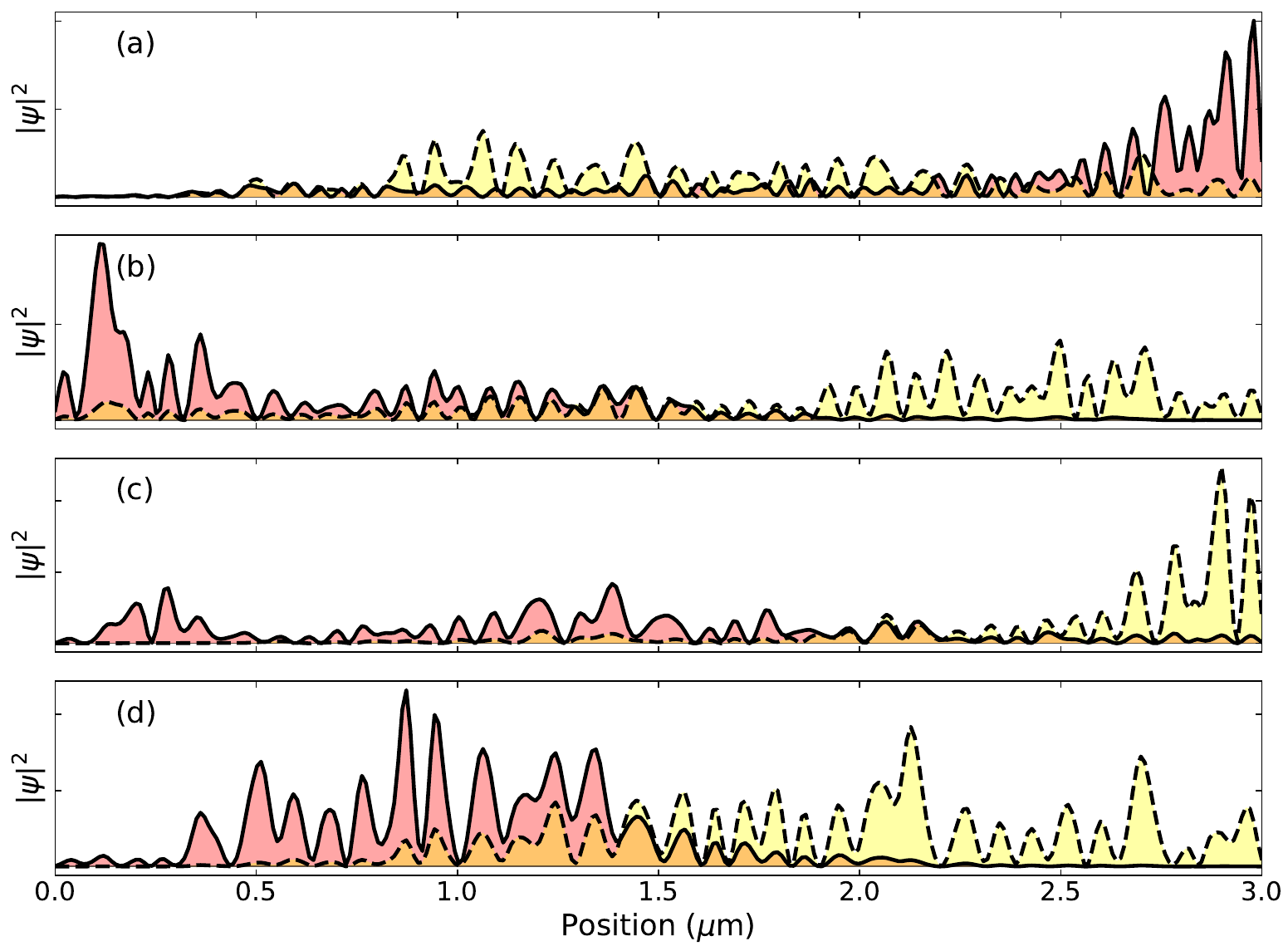}
    \caption{Spatial profiles of the wavefunctions associated with the low energy Bogoliubov-de Gennes eigenstates for the strongly disordered regime ($V_0=1.2$~meV). Panels (a)–(d) correspond to chemical potential and Zeeman field values of $(\mu,\Gamma) = (3.16,0.9822)$meV, $(2.10,1.079)$~meV, $(1.745,1.001)$~meV, and $(1.055,0.945)$~meV, respectively. In all cases, the spatial profiles correspond to partially separated, or strongly overlapping, Majorana zero energy modes.}
    \label{fig:Strong_dis_wfs}
\end{figure}

We now turn to the strongly disordered regime in order to investigate how reliably quantum capacitance and quantum inductance measurements can diagnose a fermion parity switch in the presence of fragmented topological regions. In this regime, disorder breaks the phase diagram into multiple disconnected topological islands, making the identification of parity protected crossings particularly challenging. The invariant maps in the $\mu-\Gamma$ parameter space are shown in Appendix~\ref{sec:strong_dis_reg} for reference. We select four representative pairs of chemical potential and Zeeman field values for detailed analysis, consisting of two parameter sets in the trivial regime and two in the operationally topological regime. The spatial profiles of the corresponding low energy eigenstates are shown in Fig.~\ref{fig:Strong_dis_wfs}.

\begin{figure}[H]
    \centering
    \includegraphics[width=\linewidth]{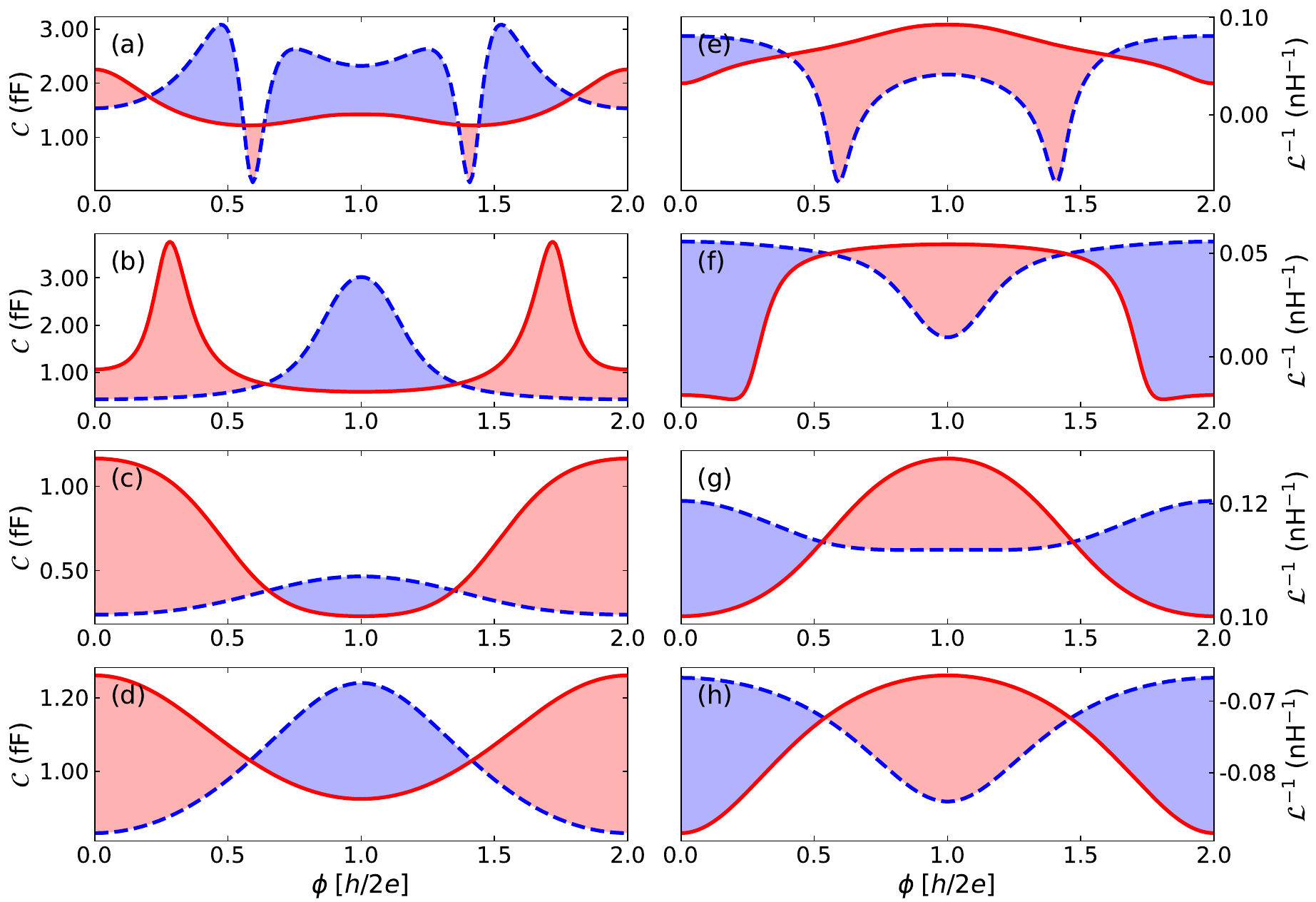}
    \caption{Flux dependent oscillations of the quantum capacitance and quantum inductance for a strongly disordered nanowire–QD system. Panels (a)–(d) show the quantum capacitance, while panels (e)–(h) show the corresponding quantum inductance, plotted as a function of magnetic flux $\Phi$. Each row corresponds to the same ($\mu,\Gamma$) values used in the corresponding row of Fig.~\ref{fig:Strong_dis_wfs}. Solid red and dashed blue curves denote the odd and even branches, respectively. The precise values of $V_{\mathrm{QD}}$ and $\lambda_{L,R}$ used in these plots are detailed in Appendix~\ref{sec:strong_dis_reg}.}
    \label{fig:Strong_dis_C_and_L_inv}
\end{figure}

\noindent In the trivial regime, the wavefunctions correspond either to strongly overlapping MZMs localized predominantly near one end of the wire [Fig.~\ref{fig:Strong_dis_wfs}(a)] or to states with substantial overlap distributed throughout the wire [Fig.~\ref{fig:Strong_dis_wfs}(d)]. In contrast, in the operationally topological regime, the wavefunctions shown in Fig.~\ref{fig:Strong_dis_wfs}(b) and (c) exhibit large characteristic length scales, reflecting the enhanced spatial extent induced by strong disorder. The representative $(\mu,\Gamma)$ values and associated choices of $\lambda$ and $V_{QD}$ were determined from a systematic parameter space analysis of the topological stability profile, low energy spectrum, and parity-resolved response functions. The corresponding spectral cuts, $\Delta\mathcal{C}_\Phi$ and $\Delta\mathcal{L}^{-1}_\Phi$ scans as a function of gate potential $V_{QD}$, are discussed in detail in Appendix~\ref{sec:strong_dis_reg}. The resulting parity-resolved oscillations as functions of magnetic flux $\Phi$ are shown in Fig.~\ref{fig:Strong_dis_C_and_L_inv}. 

We first examine the quantum capacitance responses shown in panels (a)–(d) of Fig.~\ref{fig:Strong_dis_C_and_L_inv}. The capacitance curves exhibit $h/e$ periodic oscillations for the individual parity branches, with crossings at specific flux values, although the detailed structure of these crossings varies across parameter regimes. We compare these features with the corresponding quantum inductance responses to assess whether the criteria identified in Sec.~\ref{sec:eff_model} for distinguishing true crossings from avoided or narrow double crossings remain valid in the strongly disordered regime. The quantum inductance responses, shown in panels (e)–(h) of Fig.~\ref{fig:Strong_dis_C_and_L_inv}, display $h/e$ periodic behavior, with parity crossings occurring at the same flux values as the quantum capacitance profiles.

In panel (a) of Fig.~\ref{fig:Strong_dis_C_and_L_inv}, the capacitance curves exhibit crossings between parity sectors near $\Phi\sim 0.15\,h/2e$ and $\Phi\sim 1.85\,h/2e$, which are accompanied by corresponding (slightly shifted) crossings in the quantum inductance profile shown in panel (e). In addition, narrow double crossing features appear in the capacitance near $\Phi\sim h/4e$ and $\Phi\sim 3h/4e$, and these are reflected in the quantum inductance by pronounced extrema at the same flux values. Thus, even in the presence of strong disorder, the inductance signatures associated with narrow double crossing scenarios, as identified in Sec.~\ref{sec:eff_model}, remain operative. For panels (b), (c), and (d) of Fig.~\ref{fig:Strong_dis_C_and_L_inv}, the capacitance profiles exhibit crossings between parity sectors near $\Phi\sim h/4e$ and $\Phi\sim 3h/4e$. The corresponding quantum inductance responses in panels (f), (g), and (h), respectively, also exhibit crossings at these same flux values, consistent with the presence of true crossings.

We further find that the response curves associated with parameter points identified as topologically trivial or nontrivial (refer to Appendix~\ref{sec:strong_dis_reg}) by the topological stability indicator (as defined in Eq.\ref{eq:Topo_stability}) differ significantly in how well they satisfy the criteria used in interferometric experiments\cite{MSRMicrosoft2024,arXivMicrosoft2024,PubMedMicrosoft2025} to infer a fermion parity switch. In particular, panels (a) and (c) of Fig.~\ref{fig:Strong_dis_C_and_L_inv}, corresponding to parameter points in the trivial regime, do not exhibit characteristics consistent with a robust fermion parity switch. In panel (a), the parity branches cross away from $\Phi\sim h/4e$ and $\Phi\sim 3h/4e$, while narrow double crossing regions appear at those flux values. In panel (c), $\Delta\mathcal{C}$ does not maintain a consistent magnitude as it changes sign with flux, undermining a parity-protected interpretation. By contrast, the capacitance curves associated with parameter points in the operationally topological regime [panels (b) and (d) of Fig.~\ref{fig:Strong_dis_C_and_L_inv}] exhibit well-defined crossings at $\Phi\sim h/4e$ and $\Phi\sim 3h/4e$, with $\Delta\mathcal{C}$ maintaining a consistent magnitude as $\Phi$ varies. These features are complemented by the corresponding quantum inductance responses, which indicate true crossings at the same flux values. 

Taken together, the combined analysis of quantum capacitance and quantum inductance provides a robust diagnostic for distinguishing genuine fermion parity–protected crossings from avoided or narrow double crossings. We emphasize that the appearance of pronounced extrema in the quantum inductance response, corresponding to avoided or narrow double crossing structures in the quantum capacitance profile, provides a sufficient criterion for distinguishing such false crossing scenarios from true fermion parity–protected crossings.  In conjunction with additional topological indicators such as $h/e$ periodicity of the individual parity sectors and the characteristic $h/2e$ flux shift between them, the combined analysis of quantum capacitance and quantum inductance yield a robust signature of a true fermion parity switch. However, the absence of an idealized crossing structure, or the coexistence of true crossings and narrow double crossings (as in panel (a) and (c) of Fig.~\ref{fig:Strong_dis_C_and_L_inv}), does not by itself imply that the system is trivial. Rather, in such cases the response becomes inconclusive within this diagnostic framework. Thus, while the presence of the combined quantum capacitance and inductance characteristic features constitutes strong evidence for a fermion parity–protected crossing, their absence does not definitively rule out the possibility of nontrivial topology. 

\begin{figure}
    \centering
    \includegraphics[width=\linewidth]{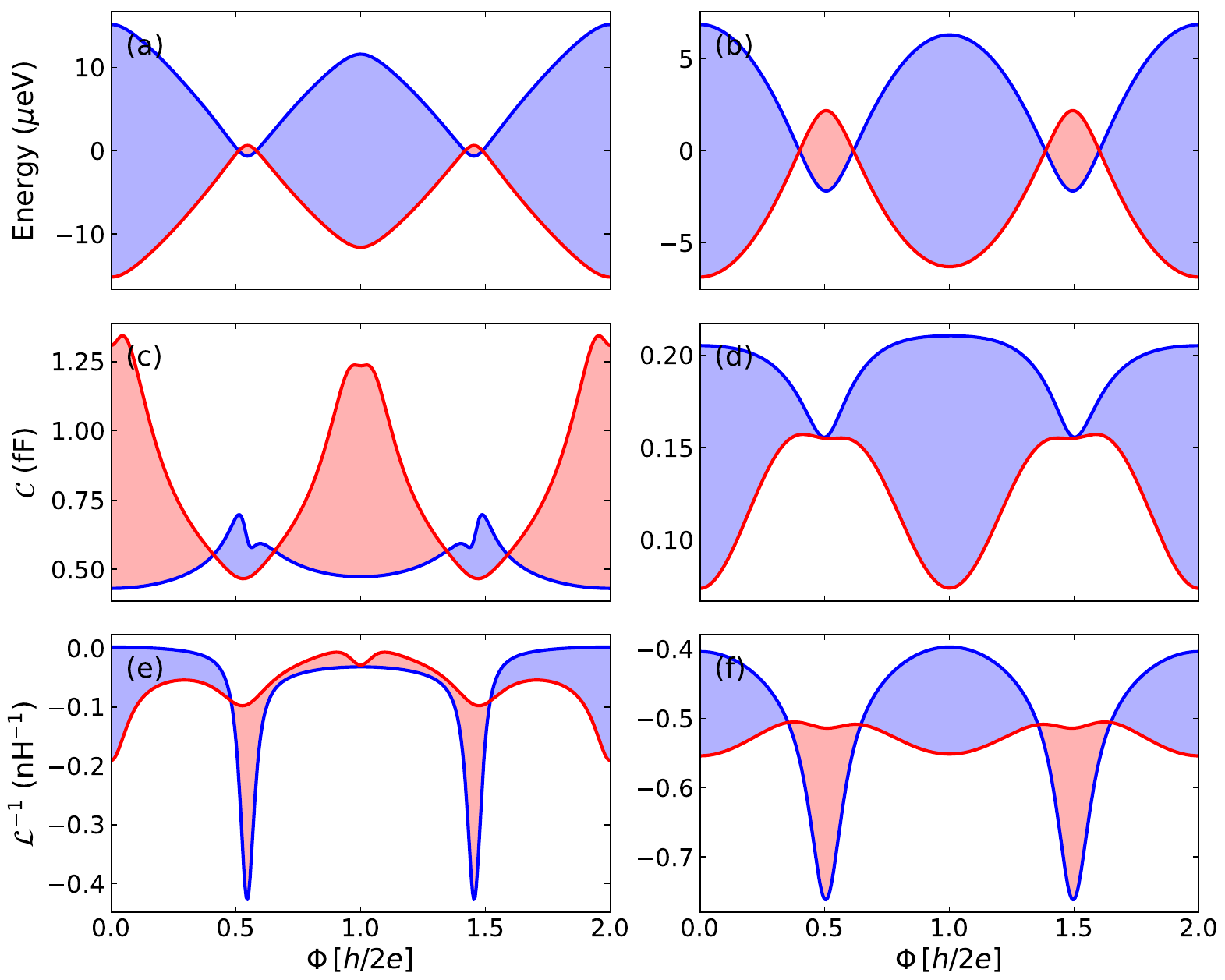}
    \caption{Flux dependent energy spectrum, quantum capacitance, and quantum inductance for a nanowire–QD system with Gaussian confinement at the wire ends. Panel (a) shows the lowest Bogoliubov-de Gennes energy eigenvalues as a function of magnetic flux $\Phi$ for a narrow double-crossing scenario, with the corresponding quantum capacitance shown in panel (c). These data correspond to the Gaussian end potential profile with an s.d. ($\sigma$) of 0.5$\mu$m at $\mu = 2.35$~meV, $\Gamma = 0.975$~meV, $\lambda_L = 0.04$, $\lambda_R = 0.05$, and $V_{\mathrm{QD}} = 2.512$~meV. Panels (b) and (d) show the corresponding flux dependent energy spectrum and quantum capacitance for an avoided crossing scenario associated with the Gaussian end potential profile with an s.d. ($\sigma$) of 0.7$\mu$m at $\mu = 1.5$~meV, $\Gamma = 0.391$~meV, $\lambda_L = \lambda_R = 0.09$, and $V_{\mathrm{QD}} = 1.961$~meV. Panels (e) and (f) show the corresponding flux dependent quantum inductance for the narrow double crossing and avoided crossing cases, respectively, exhibiting extrema near $\Phi \approx h/4e$ and $3h/4e$ in both scenarios. Solid red (dashed blue) curves denote the odd- (even-) parity branch in all panels.}
    \label{fig:Eg_avoided_double_Cap}
\end{figure}

\begin{figure*}
    \centering
    \includegraphics[width=\linewidth]{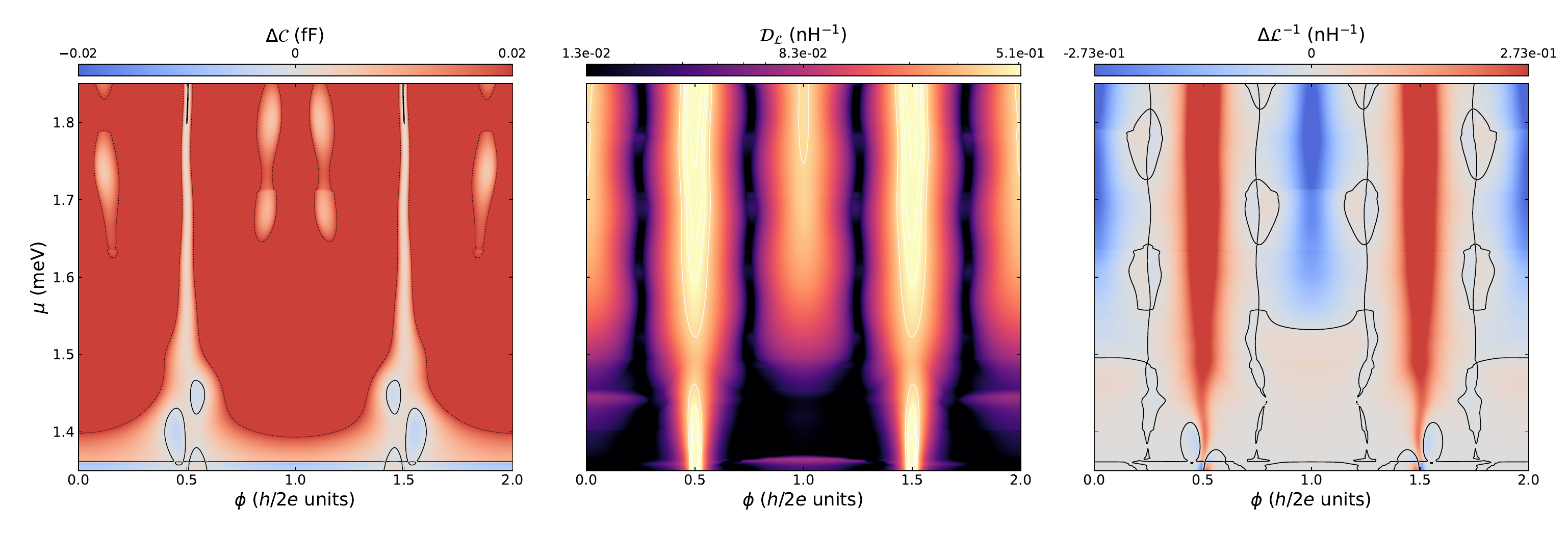}
    \caption{Three panel density plot illustrating the flux dependent parity resolved responses for a nanowire–QD system with Gaussian confinement. The left panel shows the difference in quantum capacitance $\Delta\mathcal{C}$ between the two parity branches as a function of magnetic flux $\Phi$ and chemical potential $\mu \in [1.35,1.85]$~meV. The center panel shows the corresponding trimmed deviation quantum inductance $\mathcal{D}_{\mathcal{L}}$ (see main text), highlighting extrema in the inductance response over the same range of chemical potential values. The right panel shows the difference in quantum inductance $\Delta\mathcal{L}^{-1}$ between the parity branches, indicating the presence or absence of crossings as $\mu$ is varied. All data correspond to the Gaussian confinement profile with $\sigma=0.7\mu$m, evaluated at Zeeman field $\Gamma = 0.391$~meV and tunnel coupling $\lambda = 0.09$. The quantum dot gate potential is chosen dynamically as $V_{\mathrm{QD}} = \mu + 0.461$~meV to remain close to resonance lobes for the selected range of chemical potentials.}
    \label{fig:Gauss_dis_dubious_crossing_mu_15}
\end{figure*}

\subsection{Gaussian End Potential}
\label{sec:smooth_conf_main}

In this section, we examine the role of smooth confinement in generating strongly overlapping MZMs by introducing Gaussian end potential profiles localized at the ends of the nanowire\cite{ChiuPRB2019,SciPostVuik2019}. The smooth confinement geometry provides a controlled setting in which avoided and narrow double crossing features in the quantum capacitance can be deliberately realized over extended parameter regimes. This construction enables a direct assessment of whether the quantum inductance response reliably distinguishes such false crossing scenarios from genuine fermion parity–protected crossings.

The Gaussian end potential profile used in this analysis is given by
\begin{equation}
    V_{\mathrm{Gauss}} = V_0 \left( e^{-(x-x_0)^2/2\sigma^2} + e^{-(x_N-x)^2/2\sigma^2} \right),
\end{equation}
where $x_0=0$ and $x_N=3.0~\mu$m denote the two end points of the nanowire. We consider standard deviation values $\sigma = 0.5~\mu$m and $\sigma = 0.7~\mu$m in the calculations below.

Fig.~\ref{fig:Eg_avoided_double_Cap} illustrates two representative examples of avoided and narrow double crossing scenarios. The left set of panels corresponds to a Gaussian profile with $\sigma=0.5~\mu$m at $\mu=2.4$~meV and $\Gamma=0.975$~meV, and exhibits a narrow double crossing in the capacitance parity curves shown in panel (c). The right set corresponds to a Gaussian profile with $\sigma=0.7~\mu$m at $\mu=1.5$~meV and $\Gamma=0.391$~meV, and displays an avoided crossing in the capacitance response shown in panel (d). In both cases, finite experimental resolution or limited signal-to-noise ratios (SNR) could lead to an erroneous identification of apparent crossings near $\Phi\sim h/4e$ and $\Phi\sim 3h/4e$, thereby producing false positives in capacitance-based parity detection.

The corresponding low energy spectra, shown in panels (a) and (b) of Fig.~\ref{fig:Eg_avoided_double_Cap}, demonstrate that no true fermion parity switch occurs in either case, as the energy levels both exhibit narrow double crossings rather than protected crossings. Crucially, the quantum inductance responses shown in panels (e) and (f) display pronounced extrema at the same flux values where the capacitance curves exhibit narrow double or avoided crossings. This behavior contrasts sharply with the inductance responses in genuinely topological regimes, where the parity-resolved inductance curves cross at $\Phi\sim h/4e$ and $\Phi\sim 3h/4e$, as observed previously in Fig.~\ref{fig:Diff_vqd} for the clean system, Fig.~\ref{fig:Tr_Tp_in_weak_dis}(c) for the weakly disordered topological regime, and Fig.~\ref{fig:Strong_dis_C_and_L_inv}(f)–(h) for the strongly disordered case.

Motivated by these observations, we next develop a more general characterization of avoided and narrow double crossing scenarios using density plots of the parity differences $\Delta\mathcal{C}=\mathcal{C}_e-\mathcal{C}_o$ and $\Delta\mathcal{L}^{-1}=\mathcal{L}^{-1}_e-\mathcal{L}^{-1}_o$, together with an additional quantity designed to quantify the characteristics of the pronounced extremum features in the quantum inductance responses. The first two quantities capture the raw parity contrast in the capacitance and inductance channels, respectively, while the third isolates information associated specifically with the diagnostic extremum structure. We define this diagnostic quantity as a parity-resolved trimmed deviation quantum inductance,
\begin{equation}
    \mathcal{D}_\mathcal{L}
    =
    \sum_{p=e,o}
    \left|
        \mathcal{L}^{-1}_p(\phi)
        -
        \mathrm{median}_{10\%-90\%}
        \!\left[\mathcal{L}^{-1}_p(\phi)\right]
    \right|,
    \label{eq:trim_dev_ind}
\end{equation}
where $\mathcal{L}^{-1}_{p}(\phi)$ denotes the flux-dependent quantum inductance for parity $p$, and $\mathrm{median}_{10\%-90\%}[\mathcal{L}^{-1}_p(\phi)]$ defines a parity-dependent baseline obtained by taking the median of the inductance response over the interval $\Phi\in[0,h/e]$ after excluding the upper $10\%$ and lower $10\%$ of the data to exclude the extrema. This trimmed deviation measure provides a quantitative means of identifying pronounced inductance features associated with avoided or narrow double crossings, thereby complementing the information contained in $\Delta\mathcal{C}$ and $\Delta\mathcal{L}^{-1}$.

Fig.~\ref{fig:Gauss_dis_dubious_crossing_mu_15} presents the parity-resolved transport diagnostics described above in the form of a three panel density plot as functions of magnetic flux $\Phi$ and chemical potential $\mu\in[1.35,1.85]$~meV, for fixed Zeeman field $\Gamma=0.391$~meV and a dynamically tuned quantum dot potential $V_{QD}=\mu+0.461$~meV. The left panel shows the capacitance difference $\Delta\mathcal{C}$, the center panel shows the trimmed deviation quantum inductance $\mathcal{D}_{\mathcal{L}}$, and the right panel shows the quantum inductance difference $\Delta\mathcal{L}^{-1}$. Over the entire $\mu$ range displayed, $\Delta\mathcal{C}$ does not change sign, indicating the absence of crossings between the parity sectors in the quantum capacitance response. Nevertheless, light red (white) regions appear near $\Phi\sim h/4e$ and $\Phi\sim 3h/4e$, where the parity-resolved capacitance curves approach one another closely and could be misidentified as crossings under finite experimental resolution. The center panel of Fig.~\ref{fig:Gauss_dis_dubious_crossing_mu_15} shows pronounced peaks in $\mathcal{D}_{\mathcal{L}}$ at these same flux values, coinciding with near-zero $\Delta\mathcal{C}$. These peaks indicate that the apparent capacitance crossings originate from avoided or narrow double crossings rather than true parity-protected crossings. The right panel, showing $\Delta\mathcal{L}^{-1}$, provides consistent information by displaying pronounced extremum features near $\Phi\sim h/4e$ and $\Phi\sim 3h/4e$, while also indicating that the quantum inductance responses are $h/2e$ periodic thus confirming absence of true crossing between the different parity sectors in the quantum capacitance responses.

\begin{figure}
    \centering
    \includegraphics[width=\linewidth]{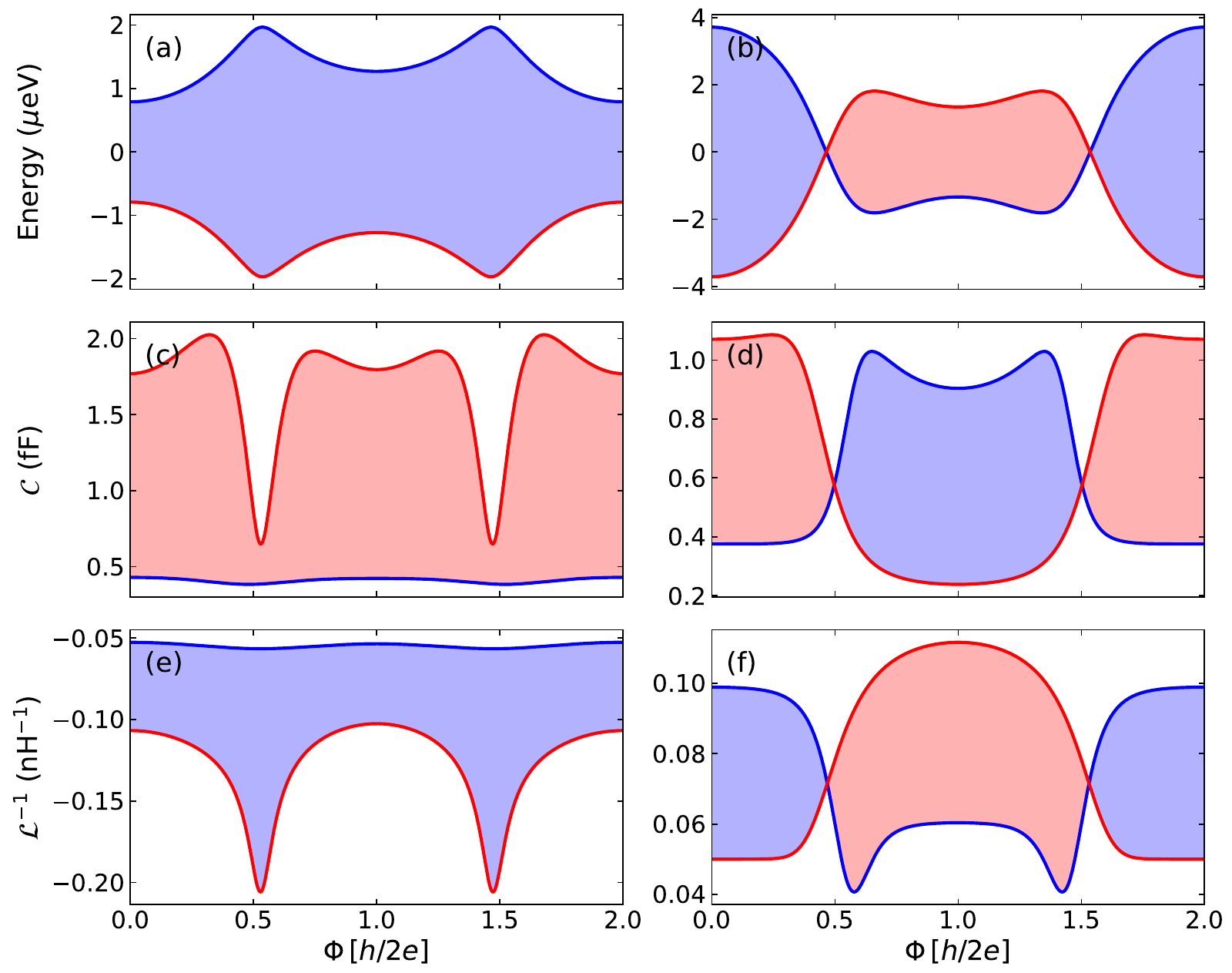}
    \caption{Flux-dependent energy spectrum, quantum capacitance, and quantum inductance for representative parameter sets in the strongly disordered regime. Panels (a), (c), and (e) correspond to an avoided crossing configuration. Panel (a) shows a gapped low energy spectrum as a function of magnetic flux $\Phi$, panel (c) shows the corresponding $h/2e$ periodic quantum capacitance exhibiting avoided crossings, and panel (e) shows the $h/2e$ periodic quantum inductance response displaying pronounced extrema near $\Phi \approx h/4e$ and $3h/4e$. Panels (b), (d), and (f) correspond to a true crossing configuration. Panel (b) shows crossings in the low energy spectrum consistent with a fermion parity switch, panel (d) shows crossings in the $h/e$ periodic quantum capacitance with consistent $\Delta\mathcal{C}$ values for the entire flux range, and panel (f) shows corresponding crossings in the $h/e$ periodic quantum inductance with consistent $\Delta\mathcal{L}^{-1}$ values throughout the entire flux range. For the selection criteria and exact values of $V_{\mathrm{QD}}$, $\lambda_{L,R}$, and other control parameters, see Appendix~\ref{sec:strong_dis_reg}.}
    \label{fig:Strongly_dis_actual_avoided_C_L_Phi}
\end{figure}

To contrast avoided and true crossings in a less fine-tuned setting, we select two representative parameter points from the strongly disordered regime (with the aid of the three-panel diagnostic density plots shown in Appendix~\ref{sec:strong_dis_reg}): (i) an avoided crossing case at $\mu=1.3589$~meV and $\Gamma=0.66$~meV, and (ii) a true crossing case at $\mu=3.11$~meV and $\Gamma=1.10$~meV. The corresponding flux-dependent energy spectra, quantum capacitance, and quantum inductance responses are shown in Fig.~\ref{fig:Strongly_dis_actual_avoided_C_L_Phi}. Panels (a), (c), and (e) in Fig.~\ref{fig:Strongly_dis_actual_avoided_C_L_Phi} correspond to the avoided crossing scenario, while panels (b), (d), and (f) correspond to the true crossing scenario.

In the avoided crossing case, the low energy spectrum in Fig.~\ref{fig:Strongly_dis_actual_avoided_C_L_Phi} panel (a) does not exhibit a true crossing, indicating the absence of a fermion parity switch. Consistently, the capacitance curves in panel (c) display avoided crossings near $\Phi\sim h/4e$ and $\Phi\sim 3h/4e$, which could be misidentified as true crossings under limited experimental resolution. The corresponding inductance response in panel (e) exhibits extrema at these same flux values, clearly signaling an avoided crossing.

In the true crossing case, Fig.~\ref{fig:Strongly_dis_actual_avoided_C_L_Phi} panel (b) shows a crossing in the low energy spectrum, indicating a fermion parity switch. The capacitance curves in panel (d) also cross at $\Phi\sim h/4e$ and $\Phi\sim 3h/4e$, with comparable $\Delta\mathcal{C}$ values throughout the entire flux range (and flux shifted by $h/2e$), consistent with the criteria established in interferometric parity detection experiments\cite{MSRMicrosoft2024,arXivMicrosoft2024,PubMedMicrosoft2025}. Crucially, the inductance response in panel (f) exhibits crossings at the same flux values, with even and odd parity curves of comparable magnitude and shifted by $h/2e$ relative to one another, rather than displaying a pair of extrema. This qualitative distinction in the inductance response provides a clear and robust separation between avoided and true crossings, demonstrating the utility of quantum inductance as a complementary diagnostic for identifying fermion parity–protected crossings in strongly disordered systems.

The microscopic BdG calculations presented above establish that the characteristic signatures identified in the effective models persist across a broad range of realistic device conditions. In particular, the combined behavior of the quantum capacitance and quantum inductance provides a reliable distinction between true fermion parity–protected crossings and avoided or narrow double crossings arising from trivial hybridization effects. Together, these results demonstrate that quantum inductance serves as a robust complementary diagnostic to quantum capacitance for identifying fermion parity switches in nanowire–QD systems, independent of whether the underlying mechanism arises from clean topological modes, disorder, or smooth confinement.

\section{Conclusion}

In this work, we have established flux-dependent quantum inductance as a robust and complementary diagnostic to quantum capacitance for identifying fermion parity switches in interferometric nanowire–QD systems. While quantum capacitance directly probes the low energy many-body spectrum variation with flux threaded through the nanowire-QD ring, we have shown that capacitance-based bimodal signatures alone are insufficient to unambiguously distinguish true fermion parity–protected crossings from avoided or narrow double crossings between different parity sectors arising from hybridization effects. In realistic devices, overlapping MZMs, disorder, or smooth confinement can generate capacitance responses that  mimic the $h/e$-periodic behavior expected of topological Majorana zero modes. Without additional information, such features can lead to false-positive interpretations of a fermion parity switch.

The central result of this work is that quantum inductance provides precisely this additional layer of discrimination. Because it probes the low energy derivatives with respect to flux induced phase of the Hamiltonian distinct from the capacitance response, it encodes complementary information about the change in curvature of the low energy spectrum. In the analytically tractable effective models (see Sec.~\ref{Transport}), we demonstrated explicitly that the flux variation of $\mathcal{C}(\Phi)$ and $\mathcal{L}^{-1}(\Phi)$ exhibits qualitatively different behavior depending on whether a genuine parity switch occurs. True parity-protected crossings produce simultaneous crossings at same flux values in both $\Delta\mathcal{C}$ and $\Delta\mathcal{L}^{-1}$ with consistent $h/e$ periodicity and $h/2e$ flux shifts between parity branches. By contrast, avoided or narrow double crossings in the quantum capacitance responses are accompanied by pronounced extrema in the quantum inductance profile at the same flux values, signaling trivial hybridization rather than a protected level crossing.

This distinction was confirmed in realistic microscopic simulations. In the smooth Gaussian confinement geometry (Sec.~\ref{sec:smooth_conf_main}), the flux dependence of the energy spectrum $E(\Phi)$, the quantum capacitance $\mathcal{C}(\Phi)$, and the quantum inductance $\mathcal{L}^{-1}(\Phi)$ revealed extended parameter regimes exhibiting avoided and narrow double crossing behavior. Although the capacitance curves alone can display apparent crossings near $\Phi\sim h/4e$ and $\Phi\sim 3h/4e$, the corresponding inductance response consistently develops pronounced extrema at these flux values, unambiguously identifying these features as trivial hybridization effects.

Similarly, in the weakly and strongly disordered regime (Sec.~\ref{sec:weak_dis_main} and \ref{sec:strong_dis_main}), we analyzed representative cases in which the flux variation of $E(\Phi)$, $\mathcal{C}(\Phi)$, and $\mathcal{L}^{-1}(\Phi)$ exhibits either avoided crossings, narrow double crossings or genuine crossings (refer to Figs.~\ref{fig:Tr_Tp_in_weak_dis} and \ref{fig:Strongly_dis_actual_avoided_C_L_Phi}). In the avoided or narrow double crossing case, the quantum inductance response develops extremum features characteristic of trivial hybridization. In contrast, in the genuine crossing case, both $\mathcal{C}(\Phi)$ and $\mathcal{L}^{-1}(\Phi)$ show crossings near $\Phi\sim h/4e$ and $\Phi\sim 3h/4e$, with no accompanying quantum inductance extrema. These results demonstrate that the quantum inductance response tracks the true topological character of the flux-driven level structure even in highly fragmented disorder landscapes.

Importantly, the combined quantum capacitance–inductance protocol should be interpreted as providing a sufficient condition for identifying a fermion parity switch within the interferometric geometry. The simultaneous presence of $h/e$ periodicity, $h/2e$ flux-shifted parity branches, and consistent crossings at same flux values in both $\mathcal{C}(\Phi)$ and $\mathcal{L}^{-1}(\Phi)$ signals a robust parity-protected crossing. Conversely, the presence of pronounced inductance extrema at putative crossing points invalidates apparent capacitance crossings. However, the absence of fully developed signatures in one diagnostic does not by itself establish triviality; rather, it identifies parameter regimes that remain inconclusive and may require optimization of the gate potential or coupling strengths. In this sense, quantum inductance serves not as a replacement for capacitance measurements, but as a decisive second tier of verification.

Taken together, our results demonstrate that quantum inductance substantially strengthens interferometric parity readout schemes by resolving ambiguities inherent in capacitance only diagnostics. The combined measurement of quantum capacitance and quantum inductance provides a systematic and experimentally accessible framework for distinguishing genuine fermion parity–protected crossings from trivial imitations across clean, smoothly confined, and strongly disordered nanowire systems. More broadly, this work highlights the importance of phase-sensitive response functions as probes of topological superconductivity and establishes a concrete methodology for implementing more reliable parity readout protocols in future Majorana-based quantum devices.

\section{ACKNOWLEDGEMENT}
S.T. and B.B.R. acknowledge support from SC Quantum and ONR-N000142312061.
J.S. acknowledges support from the Joint Quantum Institute and the Laboratory for Physical Sciences through the Condensed Matter Theory Center.
We also thank Tudor D. Stanescu for fruitful discussions.

\appendix
\section{Effective model of 2 spinful QDs}

To further elucidate how flux dependent pairing and gate tuning influence parity resolved spectra, we consider an auxiliary effective model consisting of two spinful, mutually decoupled quantum dots, each subject to induced superconducting pairing. This construction isolates the role of local pairing amplitudes and their flux dependent modulation without invoking Majorana operators explicitly. The parameters $\Delta_j$ describe the induced pairing on each dot due to proximity to a parent superconductor, while the coefficients $\zeta_j$ encode interference between direct and flux assisted tunneling paths, thereby introducing $\phi$ dependence into the effective pairing amplitude. By keeping the two dots decoupled, the model allows the many body spectrum to be constructed analytically from single dot solutions and makes the fermion parity structure transparent in the Fock basis. Fixing one dot to remain in the even parity sector effectively mimics a strongly proximitized region, while allowing the second dot to undergo parity transitions captures the essential spectral rearrangements relevant to the interferometric response. This simplified framework provides additional analytical insight into how flux modulated pairing alone can generate curvature changes and parity resolved structure in the low energy spectrum. The mutually decoupled quantum dots can be described by the Hamiltonian

\begin{align}
    H_{\mathrm{2QD}} &= (\epsilon_1 + \delta_{QD})\sum_{s} c^\dagger_sc_s + (\epsilon_2 + \delta_{QD})\sum_{s} d^\dagger_sd_s \nonumber \\
    & + \Delta_1(1+\zeta_1e^{i\phi})c_{\uparrow}c_{\downarrow} + \Delta_2(1+\zeta_2e^{i\phi})d_{\uparrow}d_{\downarrow} + h.c.
\end{align}

\noindent where $\delta_{QD}$ denotes the quantum dot gate potential and $\Delta_j$ is the induced superconducting pairing amplitude associated with the coupling of each dot to the parent superconductor. The parameters $\zeta_j$ encode the relative strength and phase of the flux dependent pairing contribution on each dot, arising from interference between direct and flux assisted tunneling processes. The superconducting phase $\phi$ enters through these terms and controls the flux dependence of the pairing amplitudes. We take dot $c$ to remain in the even parity sector throughout, representing a strongly proximitized site whose pairing is never blocked, such that fermion parity dynamics arise solely from dot $d$. Since the two dots are decoupled, we may first analyze a single dot problem to determine the many-body eigenstates and then construct the parity-resolved spectrum of the full two dot system.

We therefore begin by considering the single dot Hamiltonian

\begin{align}
    H_j = (\epsilon_j+\delta_{QD})\sum_s a^\dagger_{js}a_{js} + & \left[\Delta_j(1 + \zeta_je^{i\phi})a_{j\uparrow}a_{j\downarrow} \right. \nonumber \\ 
    & \left.+\Delta_j(1 + \zeta^*_je^{-i\phi})a^\dagger_{j\downarrow}a^\dagger_{j\uparrow}\right]
\end{align}

\noindent The Fock basis for the single dot Hamiltonian is given by $\{\ket{0},\ket{\uparrow},\ket{\downarrow},\ket{\uparrow\downarrow}\}$. In the odd parity sector spanned by $\ket{\uparrow}$ and $\ket{\downarrow}$, the superconducting pairing terms vanish identically, $a_{j\uparrow}a_{j\downarrow} = a^\dagger_{j\downarrow}a^\dagger_{j\uparrow} = 0$, yielding the eigenenergies

\begin{equation}
    E^{(odd)}_j = \epsilon_j + \delta_{QD}
\end{equation}

\noindent In the even parity sector spanned by $\ket{0}$ and $\ket{\uparrow\downarrow}$, the Hamiltonian takes the form

\begin{equation}
    H^{(even)}_j = \begin{pmatrix}
        0 & \Delta_j(1 + \zeta_je^{i\phi})\\
        \Delta_j(1 + \zeta^*_je^{-i\phi}) & 0
    \end{pmatrix}
\end{equation}

\noindent which yields the corresponding eigenenergies 

\begin{equation}
    E^{(even)}_j=\epsilon_j+\delta_{QD}\pm\sqrt{(\epsilon_j+\delta_{QD})^2+\lvert\Delta_j(1+\zeta_je^{i\phi})\rvert^2}
\end{equation} 

Since the two dots are decoupled, the many-body energies of the full system are obtained by summing the contributions from the individual dots. In the odd parity branch of the two dot system, dot $d$ ($j=2$) must be in the odd parity sector, while dot $c$ ($j=1$) remains in the even parity sector by construction. The resulting energy is therefore

\begin{align}
    E_{odd}(\phi) &= E^{(even)}_1 + E^{(odd)}_2 \nonumber \\
    & = (\epsilon_1+\delta_{QD}) + (\epsilon_2+\delta_{QD}) \nonumber \\ 
    &\pm \sqrt{(\epsilon_1+\delta_{QD})^2+\lvert\Delta_1(1+\zeta_1e^{i\phi})\rvert^2}
\end{align}

\noindent For the even parity branch, both dots occupy their respective even parity sectors, giving 

\begin{align}
    E_{even}(\phi) &= E^{(even)}_1 + E^{(even)}_2 \nonumber \\
    & = \left(\epsilon_1+\delta_{QD}\pm\sqrt{(\epsilon_1+\delta_{QD})^2+\lvert\Delta_1(1+\zeta_1e^{i\phi})\rvert^2}\right) \nonumber \\ & +\left(\epsilon_2+\delta_{QD}\pm\sqrt{(\epsilon_2+\delta_{QD})^2+\lvert\Delta_2(1+\zeta_2e^{i\phi})\rvert^2}\right)
\end{align}

\noindent The ground-state energy corresponds to the lowest-energy configuration in each parity sector and is given by

\begin{align}
    E^{gs} & = (\epsilon_1+\delta_{QD}) + (\epsilon_2+\delta_{QD})\nonumber \\ 
    &- \begin{cases}
    \displaystyle\sum_{j=1,2}\sqrt{(\epsilon_j+\delta_{QD})^2+\lvert\Delta_j(1+\zeta_je^{i\phi})\rvert^2}, & \text{even}\\
     \sqrt{(\epsilon_1+\delta_{QD})^2+\lvert\Delta_1(1+\zeta_1e^{i\phi})\rvert^2}, & \text{odd}
    \end{cases}
\end{align}

Having obtained analytical expressions for the ground-state energies, the corresponding response functions can be evaluated directly. The quantum capacitance follows from the second derivative with respect to the gate potential, $C \propto \partial^2 E / \partial \delta_{QD}^2$, while the quantum inductance is obtained from the second derivative with respect to the superconducting phase, $L^{-1} \propto \partial^2 E / \partial \phi^2$.

\begin{equation}
    C = \begin{cases}
        -\frac{\Delta^2_1\lvert A_1\rvert^2}{E^3_1} -\frac{\Delta^2_2\lvert A_2\rvert^2}{E^3_2}, & \text{for even parity}\\
        -\frac{\Delta^2_1\lvert A_1\rvert^2}{E^3_1}, & \text{for odd parity}\\
    \end{cases}
\end{equation}

\noindent where $A_j(\phi) = (1 + \zeta_j e^{i\phi})$ and $E_j = \sqrt{(\epsilon_j + \delta_{QD})^2 + \Delta_j^2 |A_j|^2}$. The quantity $|A_j(\phi)|^2$ can be expressed in the form $f(\phi) = P + Q\cos\phi + R\sin\phi$, analogous to the parametrizations introduced in Secs.~\ref{Model1} and \ref{Model2}.

\begin{equation}
    L^{-1} = \begin{cases}
        -\frac{\partial^2E_1}{\partial\phi^2} -\frac{\partial^2E_2}{\partial\phi^2}, & \text{for even parity} \\
        -\frac{\partial^2E_1}{\partial\phi^2}, & \text{for odd parity}
    \end{cases}
\end{equation}

\noindent \noindent where the second derivative of $E_j$ with respect to $\phi$ is given by

\begin{equation}
    \frac{\partial^2E_j}{\partial\phi^2} = \frac{\Delta^2_j}{2E_j}\frac{d^2f}{d\phi^2} - \frac{\Delta^4_j}{4E^3_j}\left(\frac{df}{d\phi}\right)^2
\end{equation}

\noindent Finally, we note that the flux controlled phase $\phi$ in this model is $h/2e$ periodic, reflecting its origin in Cooper-pair ($2e$) tunneling processes. This contrasts with the effective topological MZM model discussed in Sec.~\ref{Model1}, where single-electron tunneling gives rise to an effective $h/e$ flux periodicity.

\section{Derivation of quantum inductance formula}
\label{Derived_L_inverse}

We consider a Hamiltonian subject to a weak, time-dependent phase modulation induced by an applied magnetic flux. Expanding the Hamiltonian to linear order in the phase deviation, we write

\begin{equation}
    \mathcal{H}(t) = \mathcal{H}_0 + \mathcal{H}'(t)
\end{equation}

\noindent where the perturbation takes the form $\mathcal{H}'(t)=\delta\phi\cos(\omega t)\,\hat{J}$, with $\hat{J}=\partial_\phi\mathcal{H}\rvert_{\phi_0}$ the current operator evaluated at the equilibrium phase $\phi_0$. For a transition from an initial eigenstate $\ket{n}$ with energy $E_n$ to a final eigenstate $\ket{m}$ with energy $E_m$ of the unperturbed Hamiltonian $\mathcal{H}_0$, Fermi’s Golden Rule \cite{Sakurai2017} yields the total transition rate \cite{Mahan2000} for absorption of energy $\hbar\omega$ from the drive,

\begin{equation}
    W^{abs}_{n\rightarrow m} = \frac{2\pi}{\hbar}\left(\frac{\delta\phi}{2}\right)^2\sum_{n,m}\lvert\bra{n}\hat{J}\ket{m}\rvert^2\delta(E_m - E_n - \hbar\omega)
\end{equation}

\noindent while the corresponding transition rate for emission is

\begin{equation}
    W^{em}_{n\rightarrow m} = \frac{2\pi}{\hbar}\left(\frac{\delta\phi}{2}\right)^2\sum_{n,m}\lvert\bra{n}\hat{J}\ket{m}\rvert^2\delta(E_m - E_n + \hbar\omega)
\end{equation}

\noindent The net absorbed power is given by the rate of energy increase of the system,

\begin{align}
    P_{abs}(\omega) & = \sum_{m\neq0} p_n\left[(\hbar\omega)W^{abs}_{n\rightarrow m} + (-\hbar\omega)W^{em}_{n\rightarrow m}\right]
\end{align}

\noindent where $p_n$ denotes the occupation probability of the initial state. Substituting the expressions for the absorption and emission rates yields

\balance
\begin{widetext}
    \begin{equation}
        P_{abs}(\omega) =\frac{\pi\omega}{2}(\delta\phi)^2\sum_{n,m}p_n\lvert\bra{n}\hat{J}\ket{m}\rvert^2\left[\delta(E_m - E_n - \hbar\omega)-\delta(E_m - E_n + \hbar\omega)\right] 
    \end{equation}

    \noindent The emission term can be recast as an absorption process by exchanging the indices $n\leftrightarrow m$, which generates a factor $(p_n-p_m)$ and leads to

    \begin{equation}
        P_{abs}(\omega) =\frac{\pi\omega}{2}(\delta\phi)^2\sum_{n,m}(p_n-p_m)\lvert\bra{n}\hat{J}\ket{m}\rvert^2\delta(E_m - E_n - \hbar\omega)
    \end{equation}
\end{widetext}

For a linear electrical response\cite{Kubo1957}, the time-averaged absorbed power under an applied AC voltage is

\begin{equation}
    \overline{P(\omega)} = \overline{V(t)I(t)} = \frac{1}{2}\mathrm{Re}K(\omega)V^2_0
\label{eq:avg_pow}
\end{equation}

\noindent Using Josephson relations \cite{Ambegaokar1990}

\begin{equation}
    V(t) = \frac{\hbar}{2e}\frac{d\phi}{dt} = \frac{\hbar}{2e}\omega\delta\phi\cos{(\omega t)}
\end{equation}

\noindent we identify $V_0 = (\hbar/2e)\omega\delta\phi$. Substituting into Eq.~(\ref{eq:avg_pow}) gives

\begin{equation}
    \overline{P(\omega)} = \frac{1}{2}\mathrm{Re}Y(\omega)\left(\frac{\hbar}{2e}\right)^2\omega^2(\delta\phi)^2
\end{equation}

\noindent where $Y(\omega)$ is the electrical admittance, defined as $Y(\omega)=\delta I(\omega)/\delta V(\omega)$, and is related to the phase stiffness kernel (quantum inductance) $K(\omega)$ via

\begin{equation}
    Y(\omega) = \left(\frac{2e}{\hbar}\right)^2\frac{K(\omega)}{i\omega}
\label{eq:admit_impede}
\end{equation}

\noindent It follows that $\mathrm{Re}\,Y(\omega) = (2e/\hbar)^2\,\mathrm{Im}[K(\omega)/\omega]$. Equating the absorbed power obtained from linear response with the expression derived from Fermi’s Golden Rule yields

\balance
\begin{widetext}
\noindent Equating $\overline{P(t)}=P_{abs}$,
    \begin{equation}
    \frac{1}{2}\mathrm{Im}K(\omega)\omega(\delta\phi)^2 = \frac{\pi\omega}{2}(\delta\phi)^2\sum_{n,m}(p_n-p_m)\lvert\bra{n}\hat{J}\ket{m}\rvert^2\delta(E_m - E_n - \hbar\omega)
\end{equation}

\noindent which gives the imaginary part of the stiffness kernel as

\begin{equation}
    \mathrm{Im}K(\omega) = \pi\sum_{n,m}(p_n-p_m)\lvert\bra{n}\hat{J}\ket{m}\rvert^2\delta(E_m - E_n - \hbar\omega)
\label{eq:imag_k_omega}
\end{equation}

\noindent \noindent The real part of $K(\omega)$ is then obtained using the Kramers–Kronig\cite{LandauLifshitzStatPhys2} relation,

\begin{equation}
    \mathrm{Re}K(\omega) = 
    \frac{1}{\pi}\mathcal{P}\int^\infty_{-\infty}\frac{\mathrm{Im}K(\omega')}{\omega'-\omega}d\omega'
\label{eq:KK_rule}
\end{equation}

\noindent Substituting $\mathrm{Im}K(\omega)$ into Eq.~(\ref{eq:KK_rule}) yields

\begin{equation}
    \mathrm{Re}K(\omega) = \frac{1}{\pi}\mathcal{P}\int^\infty_{-\infty}d\omega'\frac{1}{\omega'-\omega}\left[\pi\sum_{n,m}(p_n-p_m)\lvert J_{nm}\rvert^2\delta(E_m - E_n - \hbar\omega')\right]
\label{eq:Int_step_1}
\end{equation}

\noindent Rewriting the delta function as,

\begin{equation}
    \delta(E_m - E_n - \hbar\omega') = \delta(\hbar(\omega_m-\omega_n)-\hbar\omega') = \frac{1}{\hbar}\delta(\omega_{mn}-\omega') = \frac{1}{\hbar}\delta(\omega'-\omega_{mn})
\label{eq:sub_1}
\end{equation}

\noindent Plugging Eq.\ref{eq:sub_1} into Eq.\ref{eq:Int_step_1},

\begin{equation}
    \mathrm{Re}K(\omega) = \frac{1}{\hbar}\left[\sum_{n,m}(p_n-p_m)\lvert J_{nm}\rvert^2\mathcal{P}\int^\infty_{-\infty}d\omega'\frac{\delta(\omega'-\omega_{mn})}{\omega'-\omega}\right]
\end{equation}

\noindent Evaluating the integral then yields

\begin{equation}
    \mathrm{Re}K(\omega) = \sum_{n,m}(p_n-p_m)\lvert J_{nm}\rvert^2\mathcal{P}\left[\frac{1}{\hbar\omega_{mn}-\hbar\omega}\right]
\end{equation}

\noindent Symmetrizing the expression using $\omega_{mn}=\omega_{nm}$, $|J_{nm}|^2=|J_{mn}|^2$, and $(p_m-p_n)=-(p_n-p_m)$ leads to

\begin{equation}
    \mathrm{Re}K(\omega) = \frac{1}{2}\sum_{n,m}(p_n-p_m)\lvert J_{nm}\rvert^2\left[\mathcal{P}\frac{1}{\hbar\omega_{mn}-\hbar\omega}+\mathcal{P}\frac{1}{\hbar\omega_{mn}+\hbar\omega}\right]
\end{equation}

\noindent resulting in the final expression \cite{Kos2013}

\begin{equation}
    \mathrm{Re}K(\omega) = \sum_{n,m}(p_n-p_m)\lvert J_{nm}\rvert^2\mathcal{P}\left[\frac{\hbar\omega_{mn}}{(\hbar\omega_{mn})^2-(\hbar\omega)^2}\right]
\end{equation}

\noindent Combining the real and imaginary components yields the paramagnetic contribution to the phase stiffness kernel,

\begin{equation}
    K_{\mathrm{para}}(\omega) = -2\sum_{n,m}(p_n-p_m)\lvert J_{nm}\rvert^2\left[\frac{E_m-E_n}{(E_m-E_n)^2-(\hbar(\omega+i\eta))^2}\right]
\end{equation}

\noindent where the indices $n,m$ satisfy the same constraints specified in Sec.~\ref{Transport} and $\eta$ is a finite phenomenological broadening. Since the Hamiltonian depends nonlinearly on $\phi$, there is also a diamagnetic contribution,

\begin{equation}
    Y_{\mathrm{dia}}(\omega) = \frac{2e}{\hbar}\frac{\langle \widehat{D}\rangle}{i\omega}
\end{equation}

\noindent where $\widehat{D}\equiv\partial_\phi^2\mathcal{H}$. Using Eq.~(\ref{eq:admit_impede}), the full dynamic stiffness kernel can be written as

\begin{equation}
    K(\omega) = \langle \widehat{D}\rangle_n-2\sum_{n,m}(p_n-p_m)\lvert J_{nm}\rvert^2\left[\frac{E_m-E_n}{(E_m-E_n)^2-(\hbar(\omega+i\eta))^2}\right]
\end{equation}

\noindent The corresponding dynamic quantum inductance is therefore

\begin{equation}
    \mathcal{L}^{-1} = \left(\frac{2e}{\hbar}\right)^2K(\omega)
\end{equation}

\end{widetext}

\section{Topological stability metric}
\label{sec:T_s_dependence}

\begin{figure}[b]
    \centering
    \includegraphics[width=\linewidth]{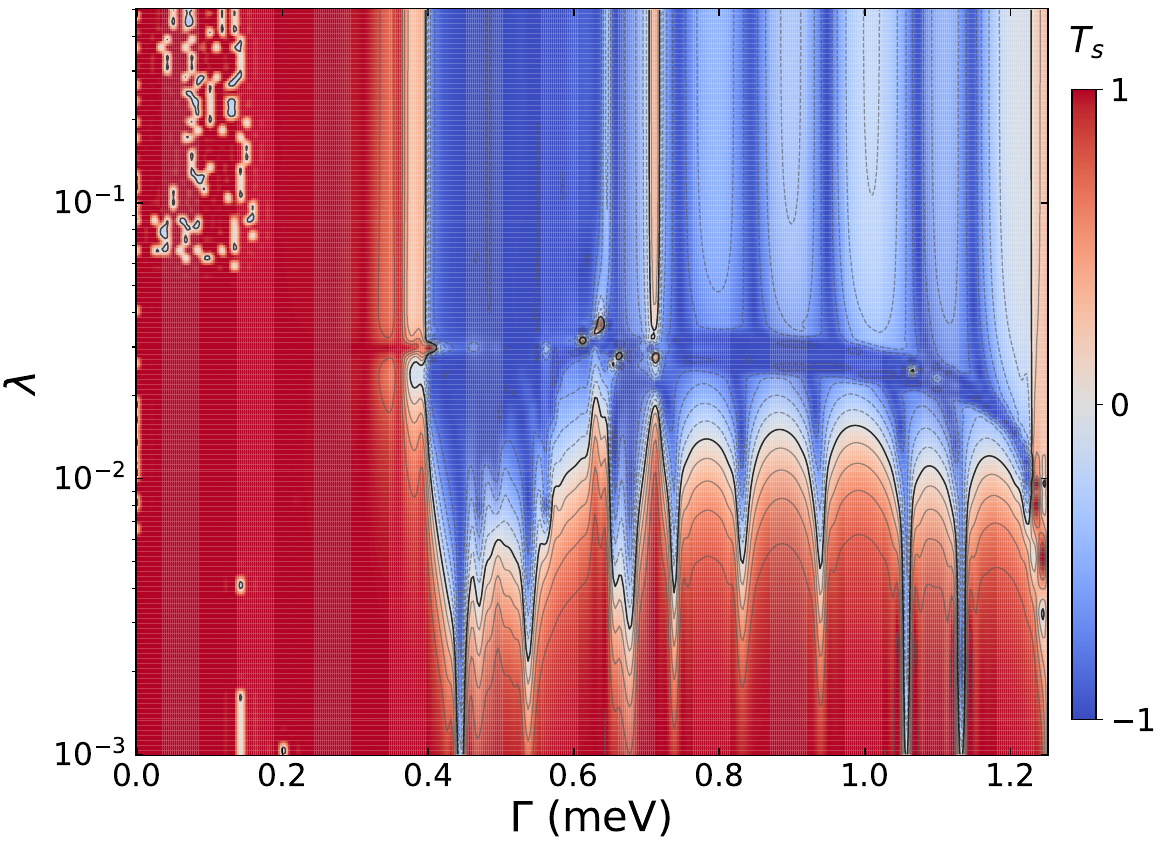}
    \caption{Dependence of the topological stability $T_s$ (refer to Eq.~\ref{eq:Topo_stability}) on the tunnel coupling strength $\lambda$ between the wire ends and the quantum dot for a weakly disordered system. The horizontal axis denotes the Zeeman field $\Gamma$, varied over the range $\Gamma \in [0,1.25]$~meV, while the vertical axis shows the coupling strength $\lambda$, plotted on a logarithmic scale from $0.001$ to $0.5$~meV. The chemical potential and quantum dot gate potential are fixed at $\mu = 0.5$~meV and $V_{\mathrm{QD}} = 0.551$~meV, respectively. Blue (red) regions correspond to negative (positive) values of $T_s$, indicating operationally topological (trivial) regimes.}
    \label{fig:Ts_sweep_lambda}
\end{figure}

The purpose of this analysis is to quantify how the coupling to the quantum dot modifies the operationally topological character of the system. Fig.~\ref{fig:Ts_sweep_lambda} shows the dependence of the topological stability $T_s$ (see Eq.~\ref{eq:Topo_stability}) on the tunnel coupling strength $\lambda$ between the nanowire ends and the quantum dot for a weakly disordered system. The horizontal axis corresponds to the Zeeman field $\Gamma$, varied over the range $\Gamma \in [0,1.25]$~meV, while the vertical axis shows the coupling strength $\lambda$ on a logarithmic scale from $0.001$ to $0.5$. The chemical potential and quantum dot gate potential are fixed at $\mu = 0.5$~meV and $V_{\mathrm{QD}} = 0.551$~meV, respectively. Blue (red) regions denote negative (positive) values of $T_s$, corresponding to operationally topological (trivial) regimes. Three distinct regimes can be identified: (i) an effectively uncoupled regime $\lambda \ll 0.01$, (ii) a weakly coupled regime $\lambda \sim 0.01$, and (iii) a strongly coupled regime $\lambda \gg 0.01$.

In the effectively uncoupled regime, the absence of quasiparticle transfer across the quantum dot prevents changes in fermion parity, and the Pfaffian invariant \cite{PUKitaev2001} remains positive across the Zeeman field range considered. In the weakly coupled regime, $\lambda$ strongly influences the hybridization between end states. In the presence of disorder or finite-size effects, insufficient spatial separation of these states can lead to hybridization induced quasiparticle exchange, thereby suppressing a fermion parity switch. In the strongly coupled regime, the sensitivity to $\lambda$ weakens again, as the quantum dot no longer introduces additional hybridization beyond that set by the local nanowire spectrum and effectively behaves as an additional site in the wire. All results presented in the main text correspond to the weakly coupled regime, consistent with the objective of using the quantum dot as a minimally invasive probe of the transport response.

\begin{figure*}[ht]
    \centering
    \includegraphics[width=\linewidth]{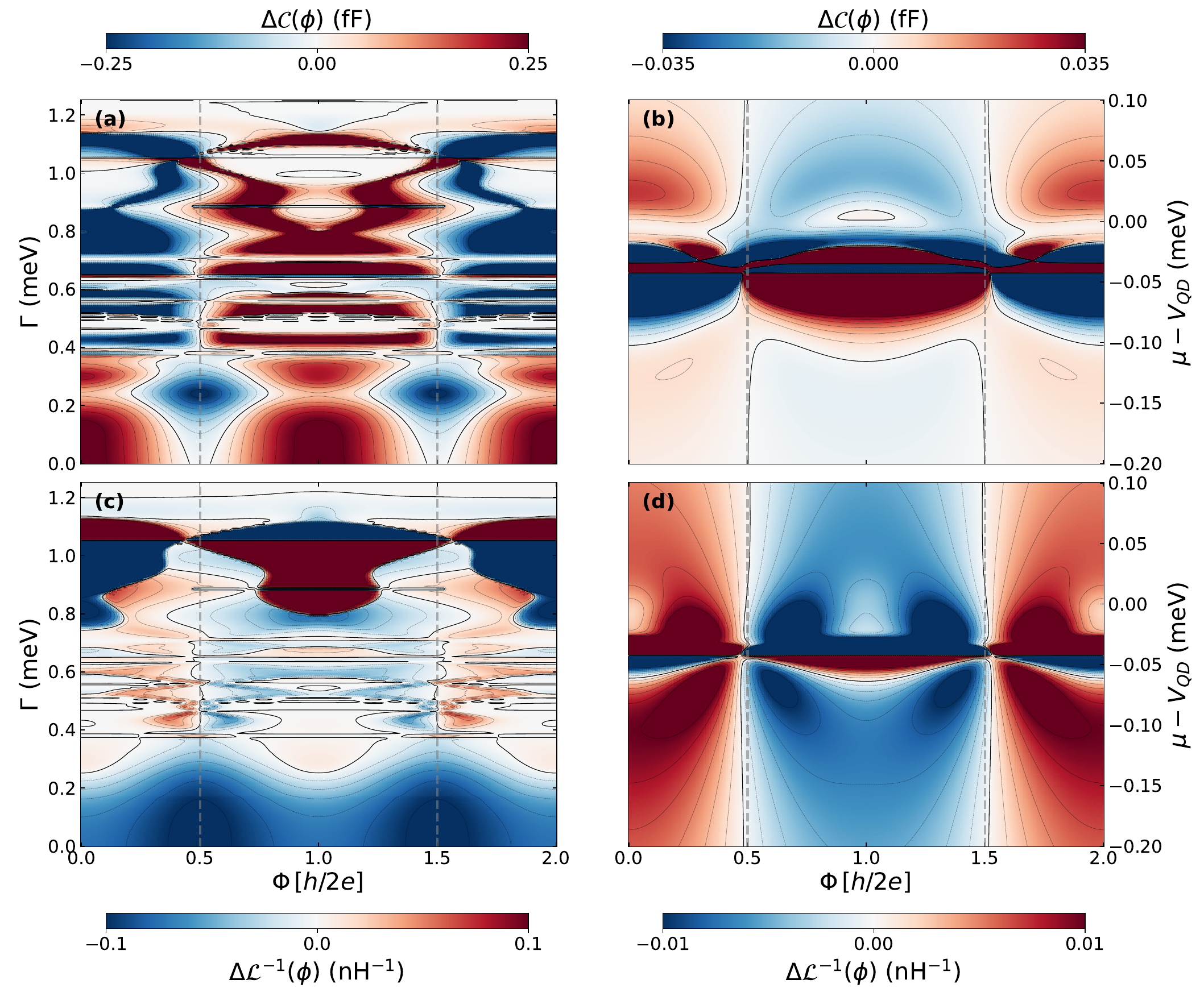}
    \caption{Density plots of the parity resolved differences in quantum capacitance and quantum inductance for a weakly disordered nanowire–QD system at a fixed chemical potential. Panels (a) and (b) show $\Delta\mathcal{C} = C_e - C_o$ as a function of magnetic flux $\Phi$ for varying (a) Zeeman field $\Gamma$ and (b) quantum dot gate potential $V_{\mathrm{QD}}$. Panels (c) and (d) show the corresponding quantum inductance difference $\Delta\mathcal{L}^{-1} = L^{-1}_e - L^{-1}_o$ plotted over the same ranges of $\Phi$ with (c) varying $\Gamma$ and (d) varying $V_{\mathrm{QD}}$. All panels correspond to a weakly disordered system with disorder strength $V_0 = 0.3$~meV, chemical potential $\mu = 0.5$~meV, and tunnel coupling $\lambda = 0.025$ between the nanowire and the quantum dot. For panels (b) and (d), the Zeeman field and gate potential are fixed at $\Gamma = 0.66$~meV and $V_{\mathrm{QD}} = 0.551$~meV, respectively.}
    \label{fig:Delta_C_L_density_plot}
\end{figure*}

We further note that the topological stability $T_s$ also depends explicitly on the quantum dot gate potential $V_{\mathrm{QD}}$ through its inclusion in the full Hamiltonian. However, $V_{\mathrm{QD}}$ is not treated as an independent control parameter in the present analysis. For each selected nanowire parameter set, the gate potential is chosen based on the location of the resonance in the parity-resolved responses $\Delta\mathcal{C}$ and $\Delta\mathcal{L}^{-1}$. Importantly, the resonant value of $V_{\mathrm{QD}}$ itself depends on the coupling strength $\lambda$: increasing $\lambda$ shifts the resonance away from the chemical potential. Consequently, just examining the dependence of $T_s$ on $\lambda$ is sufficient for the purposes of this study. Separate scans of $T_s$ as a function of $V_{\mathrm{QD}}$ are therefore not performed, since the capacitance and inductance diagnostics are evaluated at the gate potentials that optimize their visibility.

\section{Parameter space dependence of $\Delta\mathcal{C}(\Phi)$ and $\Delta\mathcal{L}^{-1}(\Phi)$}
\label{sec:ps_C_L_inv_phi}

\begin{figure}[b]
    \centering
    \includegraphics[width=\linewidth]{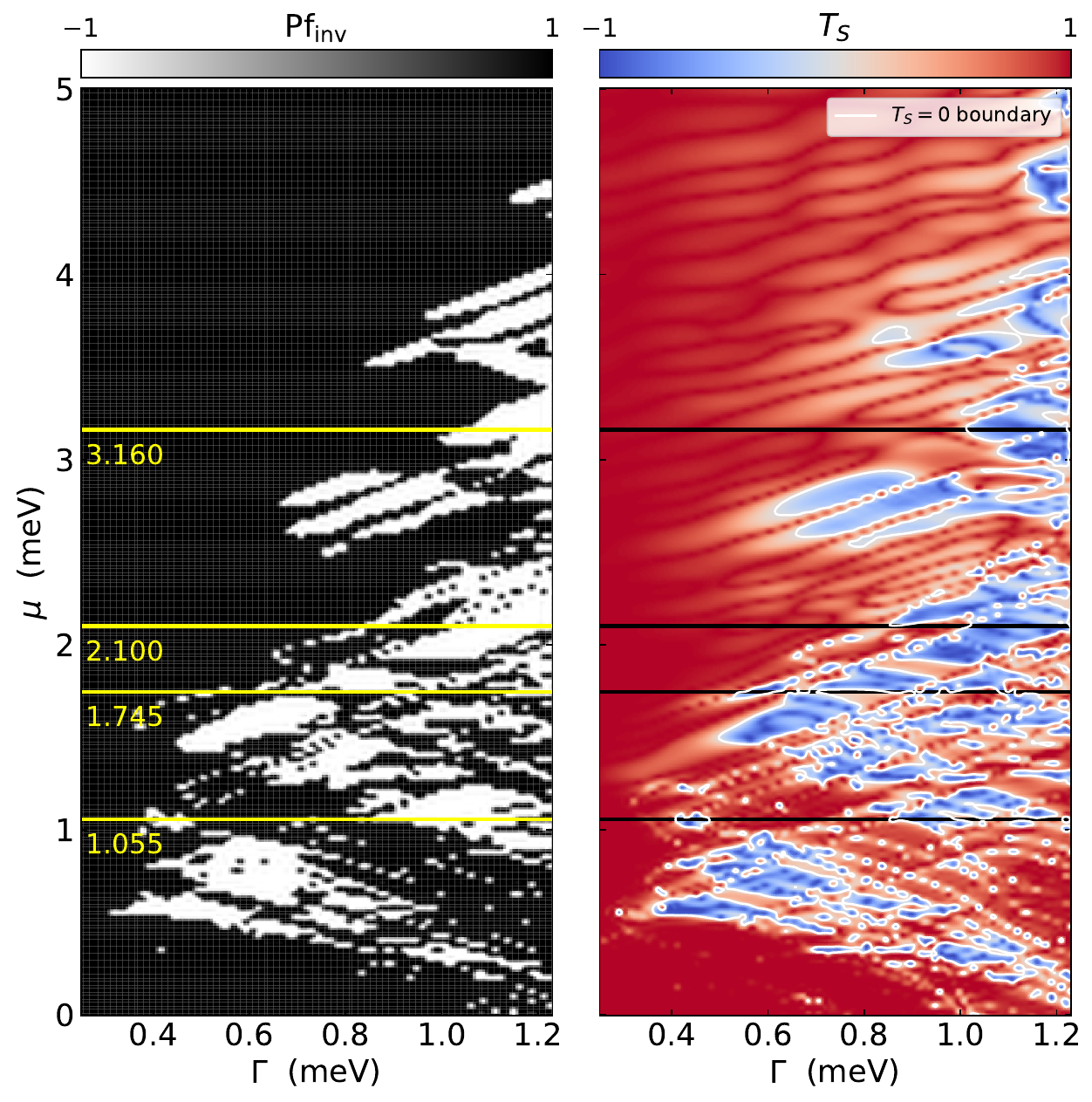}
    \caption{Invariant maps for a strongly disordered nanowire–QD system with disorder strength $V_0 = 1.2$~meV. The left panel shows the Pfaffian invariant (refer to Eq.~\ref{eq:pfaff}) $\mathcal{Q}$, where white (black) regions correspond to $\mathcal{Q} = -1$ ($+1$), indicating operationally topological (trivial) regimes. The right panel shows the corresponding topological stability (refer to Eq.~\ref{eq:Topo_stability}) $T_s \in [-1,1]$ plotted as a color density map. The chemical potential $\mu$ is varied along the vertical axis from $0$ to $5.0$~meV and the Zeeman field $\Gamma$ along the horizontal axis from $0.25$ to $1.25$~meV. The quantum dot gate potential is chosen to track the chemical potential as $V_{\mathrm{QD}} = \mu + 0.045$, and the tunnel coupling between the nanowire and the quantum dot is fixed at $\lambda = 0.025$.}
    \label{fig:INV_maps}
\end{figure} 

In this section, we examine how the parity-resolved quantum capacitance and quantum inductance depend on the quantum dot gate potential $V_{QD}$ and the Zeeman field $\Gamma$. Because these observables probe different derivatives of the flux-dependent energy spectrum, their visibility and diagnostic strength can vary across parameter space. A systematic analysis of their behavior as functions of $\Phi$, $\Gamma$, and $V_{QD}$ is therefore necessary to assess how reliably they identify fermion parity switches and to determine optimal gate settings for resolving topological features.

Fig.~\ref{fig:Delta_C_L_density_plot} shows the parity-resolved differences in quantum capacitance and quantum inductance for a weakly disordered nanowire–QD system. Panels (a) and (b) display $\Delta\mathcal{C}=C_e-C_o$ as functions of magnetic flux $\Phi$ while varying the Zeeman field $\Gamma$ and the quantum dot gate potential $V_{\mathrm{QD}}$, respectively. Panels (c) and (d) show the corresponding quantum inductance difference $\Delta\mathcal{L}^{-1}=L^{-1}_e-L^{-1}_o$ over the same parameter ranges.

Fig.~\ref{fig:Delta_C_L_density_plot}(a) shows $\Delta\mathcal{C}$ as a function of $\Phi$ with increasing $\Gamma$ for fixed $\mu=0.5$~meV and $V_{QD}=0.551$~meV. Referring to the topological quantum phase transition identified in Sec.~\ref{sec:weak_dis_main} in Fig.~\ref{fig:Weak_dis_E_spec}, for $\Gamma \lesssim 0.4$~meV the capacitance difference exhibits $h/2e$-periodic behavior in $\Phi$, consistent with a topologically trivial regime and in agreement with the corresponding $T_s$ density plot. For $\Gamma \gtrsim 0.4$~meV, $\Delta\mathcal{C}$ becomes $h/e$ periodic over a broad Zeeman-field range extending to $\Gamma\sim1.2$~meV, beyond which the parent superconducting gap collapses. In this regime, capacitance crossings occur near $\Phi\sim h/4e$ and $\Phi\sim 3h/4e$, indicated by sign changes from blue (negative $\Delta\mathcal{C}$) to red (positive $\Delta\mathcal{C}$), consistent with a fermion parity switch and an operationally topological phase.

The dependence of $\Delta\mathcal{C}$ on the quantum dot gate potential is shown in Fig.~\ref{fig:Delta_C_L_density_plot}(b) for fixed $\mu=0.5$~meV and $\Gamma=0.66$~meV. A pronounced sign change in $\Delta\mathcal{C}$ occurs near $V_{QD}\sim0.54$~meV, corresponding to a horizontal cut through the density plot where the response amplitude is largest. This enhanced response arises when $V_{QD}$ lies close to the nanowire chemical potential, consistent with the resonance-lobe structure observed in Fig.~\ref{fig:Clean_sys_topo_regime}(c). Apparent discontinuities in the vertical direction reflect limitations of parity tracking in a multidimensional parameter sweep. In the present calculations, parity is tracked as $\Phi$ is varied from $0$ to $h/e$ for each fixed value of $V_{QD}$; continuous tracking across all parameter directions would be computationally prohibitive.

\begin{figure}[b]
    \centering
    \includegraphics[width=\linewidth]{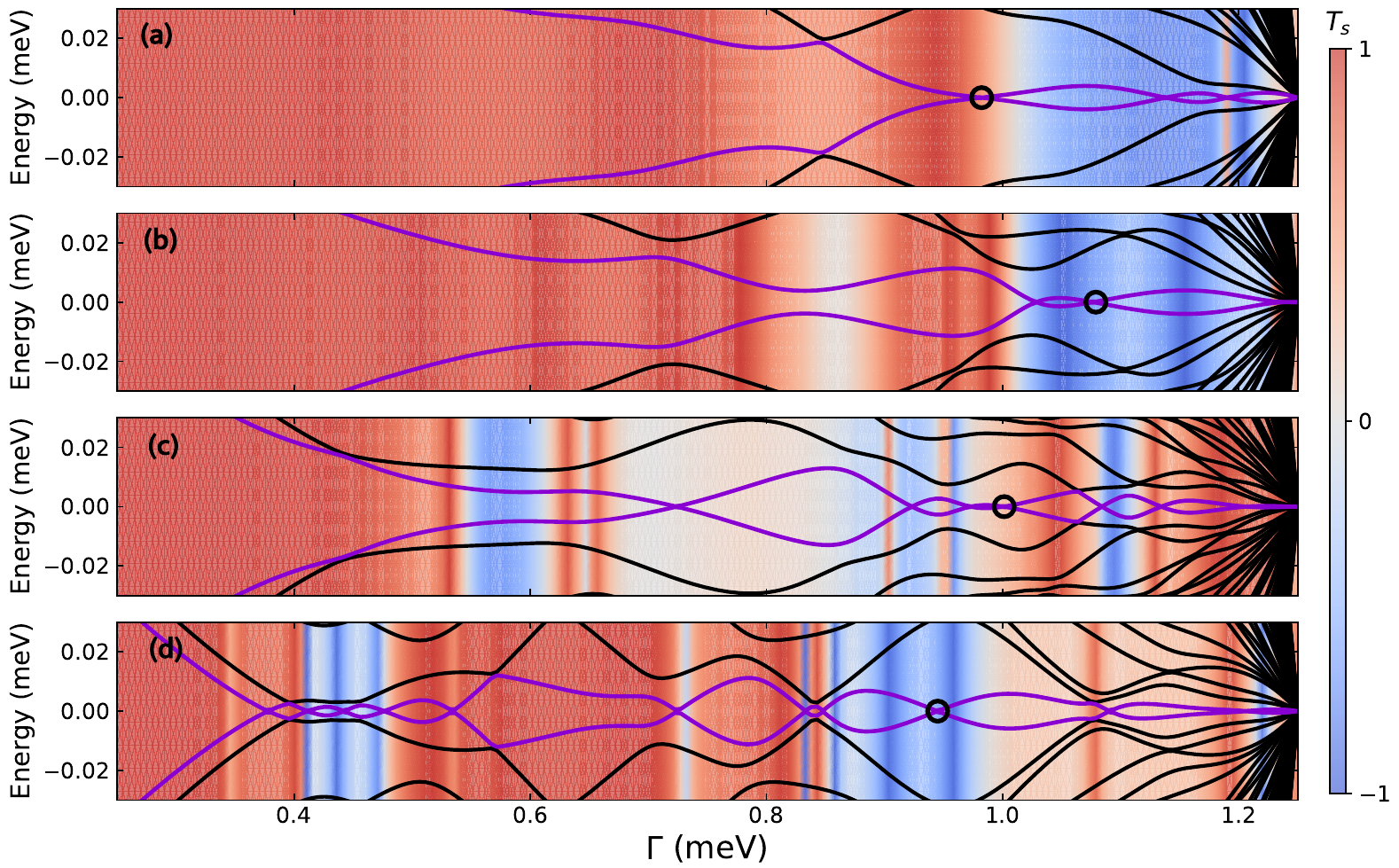}
    \caption{Low energy spectra corresponding to horizontal cuts of the invariant maps shown in Fig.~\ref{fig:INV_maps}. Panels (a)–(d) show the energy spectrum as a function of Zeeman field $\Gamma$ for fixed chemical potentials $\mu = 3.16$, $2.10$, $1.745$, and $1.055$~meV, respectively. The background color denotes the topological stability $T_s$ (refer to Eq.~\ref{eq:Topo_stability}) evaluated at the corresponding chemical potential with tunnel coupling $\lambda = 0.025$ and a dynamic gate potential chosen as $V_{\mathrm{QD}} = \mu + 0.045$~meV. Purple curves highlight the two lowest Bogoliubov-de Gennes eigenvalues that constitute the MZMs in the topological regime.}
    \label{fig:Strongly_dis_Energy_spectrum}
\end{figure}

Fig.~\ref{fig:Delta_C_L_density_plot}(c) shows the corresponding density plot of $\Delta\mathcal{L}^{-1}$ as a function of $\Phi$ and $\Gamma$ for the same $\mu$ and $V_{QD}$ values as in panel (a). Although the $h/e$ periodic structure characteristic of the topological regime is present, it becomes clearly visible only for $\Gamma \gtrsim 0.75$~meV, highlighting the sensitivity of the quantum inductance response to the gate potential. This behavior is clarified by Fig.~\ref{fig:Delta_C_L_density_plot}(d), which shows $\Delta\mathcal{L}^{-1}$ as a function of $V_{QD}$ at fixed $\mu=0.5$~meV and $\Gamma=0.66$~meV. At $V_{QD}=0.551$~meV, the system lies slightly outside the resonance window $V_{QD}\in[\mu+0.025,\mu+0.05]$~meV for the inductance response, whereas the corresponding capacitance response remains sizable over a broader interval $V_{QD}\in[\mu+0.025,\mu+0.1]$~meV. Consequently, the $h/e$ periodic structure is more readily resolved in $\Delta\mathcal{C}$ than in $\Delta\mathcal{L}^{-1}$ at lower Zeeman fields.

In contrast, near $\Gamma\sim1.0$~meV, a comparison of Fig.~\ref{fig:Delta_C_L_density_plot}(a) and panel (c) shows that the $h/e$ periodicity is captured more clearly in the quantum inductance response than in the quantum capacitance response. Fig.~\ref{fig:Delta_C_L_density_plot}(d) further shows a clear sign change in $\Delta\mathcal{L}^{-1}$ on either side of $V_{QD}\sim0.54$~meV, consistent with the resonance features observed in the clean system inductance response in Fig.~\ref{fig:Clean_sys_topo_regime}(d).

\begin{figure}[t]
    \centering
    \includegraphics[width=\linewidth]{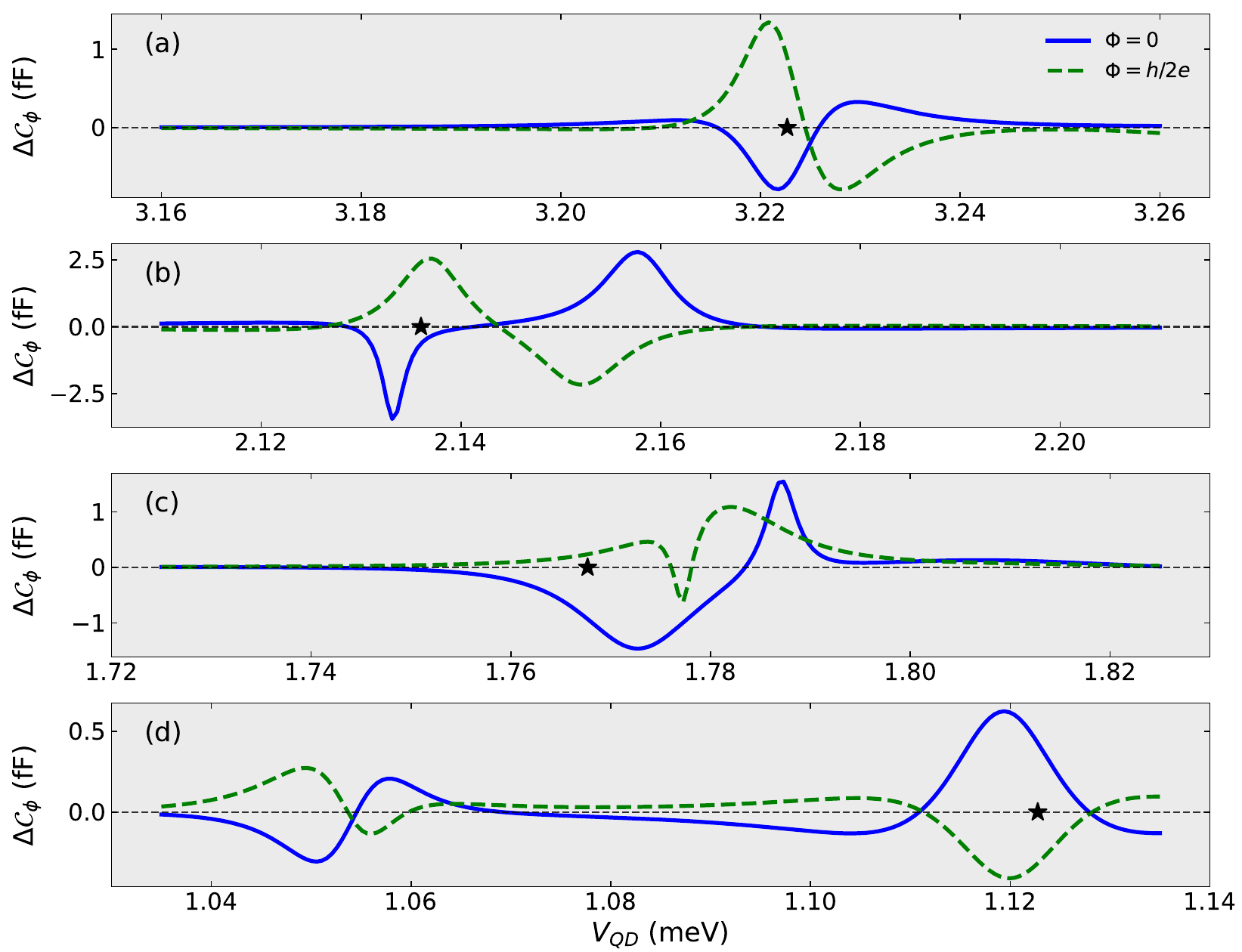}
    \caption{Parity resolved quantum capacitance difference $\Delta\mathcal{C}_\Phi$ as a function of the quantum dot gate potential $V_{\mathrm{QD}}$ for a strongly disordered nanowire–QD system. The control parameters for each panel correspond to those used in the associated wavefunction plots shown in Sec.~\ref{sec:strong_dis_main} in Fig.~\ref{fig:Strong_dis_wfs}. Solid blue and dashed green curves denote $\Phi = 0$ and $\Phi = h/2e$, respectively. Black stars mark the values of $V_{\mathrm{QD}}$ at which the flux dependent quantum capacitance and quantum inductance oscillations are evaluated in Sec.~\ref{sec:strong_dis_main} in Fig.~\ref{fig:Strong_dis_C_and_L_inv}. The parameters for the marked points are (a) $\lambda_L = 0.04$, $\lambda_R = 0.015$, $V_{\mathrm{QD}} = 3.223$~meV; (b) $\lambda_L = 0.02$, $\lambda_R = 0.025$, $V_{\mathrm{QD}} = 2.136$~meV; (c) $\lambda_L = 0.03$, $\lambda_R = 0.02$, $V_{\mathrm{QD}} = 1.768$~meV; and (d) $\lambda_L = 0.035$, $\lambda_R = 0.025$, $V_{\mathrm{QD}} = 1.123$~meV.}
    \label{fig:Strong_dis_delta_C}
\end{figure}

Taken together, these results do not indicate distinct Zeeman field regions in which one diagnostic is intrinsically superior to the other. Rather, they demonstrate that quantum capacitance and quantum inductance probe different aspects of the flux-dependent energy structure and therefore exhibit different sensitivities to the choice of gate potential. The apparent prominence of $h/e$ periodicity in one response channel over a limited Zeeman field range reflects the specific choice of $V_{QD}$, rather than a fundamental limitation of the complementary diagnostic. Importantly, the presence of characteristic topological features such as $h/e$ periodicity, $h/2e$ flux-shifted parity branches, and crossings near $\Phi\sim h/4e$ and $\Phi\sim 3h/4e$ in only one diagnostic is not sufficient to establish a fermion parity switch. In such cases, the parameter point remains inconclusive within the present framework. Instead, regions in which one response channel exhibits partial or suggestive topological features motivate further optimization of $V_{QD}$ to determine whether the complementary diagnostic can display consistent behavior at a different gate setting. Because quantum capacitance and quantum inductance probe distinct derivatives of the flux-dependent energy spectrum, their combined analysis provides a more comprehensive mapping of the parameter space. In particular, regions where one diagnostic reveals $h/e$ periodicity and parity crossings serve as indicators of potentially relevant parameter sets that may otherwise be overlooked. By retuning $V_{QD}$ and re-evaluating both $\Delta\mathcal{C}$ and $\Delta\mathcal{L}^{-1}$, one can more reliably identify parameter regimes consistent with a true fermion parity switch.

\section{Strongly disordered regime}
\label{sec:strong_dis_reg}

\begin{figure}[b]
    \centering
    \includegraphics[width=\linewidth]{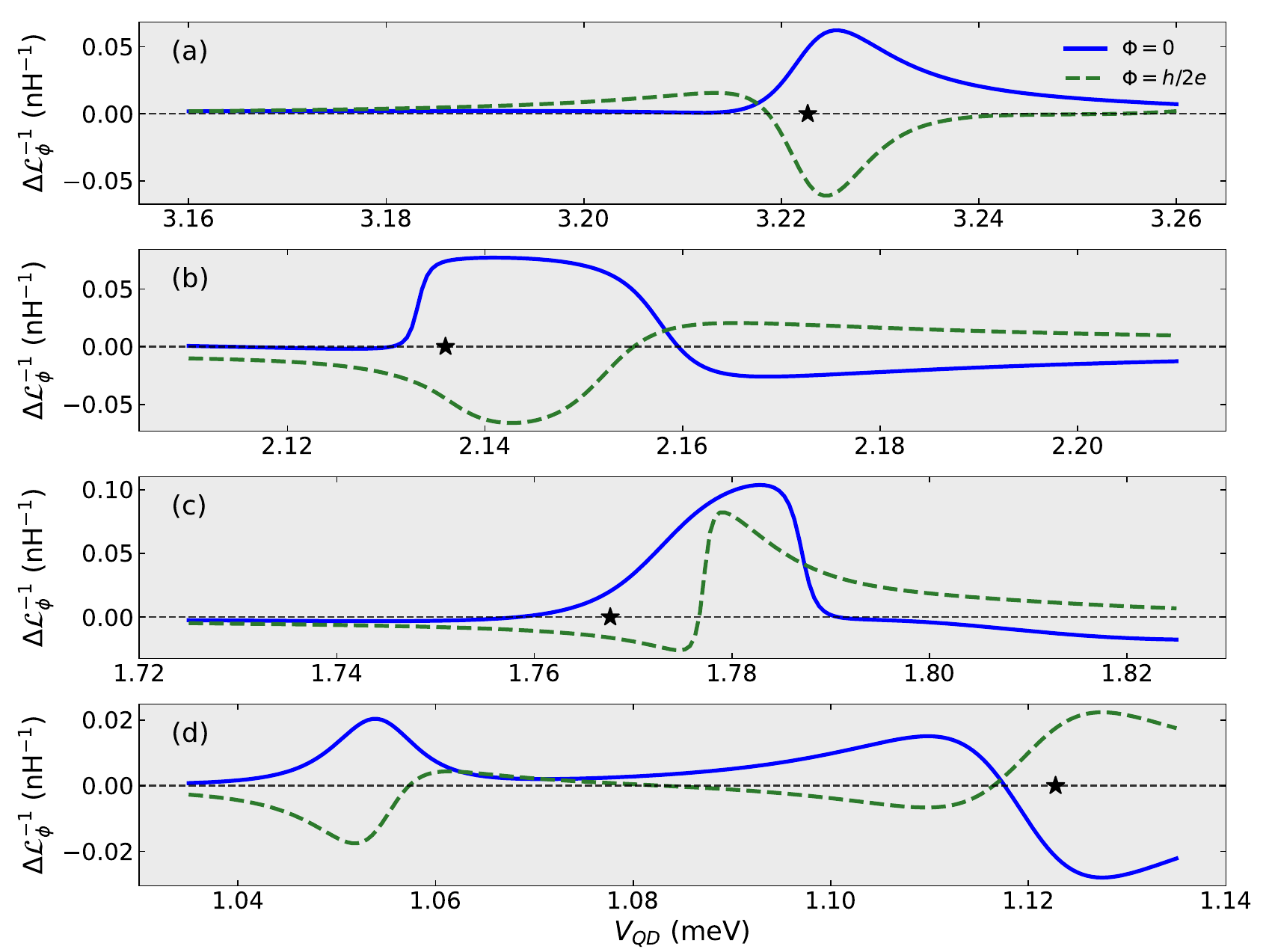}
    \caption{Parity resolved quantum inductance difference $\Delta\mathcal{L}_\Phi^{-1}$ as a function of the quantum dot gate potential $V_{\mathrm{QD}}$ for a strongly disordered nanowire–QD system. The control parameters for each panel correspond to those used in the associated wavefunction plots shown in Sec.~\ref{sec:strong_dis_main} in Fig.~\ref{fig:Strong_dis_wfs}. Solid blue and dashed green curves denote $\Phi = 0$ and $\Phi = h/2e$, respectively. Black stars mark the values of $V_{\mathrm{QD}}$ at which the flux dependent quantum capacitance and quantum inductance oscillations are evaluated. The parameters for the marked points are the same as specified in Fig.\ref{fig:Strong_dis_delta_C}.}
    \label{fig:Strong_dis_delta_L_inv}
\end{figure}

\begin{figure*}[t!]
    \centering
    \includegraphics[width=\linewidth]{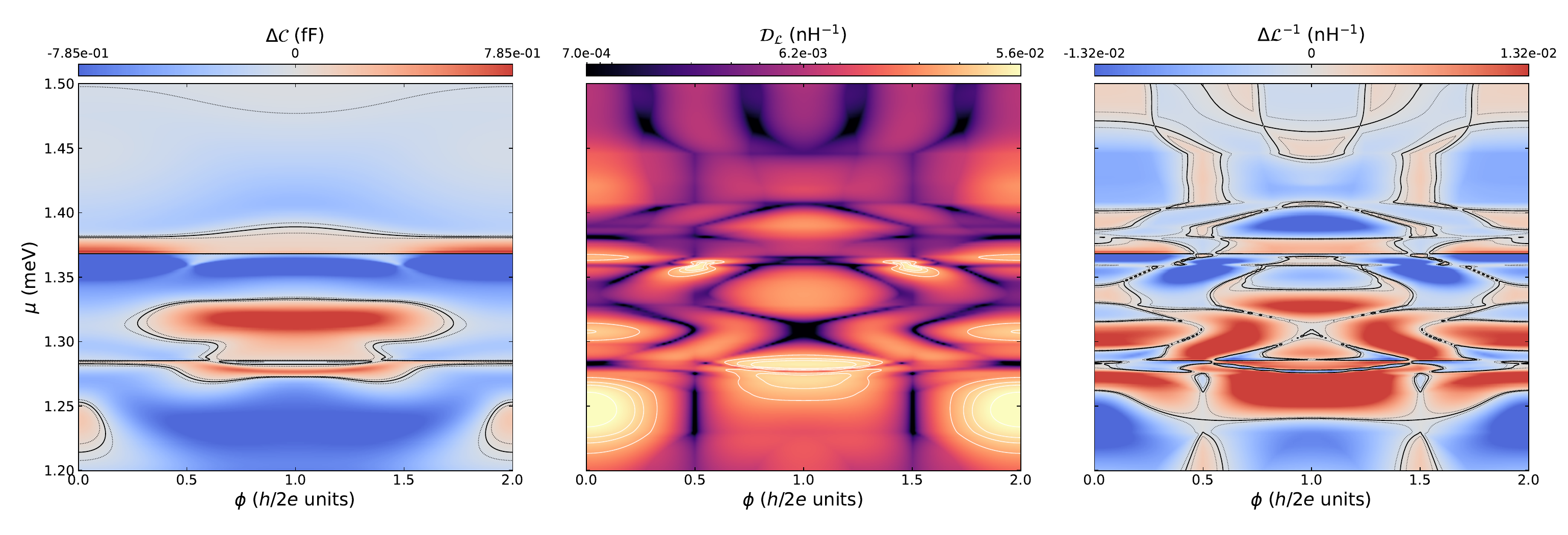}
    \caption{Three panel density plot illustrating the flux dependent parity resolved responses for a strongly disordered nanowire–QD system. The left panel shows the difference in quantum capacitance $\Delta\mathcal{C}$ between the two parity branches as a function of magnetic flux $\Phi$ and chemical potential $\mu \in [1.2,1.5]$~meV. The center panel shows the corresponding trimmed deviation quantum inductance $\mathcal{D}_{\mathcal{L}}$ (see main text), highlighting extrema in the inductance response over the same range of chemical potential values. The right panel shows the difference in quantum inductance $\Delta\mathcal{L}^{-1}$ between the parity branches as $\mu$ is varied. All data correspond to a vertical cut at Zeeman field $\Gamma = 0.66$~meV for a strongly disordered system with disorder strength $V_0 = 1.2$~meV and tunnel coupling $\lambda = 0.04$. The quantum dot gate potential is chosen dynamically as $V_{\mathrm{QD}} = \mu + 0.091$~meV.}
    \label{fig:Avoided_crossing_density_plot}
\end{figure*} 

\begin{figure}[h]
    \centering
    \includegraphics[width=\linewidth]{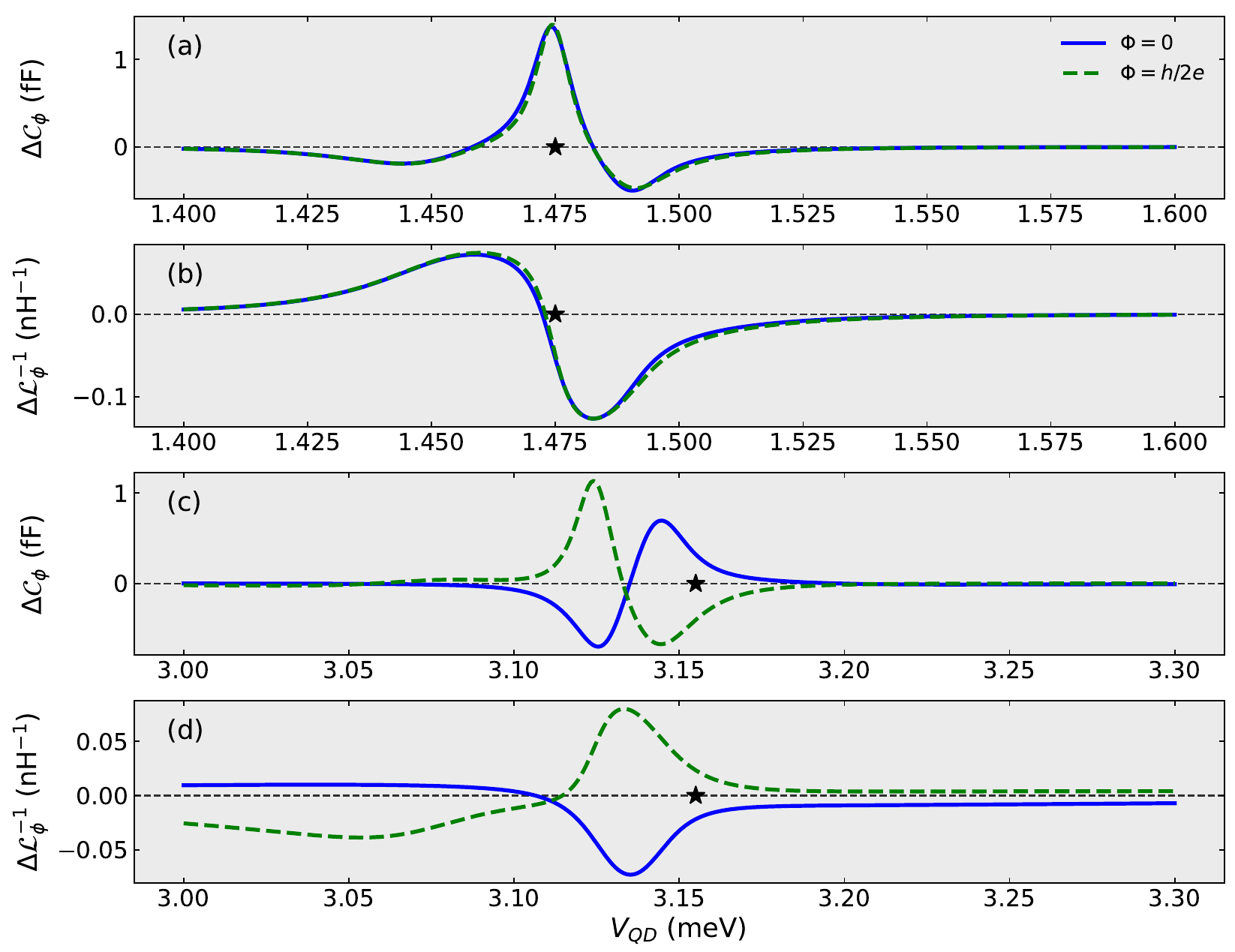}
    \caption{Flux dependent parity resolved transport differences for representative horizontal cuts of the density plots shown in Figs.~\ref{fig:Avoided_crossing_density_plot} and \ref{fig:Actual_crossing_density_plot}. Panels (a) and (b) show the quantum capacitance difference $\Delta\mathcal{C}_\Phi$ and quantum inductance difference $\Delta\mathcal{L}_\Phi^{-1}$, respectively, for a cut at chemical potential $\mu = 1.3589$~meV. Panels (c) and (d) show the corresponding $\Delta\mathcal{C}$ and $\Delta\mathcal{L}^{-1}$ for a cut at $\mu = 3.11$~meV.}
    \label{fig:Strongly_dis_actual_avoided_delta_C_L}
\end{figure}

We examine the strongly disordered regime by mapping the Pfaffian invariant\cite{PUKitaev2001} and topological stability $T_s$ (refer to Eq.~\ref{eq:Topo_stability}) over the $(\mu,\Gamma)$ parameter space. The resulting invariant maps provide a global view of the operationally trivial and topological regions in the presence of strong disorder. Representative chemical potentials are identified through horizontal cuts in $\mu$, and for each selected value the corresponding low energy spectra are analyzed as functions of the Zeeman field. This allows us to isolate parameter points displaying distinct spectral characteristics, including regimes with strongly overlapping end states as well as regimes exhibiting more separated low energy modes. For each chosen $(\mu,\Gamma)$ pair, we then compute the parity-resolved response functions $\Delta\mathcal{C}_\Phi$ and $\Delta\mathcal{L}_\Phi^{-1}$ as functions of the quantum dot gate potential $V_{QD}$. The gate potential is selected near resonance values where the transport responses are clearly resolved. These parameter sets form the basis for the flux-dependent energy, capacitance, and inductance profiles discussed in the main text in Sec.~\ref{sec:strong_dis_main}.

\begin{figure*}[t!]
    \centering
    \includegraphics[width=\linewidth]{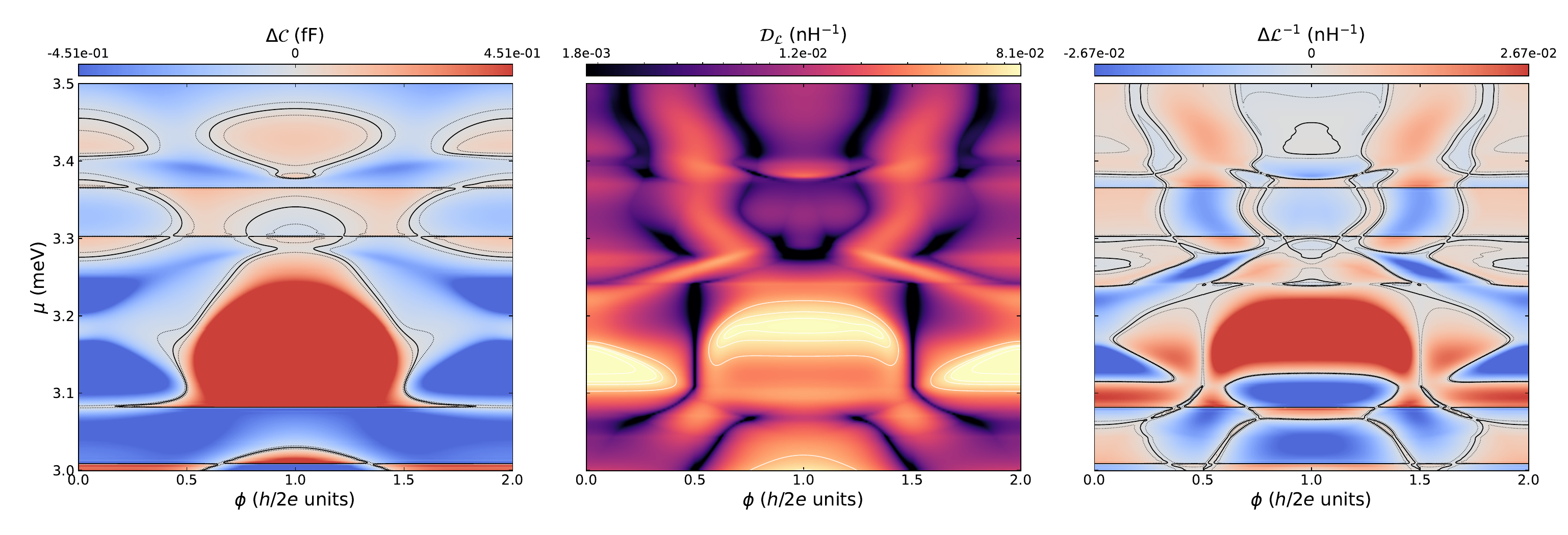}
    \caption{Three panel density plot illustrating the flux dependent parity resolved responses for a strongly disordered nanowire–QD system. The left panel shows the difference in quantum capacitance $\Delta\mathcal{C}$ between the two parity branches as a function of magnetic flux $\Phi$ and chemical potential $\mu \in [3.0,3.5]$~meV. The center panel shows the corresponding trimmed deviation quantum inductance $\mathcal{D}_{\mathcal{L}}$ (see Sec.~\ref{sec:smooth_conf_main}), highlighting extrema in the inductance response over the same range of chemical potential values. The right panel shows the difference in quantum inductance $\Delta\mathcal{L}^{-1}$ between the parity branches as $\mu$ is varied. All data correspond to a vertical cut at Zeeman field $\Gamma = 1.10$~meV for a strongly disordered system with disorder strength $V_0 = 1.2$~meV and tunnel coupling $\lambda = 0.025$. The quantum dot gate potential is chosen dynamically as $V_{\mathrm{QD}} = \mu + 0.045$~meV to inspect the topological island shown in Fig.~\ref{fig:INV_maps}.}
    \label{fig:Actual_crossing_density_plot}
\end{figure*}

Fig.~\ref{fig:INV_maps} shows the invariant maps for the strongly disordered nanowire–QD system with disorder strength $V_0=1.2$~meV. The left panel displays the Pfaffian invariant $\mathcal{Q}$ (refer to Eq.~\ref{eq:pfaff}) evaluated for the nanowire without the quantum dot, while the right panel shows the corresponding topological stability $T_s$ (refer to Eq.~\ref{eq:Topo_stability}) computed for the full nanowire–QD system. The chemical potential $\mu$ and Zeeman field $\Gamma$ span the ranges are indicated in the caption of Fig.~\ref{fig:INV_maps}, with the gate potential chosen dynamically as $V_{QD}=\mu+0.045$~meV and the tunnel coupling fixed at $\lambda=0.025$. As seen in Fig.~\ref{fig:INV_maps}, the dominant topological islands are shifted toward larger chemical potential and Zeeman field values relative to the weakly disordered case, consistent with the effects of strong disorder. The Pfaffian invariant\cite{PUKitaev2001} and $T_s$\cite{PRBSau2025} show good qualitative agreement in identifying the principal topological regions, reflecting the weak coupling regime of the nanowire–QD interface established in Fig.~\ref{fig:Ts_sweep_lambda}. In this limit, the quantum dot primarily serves as a probe and does not significantly alter the underlying topological structure of the nanowire.

To identify representative parameter points for detailed transport analysis, we consider horizontal cuts of the invariant maps at fixed chemical potentials. The selected values are $\mu=1.055$~meV, intersecting the lower boundary of a large topological island near $\Gamma\sim0.9$~meV; $\mu=1.745$~meV, traversing a fragmented region; $\mu=2.100$~meV, intersecting a large topological island at $\Gamma\gtrsim1.0$~meV; and $\mu=3.16$~meV, intersecting another extended topological region at $\Gamma\gtrsim0.94$~meV. The corresponding low energy spectra are shown in Fig.~\ref{fig:Strongly_dis_Energy_spectrum}, where the two lowest Bogoliubov–de Gennes eigenvalues are highlighted in purple. These spectra are computed without coupling to the quantum dot in order to clearly identify the emergence of low energy modes in the nanowire.

Specific Zeeman field values, marked by black circles in Fig.~\ref{fig:Strongly_dis_Energy_spectrum}, are selected to sample both trivial and topological regions as indicated by $T_s$, while also satisfying the topological gap criterion. For each selected $(\mu,\Gamma)$ pair, the tunnel couplings $\lambda_L$ and $\lambda_R$ are chosen to optimize overlap between the quantum dot and the corresponding low energy wavefunctions (shown in Fig.~\ref{fig:Strong_dis_wfs} in Sec.~\ref{sec:strong_dis_main}), ensuring effective probing of the relevant states.

With the couplings fixed, we next examine the dependence of the parity-resolved transport responses on the gate potential. Figs.~\ref{fig:Strong_dis_delta_C} and \ref{fig:Strong_dis_delta_L_inv} show the capacitance difference $\Delta\mathcal{C}_\Phi$ and quantum inductance difference $\Delta\mathcal{L}_\Phi^{-1}$, respectively, as functions of $V_{QD}$ for the selected parameter sets. In Fig.~\ref{fig:Strong_dis_delta_C}, the capacitance difference reaches values of order $1$~fF and up to $2.5$~fF in panel (b), with narrow resonance lobes near the chemical potential determined by the chosen tunnel couplings. Fig.~\ref{fig:Strong_dis_delta_L_inv} shows that the corresponding inductance differences attain magnitudes of $0.05$–$0.10$~nH$^{-1}$ across panels (a)–(d), indicating that the inductance response remains sizable even in the presence of strong disorder.

For each parameter set, the gate potentials marked by black stars in Figs.~\ref{fig:Strong_dis_delta_C} and \ref{fig:Strong_dis_delta_L_inv} are selected near resonance-lobe regions. These values are used in the main text (see Sec.~\ref{sec:strong_dis_main}) to compute the flux-dependent energy spectra and transport responses.

Although strongly disordered systems generally make it difficult to anticipate ambiguous crossing scenarios a priori, we do identify an illustrative example in Fig.~\ref{fig:Avoided_crossing_density_plot} (see Sec.~\ref{sec:smooth_conf_main} of the main text for definitions of the quantities shown in the density plots) and contrast it with an actual crossing scenario in Fig.~\ref{fig:Actual_crossing_density_plot}. To examine in detail a parameter region where ambiguous capacitance features arise, we consider a vertical cut through the strongly disordered phase diagram at fixed low Zeeman field $\Gamma=0.66$~meV. The chemical potential is varied over the interval $\mu\in[1.2,1.5]$~meV, with tunnel coupling $\lambda=0.04$ and a dynamically tuned gate potential $V_{QD}=\mu+0.091$~meV. The resulting three panel density map, shown in Fig.~\ref{fig:Avoided_crossing_density_plot}, displays the parity-resolved capacitance difference $\Delta\mathcal{C}$, the trimmed deviation quantum inductance $\mathcal{D}_{\mathcal{L}}$, and the inductance difference $\Delta\mathcal{L}^{-1}$ as functions of magnetic flux and chemical potential. The $\Delta\mathcal{C}$ response (in left panel of Fig.~\ref{fig:Avoided_crossing_density_plot}) near $\mu\sim1.359$~meV exhibits avoided crossing features (indicated by the light blue or white regions) around $\Phi\sim h/4e$ and $\Phi\sim3h/4e$ that could be misinterpreted as crossings. The corresponding quantum inductance diagnostics, however, show pronounced peaks in the trimmed deviation response (in the center panel of Fig.~\ref{fig:Avoided_crossing_density_plot}) and no sign changes in $\Delta\mathcal{L}^{-1}$ (in the right panel of Fig.~\ref{fig:Avoided_crossing_density_plot}), confirming that these capacitance features originate from an avoided crossing rather than a true fermion parity switch.

For comparison, we consider a vertical cut through a large topological island identified in Fig.~\ref{fig:INV_maps}, at fixed Zeeman field $\Gamma=1.10$~meV. In this case the chemical potential is varied over $\mu\in[3.0,3.5]$~meV with tunnel coupling $\lambda=0.025$ and a dynamically chosen gate potential $V_{QD}=\mu+0.045$~meV. The corresponding three panel density map in Fig.~\ref{fig:Actual_crossing_density_plot} shows $\Delta\mathcal{C}$, $\mathcal{D}_{\mathcal{L}}$, and $\Delta\mathcal{L}^{-1}$ over this parameter range. The left panel shows $h/2e$ flux shifted extended regions of sign change in $\Delta\mathcal{C}$ over $\mu\approx3.09$–$3.25$~meV. The center panel displays the trimmed deviation inductance $\mathcal{D}_{\mathcal{L}}$, which remains close to its baseline near $\Phi\sim h/4e$ and $\Phi\sim3h/4e$, indicating the absence of extrema. The right panel shows $h/e$ periodic $\Delta\mathcal{L}^{-1}$ exhibiting parity crossings whose flux locations coincide with that of the $\Delta\mathcal{C}$ response thus indicating a true crossing. An illustrative point in this region is $\mu=3.11$~meV.

To make the distinction between avoided and true crossing regimes more explicit, we extract representative horizontal cuts from the density plots shown in Figs.~\ref{fig:Avoided_crossing_density_plot} and \ref{fig:Actual_crossing_density_plot}. The resulting flux-dependent parity-resolved transport differences are shown in Fig.~\ref{fig:Strongly_dis_actual_avoided_delta_C_L}. Panels (a) and (b) display the quantum capacitance difference $\Delta\mathcal{C}_\Phi$ and quantum inductance difference $\Delta\mathcal{L}^{-1}_\Phi$, respectively, for a cut at $\mu=1.3589$~meV within the avoided crossing regime. Panels (c) and (d) show the corresponding quantities for $\mu=3.11$~meV within the genuine-crossing regime.

For $\mu=1.3589$~meV, neither $\Delta\mathcal{C}$ nor $\Delta\mathcal{L}^{-1}$ changes sign as a function of flux, while the trimmed deviation inductance (see Fig.~\ref{fig:Avoided_crossing_density_plot}) exhibits pronounced peaks near $\Phi\sim h/4e$ and $\Phi\sim3h/4e$, consistent with an avoided crossing. In contrast, for $\mu=3.11$~meV, both $\Delta\mathcal{C}$ and $\Delta\mathcal{L}^{-1}$ change sign near $\Phi\sim h/4e$ and $\sim3h/4e$, and no extrema appear in the trimmed deviation response (see Fig.~\ref{fig:Actual_crossing_density_plot}), consistent with a true level crossing.

Together, these results document the procedure used to identify representative trivial, avoided crossing, and genuine crossing regimes within the strongly disordered phase diagram and provide the supporting diagnostics for the flux-dependent analyses presented in the main text.

\newpage
\bibliography{bibliography}

\end{document}